\documentclass[12pt,preprint]{aastex}

\usepackage{natbib}
\usepackage{graphicx}

\bibpunct{(}{)}{;}{a}{}{,}

\begin{document}                          

\title{Star Formation in Isolated Disk Galaxies. \\ 
I. Models and Characteristics of Nonlinear Gravitational Collapse}

\author{Yuexing Li\altaffilmark{1,2}, Mordecai-Mark Mac Low\altaffilmark{2,1}
and Ralf S. Klessen\altaffilmark{3}} 

\affil{$^{1}$Department of Astronomy, Columbia University, New York,
NY 10027, USA}
\affil{$^{2}$Department of Astrophysics, American Museum of Natural
History, 79th Street at Central Park West, New York, NY 10024-5192, USA}
\affil{$^{3}$Astrophysikalisches Institut Potsdam, An der Sternwarte
16, D-14482 Potsdam, Germany} 
\email{yxli@astro.columbia.edu, mordecai@amnh.org, rklessen@aip.de} 

\begin{abstract}
  
  We model gravitational collapse leading to
  star formation in a wide range of isolated disk galaxies
  using a three-dimensional, smoothed particle hydrodynamics code.
  The model galaxies include a dark matter halo and a disk of stars
  and isothermal gas. Absorbing sink particles are used to directly measure
  the mass of gravitationally collapsing gas. They reach masses characteristic
  of stellar clusters. In this paper, we describe our galaxy models and
  numerical methods, followed by an investigation of the gravitational
  instability in these galaxies.  Gravitational collapse forms star
  clusters with correlated positions and ages, as observed, for example, in
  the Large Magellanic Cloud.  Gravitational instability alone acting in
  unperturbed galaxies appears sufficient to produce flocculent spiral
  arms, though not more organized patterns. Unstable galaxies show
  collapse in thin layers in the galactic plane; associated dust will
  form thin dust lanes in those galaxies, in agreement with
  observations.  We find an exponential relationship between the
  global collapse timescale and the minimum value in a galaxy of
  the Toomre instability parameter for a combination of stars and gas
  $Q_{\rm sg}$.  Furthermore, collapse occurs only in regions with
  $Q_{\rm sg} < 1.6$. Our results suggest that vigorous starbursts
  occur where $Q_{\rm sg} \ll 1$, while slow star formation takes
  place at higher values of $Q_{\rm sg}$ below 1.6. Massive, or
  gas-rich, galaxy has low initial $Q_{\rm sg}$, giving high star
  formation rate, while low-mass, or gas-poor galaxy has high
  initial $Q_{\rm sg}$, giving low star formation rate.

\end{abstract}


\section{INTRODUCTION}
Stars are the fundamental building blocks of galaxies. They form at
widely varying rates in different galaxies \citep{kennicutt98a}. The
mechanisms that control star formation in galaxies continue to be
debated \citep[e.g.,][]{shu87, elmegreen02, larson03, mk04}.
Gravitational collapse is opposed by gas pressure, supersonic
turbulence, magnetic fields, and rotational shear.  Gas pressure in
turn is regulated by radiative cooling and stellar and turbulent
heating; turbulence is driven by supernova explosions, spiral density
waves, and magnetorotational instabilities; and magnetic fields are
generated by galactic dynamos.

Despite this complexity, disk galaxies follow two simple empirical
laws. First, the Schmidt law, which relates the global star formation rate to
the total gas surface density (\citealt{schmidt59, kennicutt89,
kennicutt98b}). Second, stars form far more vigorously above a critical gas
surface density threshold \citep{martin01}.  These two observations appear to
both be explainable by the action of large-scale gravitational
instability \citep*{li05}. The action of gravitational instability is also
suggested by the observation that thin dust lanes in galaxies only form in
gravitationally unstable regions (\citealt*{dalcanton04}).  

How does gravitational instability control star formation in galaxies?
The nonlinear development of gravitational instability requires
numerical modeling to understand.  There have been many simulations of
disk galaxies, including isolated galaxies, galaxy mergers, and
galaxies in a cosmological context with different assumptions about
the nature and distribution of dark matter (e.g., \citealt*{katz91,
navarro91, katz92, friedli93, friedli94, steinmetz94, mihos94,
nfw95, barnes96, sommer-larsen99, steinmetz99, springel00,
sommer-larsen01, barnes02, sommer-larsen03, springel03,
governato04}; see \citealt{robertson04} for a recent review as well
as further results). However, in these simulations, gravitational
collapse and star formation are either not resolved, or are followed
with empirical recipes tuned to reproduce the observations \textit{a
  priori}. The mechanisms that produce the empirical star formation
laws have remained uncertain. Recent cosmological simulations by
\citet{kravtsov03} show that the global Schmidt law occurs naturally
in self-consistent galaxy models, more or less independent of the
strength of feedback, and that the Schmidt law is related to the
overall density distribution of the interstellar medium. However, the
strength of gravitational instability was not measured in the galaxies
formed in Kravtsov's models, so the relationship between gravitational
instability and star formation was not made explicit.

Gravitational stability against local axisymmetric perturbations in a
thin, differentially rotating disk of collisionless particles can be
shown with a linear perturbation analysis \citep{toomre64} to require that 
(\citealt{toomre64}): 
\begin{equation}
\label{eq_toomre}
Q = \frac{\kappa \sigma}{3.36 G \Sigma} > 1,
\end{equation}
where $\kappa$, $\sigma$ and $\Sigma$ are the epicyclic frequency, velocity
dispersion, and surface density of the disk, respectively. In the case of gas,
the constant 3.36 is replaced by $\pi$ \citep{safronov60, goldreich65}. These
early studies focused on a single-component of the disk (see also
\citealt{quirk72}). \citet{jog84} extended this to a two-fluid treatment in
order to introduce different velocity dispersions for gas and stars, but used
the collisional formalism for both. \citet{rafikov01} derives a stability
criterion treating stars as a collisionless population and the gas as
collisional. He shows that the dispersion relation in this case differs from
that for single component or two-fluid models. The combined stability
criterion for stars and gas is $Q_{\rm sg} > 1$. \citet{rafikov01} shows that
\begin{equation}
\label{eq_qsg}
\frac{1}{Q_{\rm {sg}}} = \frac{2}{Q_{\rm s}}\frac{1}{q}\left[1 - e^{-q^2}
I_0(q^2)\right] + \frac{2}{Q_{\rm g}}\xi\frac{q}{1+q^2\xi^2} \,,
\end{equation}
where the parameters for stars and gas individually are
\begin{equation}
Q_{\rm s} = \frac{\kappa \sigma_{\rm s}}{\pi G \Sigma_{\rm s}}  \mbox{
  and } 
Q_{\rm g} = \frac{\kappa      c_{\rm g}}{\pi G \Sigma_{\rm g}} \,,
\end{equation}
the quantities $q = k\sigma_{\rm s}/\kappa$ and $\xi=c_{\rm
  g}/\sigma_{\rm s}$, $\sigma_{\rm s}$ is the radial velocity
dispersion of stars, $c_{\rm g}$ is the sound speed of the gas,
$\Sigma_{\rm s}$ and $\Sigma_{\rm g}$ are the surface densities for stars
and gas, $I_0$ is the Bessel function of order 0, and $k$ is the
dimensionless wavenumber of the perturbation. (Note that for
notational convenience \citet{rafikov01} defined $Q_{\rm s} = 3.36/\pi
Q$ for the stars.)

In order to investigate gravitational instability in disk galaxies and
consequent star formation, we model isolated galaxies with a wide range of
masses. Each galaxy is composed of a dark matter halo, and a disk of stars,
with different initial fractions of isothermal gas. In this paper, we present
the galaxy models and computational methods, and the star formation morphology
associated with gravitational instability. Preliminary results on the global
Schmidt law and star formation thresholds were presented by \citet{li05}.  In
future work we will give more detailed results on them, as well as discussing
local Schmidt laws and cluster mass spectra. In \S~\ref{sec_com} we describe
our computational methods, galaxy models and parameters. In \S~\ref{sec_qsg}
we investigate the gravitational instability in these galaxies. Results are
then presented for collapse morphology (\S~\ref{sec_sf}), time evolution
(\S~\ref{sec_tsf}), spatial distribution of gravitational collapse and
star formation (\S~\ref{sec_dist}), and correlations between distance and age
separation (\S~\ref{sec_corr}).  In \S~\ref{sec_dis} we discuss the
assumptions used in the simulations and the observational implications
of our work. We summarize in \S~\ref{sec_sum}.

\section{COMPUTATIONAL METHODS}
\label{sec_com}
\subsection{Code}
\label{subsec_code}
In our models, we follow collisionless stars and dark matter with a
tree-based N-body algorithm, and collisional, dissipative gas dynamics with
smoothed particle hydrodynamics (SPH). The Lagrangian SPH algorithm follows 
the local mass density \citep{monaghan92}. Fluid properties are sampled
by an ensemble of particles, and flow quantities are obtained by averaging
over neighboring particles with a smoothing kernel. Its Lagrangian nature
allows a locally changing resolution to follow the mass density as particles
freely move. SPH can resolve large density contrasts in star-forming regions.  
We use the publicly available, three-dimensional, parallel, N-body/SPH
code GADGET v1.1 (\citealt*{springel01}). This code follows gravity
using the Barnes-Hut \citeyear{barnes86} tree algorithm. It uses
individual and adaptive timesteps for all particles, and combines this
with a scheme for dynamical tree updates. It is parallelized using the
orthogonal, recursive, bisection algorithm \citep{dubinski96} for domain
decomposition. It has been successfully used in simulations ranging
from grand cosmological scale to galactic scale. GADGET was originally
designed to model interacting galaxies (its name is a rough acronym
for Galaxies with Dark matter and Gas intEracT; \citealt{springel01}),
and it has excellent properties for this problem. 

\subsection{Sink Particle Implementation}

\label{subsec_sink}
We have modified GADGET v1.1 to include absorbing sink particles that
replace high density gas regions \citet{jappsen04}. In SPH
simulations, the evolution of dense, highly resolved regions can take
prohibitively long due to the very small timesteps required by the
Courant condition. Sink particles are designed to circumvent this
problem \citep*{bate95}. The implementation of sink particles in the
code includes two parts: creation and accretion. We follow the
procedures of \citep{bate95}. A region of gravitationally bound,
converging gas above some critical density is replaced by a single,
massive particle that interacts gravitationally and inherits the mass,
and linear and angular momentum of the gas.  It accretes surrounding
gas particles that pass within its accretion radius and are
gravitationally bound.  The accretion radius is chosen to be
comparable to the Jeans radius of cores at the critical density and
remains fixed throughout.  Boundary conditions at the surface of the
sink particle are computed by extrapolating hydrodynamic properties to
get the correct pressure gradients.

In additon to the above procedures used in a serial code as in
\citet{bate95}, our implementation of sink particles in a parallel
code takes into account the domain boundaries, as described in detail
in \citet{jappsen04}. The parallel version of GADGET distributes the
SPH particles onto the individual processors using a spatial domain
decomposition, so that each processor hosts a rectilinear 
piece of the computational volume. If the position of a sink particle
is near the boundary of a domain, the accretion radius overlaps with
domains on other processors. Following the paradigm for SPH particles
in GADGET v1.1, the sink particle data is then broadcast to all
processors. Every processor searches for gas particles in its domain
that lie within the accretion radius of the sink particle and processes them,
passing data back to the processor holding the sink. 

We have done a standard test of the collapse of a rotating, isothermal sphere 
with a small $m = 2$ perturbation \citep{boss79, bate95}. As gravitational
collapse proceeds, a rotationally supported, high-density bar forms, embedded
in a disk-like structure. The two ends of the bar become gravitationally
unstable, resulting in the formation of a binary system. Sink particles form
as the binary cores collapse. We see no further subfragmentation (see also
\citealt{truelove97}). 

This test supports the validity of our implementation of the sink
particles. However, there are small deviations of the sink particle
parameters such as creation time, mass, and positions as we vary the
number of the processors. The unmodified GADGET v1.1 also shows such small
deviations as processor number is changed. These variations are due to the
differences in the extent of the domain on each processor. When the force on a
particular particle is computed, the force exerted by distant groups of
particles is approximated by their lowest multipole moments.  Each processor
constructs its own Barnes-Hut \citeyear{barnes86} tree, though, so
small differences in the tree walks result in small differences in the
force calculations.  Also, the parallel code allows a fluctuation of
the number of neighbors for the smoothing kernel from 40 to 45, which
also contributes to the variations. Nevertheless, \citet{jappsen04}
show that these differences are only at the 0.1\% level, and the
overall results are not influenced by change in processor number.

Each sink particle defines a control volume with a
fixed radius of 50 pc.  They are created when the gas density exceeds
1000 cm$^{-3}$ in a converging flow, and the gas is gravitationally
bound.  The choice of critical density was ultimately determined by
the limitations of available computational resources.  However, we do
note that in our isothermal models, the pressure at the critical
density $P/k \simeq 10^7$~K. Radiative cooling is certain at these
temperatures and pressures, If we allow for cooling down to the
10--100~K temperatures typical of molecular clouds, this pressure corresponds
to densities of $10^5$--$10^6$~cm$^{-3}$. Stars form rapidly from
gravitationally bound gas at this density. Use of sink particles allows
us to directly follow the gravitationally collapsing gas and
accurately measure the collapsed mass available for star formation.

\subsection{Sink Particle Interpretation}

In the present simulations, 
we interpret the creation of sink particles as representing the formation of 
molecular gas and stellar clusters.  
We neglect recycling of gas from molecular clouds back into the warm
atomic and ionized medium represented by SPH particles in our
simulation, however, an important limitation that will have to be
addressed in future work.

To quantify star formation, we assume that individual sinks represent dense
molecular clouds that form stars at some efficiency. Observations by
\citet{rownd99} suggest that the {\em local} star formation efficiency (SFE)
in molecular clouds remains roughly constant. \citet{kennicutt98b} found SFE
of 30\% for starburst galaxies that \citet{wong02} showed are dominated by
molecular gas.  We take this global SFE to be a measure of the SFE in
individual molecular clouds as it appears most gas in these galaxies has
already become molecular. We therefore adopt a fixed local SFE of $\epsilon$ =
30\% to convert the mass of sinks to stars, while making the simple
approximation that the remaining 70\% of the sink particle mass remains in
molecular form.  

The long-term evolution of the mass included in sink particles in our
simulations is not simulated.  The gas that does not form stars will
probably not remain in molecular form for long, as molecular cloud
lifetimes may be as short as 10~Myr \citep*{fukui99,bhv99}.  However,
so long as that gas remains in a region subject to gravitational
instability, it will be subject to further episodes of collapse and
star formation, particularly after the short-lived massive stars
formed in each episode cease producing ionizing radiation and stellar
winds.  The star clusters that form in sink particles, in turn, will
lose stars by tidal stripping and similar effects
(e.g. \citealt{fall01}). Both of these effects should be treated in future 
work.

\subsection{Galaxy Models}
\label{subsec_gal}

Our galaxy models consist of a dark matter halo, and an initially exponential
disk composed of stars and isothermal gas. In these models we include no
bulges. The galaxy structure follows the analytical work by \citet*{mo98}  
as implemented numerically by \citet{springel99} and \citet{springel00}.  The
galaxy models and parameters are listed in Table 1. We characterize our models
by $V_{200}$, the rotational velocity at the virial radius $R_{200}$ where the
overdensity above the cosmic average is 200.  We have run models of galaxies
with rotational velocity $V_{200}$ = 50--220~km~s$^{-1}$, with gas fractions
of 20--90\% of the disk mass for each velocity.   

A Milky Way-size galaxy has $V_{200}$ = 160~km~s$^{-1}$ at a virial radius
$R_{200} \approx$ 230 kpc, disk mass fraction $m_{\rm d} \approx 6\%$ , and
gas fraction $f_{\rm g} \approx 7\%$ of the disk mass (\citealt*{dubinski96,
dehnen98, springel00, dubinski05}). The Milky Way also has a bulge and is 
interacting with the Magellanic Clouds, so none of the simulations presented
here exactly reproduces it. Instead, our goal is to do a parameter study
to investigate the dominant physics that controls gravitational collapse and
star formation in galaxies.

We adopt standard values for the Hubble constant $H_0 = 70$ km s$^{-1}$
Mpc$^{-1}$, the halo concentration parameter $c = 5$ and spin
parameter $\lambda = 0.05$.  The spin parameter used is a typical one
for galaxies subject to the tidal forces of the cosmological
background \citep{springel00}. \citet{reed03} suggest a wide range of
$c$ for galaxy-size halos. However, this parameter is based on a simple model
of the halo formation time \citep*{nfw97}, which has a poorly known
distribution \citep{mo98}. \citet{springel99} suggest that $c = 5$ is
theoretically expected for flat, low-density universes. In order to study the
effects of the velocity dispersion of the gas in the galaxy, we choose two
values of the effective sound speed, one suggested by observations of $c_s =
6$ km s$^{-1}$ (low temperature models), and a higher value $c_s = 15$ km
s$^{-1}$ (high temperature models). 

\subsection{Computational Parameters}
\label{subsec_num}

Resolution is important in numerical simulations. In N-body/SPH
simulations, three numerical criteria are known that must be satisfied
to prevent numerical artifacts and produce results with physical
meaning: \textit{(i) Jeans criterion.} In a self-gravitating medium,
the local thermal Jeans mass at each density must be resolved
by the SPH kernel (\citealt{bate97, whitworth98}). \citet{truelove97}
enunciated the equivalent condition for grid-based algorithms.
\textit{(ii) Gravity-hydro balance criterion.} In SPH simulations,
gravity and hydrodynamic forces must have equal resolution, which
requires that the gravitational softening length must not fall below
the minimum hydrodynamic smoothing length $h$ \citep{bate97}.
\textit{(iii) Equipartition criterion.} If the masses of two
populations of particles are unequal, the more massive particles will
heat up the less massive ones by two-body interactions
(\citealt{steinmetz97}). To avoid the two-body heating problem, every
particle type must have mass $m_{\rm p}$ and gravitational softening
length $\epsilon_{\rm p}$ related by $(Gm_{\rm p}/\epsilon_{\rm
  p})^{1/2} < V_{\rm th}$, where $V_{\rm th}$ is the thermal velocity
dispersion.

We set up our simulations to satisfy the above three numerical
criteria, with the computational parameters listed in Table~2 
and Table~3. We choose the particle number for each model such that they not
only satisfy the criteria, but also all runs have at least $10^6$
total particles. The gas, halo and stellar disk particles are distributed with
number ratio $N_{\rm g}$ : $N_{\rm h}$ : $N_{\rm d}$ = 5 : 3 : 2. The
gravitational softening lengths of the halo $\epsilon_{\rm h} = 40$~pc
and disk $\epsilon_{\rm d}= 10$~pc, while the softening length of the
gas $\epsilon_{\rm g}$ is given in Table~2 for each high $T$ model and in
Table~3 for the low $T$ case studies that we ran. The kernel number is the
number of kernel masses used to resolve a Jeans mass $M_{\rm J}$, $N_{\rm k} =
  M_{\rm J}/(N_{\rm  ngb}m_{\rm g})$, where $m_{\rm g}$ is again the gas
particle mass, and $N_{\rm ngb} \sim 40$ is the number of neighbors in a
smoothing kernel. The minimum spatial and mass resolutions in the gas are
given by the gravitational softening length $\epsilon_{\rm g}$ and twice the
kernel mass ($\sim 80 m_g$). The most unstable value of the Toomre criterion
for gravitational instability that couples stars and gas, $Q_{\rm sg}$
is calculated according to equation (\ref{eq_qsg}), using both the low
$T$ and high $T$ sound speeds. This minimum value is derived using
the wavenumber $k$ that gives the lowest $Q_{sg}$ at each radius, and then
taking the overall minimum. 

\subsection{Resolution Study}
\label{subsec_res}

Resolution of the Jeans length is vital for simulations of gravitational
collapse \citep{truelove97,bate97}. Exactly how well the Jeans length must be
resolved remains a point of controversy, however. We have found the
\citet{bate97} criterion sufficient in a study of gravoturbulent fragmentation
in a uniform medium with a resolution study using up to $10^7$ SPH particles
\citep{jappsen04}. The question for SPH models is the value of the kernel
number $N_{\rm k}$ sufficient to resolve collapse. \citet{bate97} suggested
that $N_{\rm k} \ge 2$ is sufficient. 

In order to directly test the resolution needed for our problem, we
have carried out a resolution study on low T models G100-1 and G220-1,
with three resolution levels having total particle numbers $N_{\rm
  tot} = 10^5$ (R1), $8\times 10^5$ (R8) and $6.4\times 10^6$ (R64).
The particle numbers are chosen such that the maximum {\em spatial}
resolution increases by a factor of two between each pair of runs. The
parameters of the resolution study are listed in Table~4. In the discussion of
our results, we also describe the results of this resolution study as
applicable. 

\section{GRAVITATIONAL INSTABILITY IN DISK GALAXIES}
\label{sec_qsg}
\begin{figure}
\begin{center}
\includegraphics[width=3.0in]{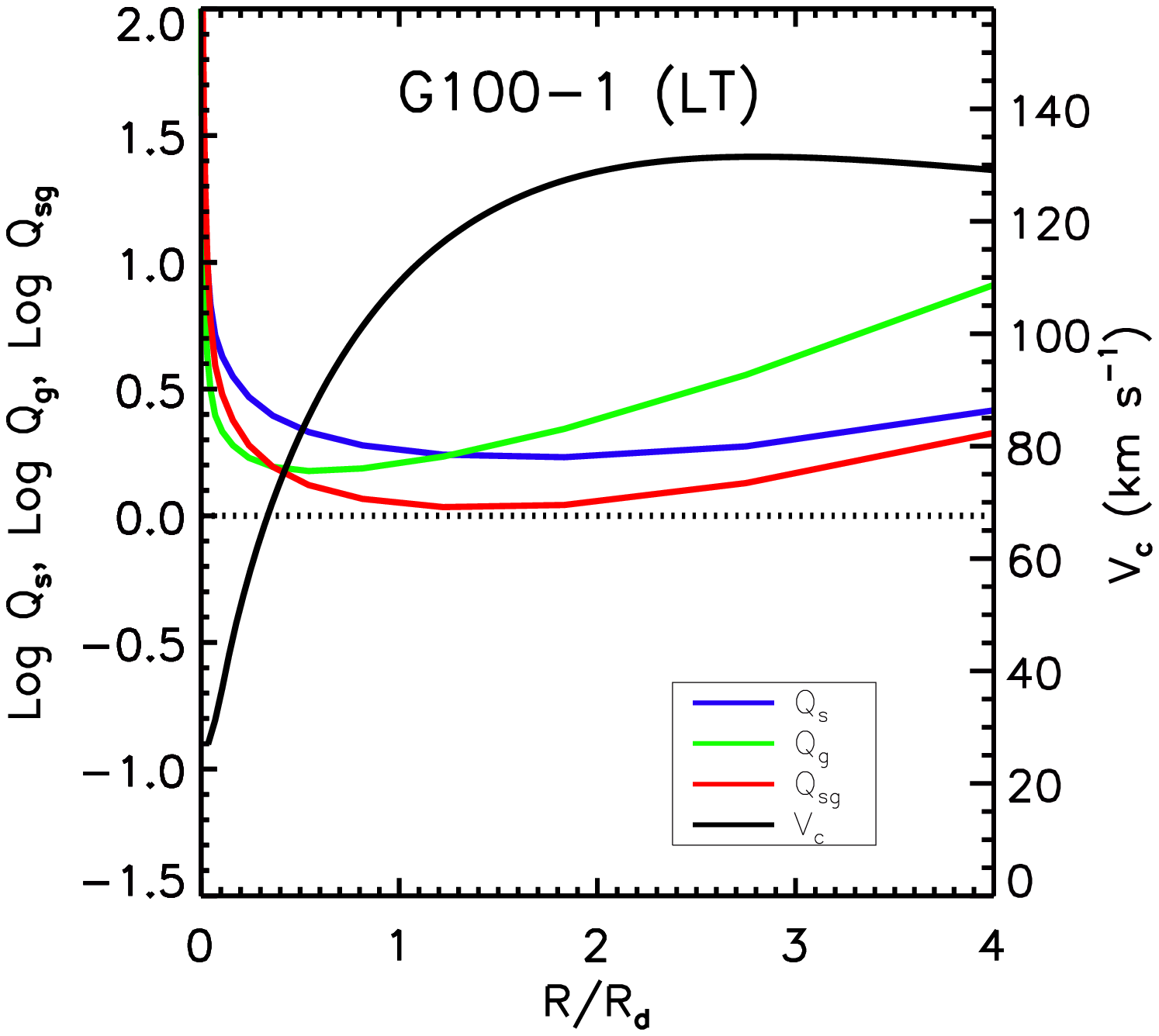}
\hspace{0.1in}
\includegraphics[width=3.0in]{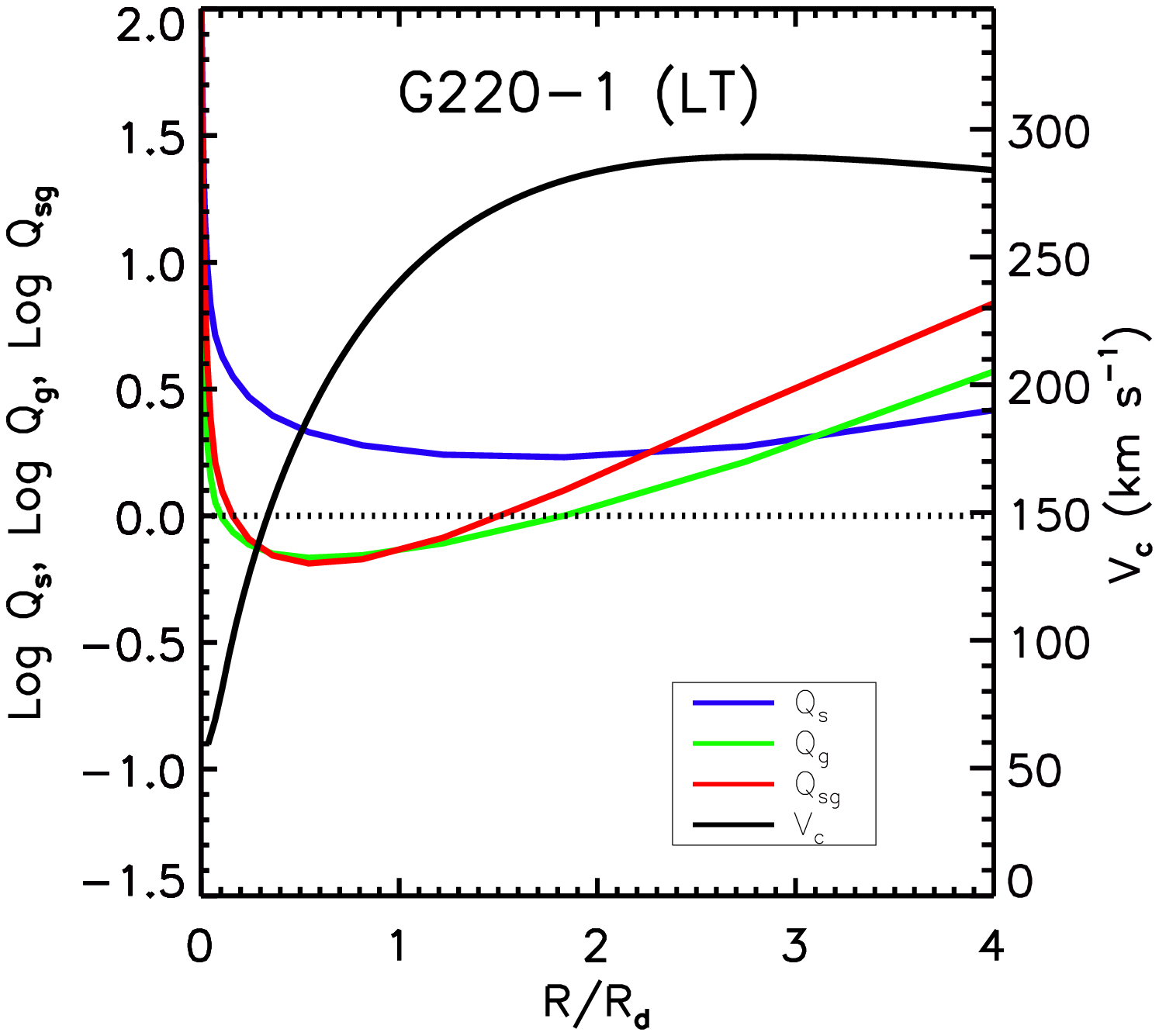}\\
\includegraphics[width=3.0in]{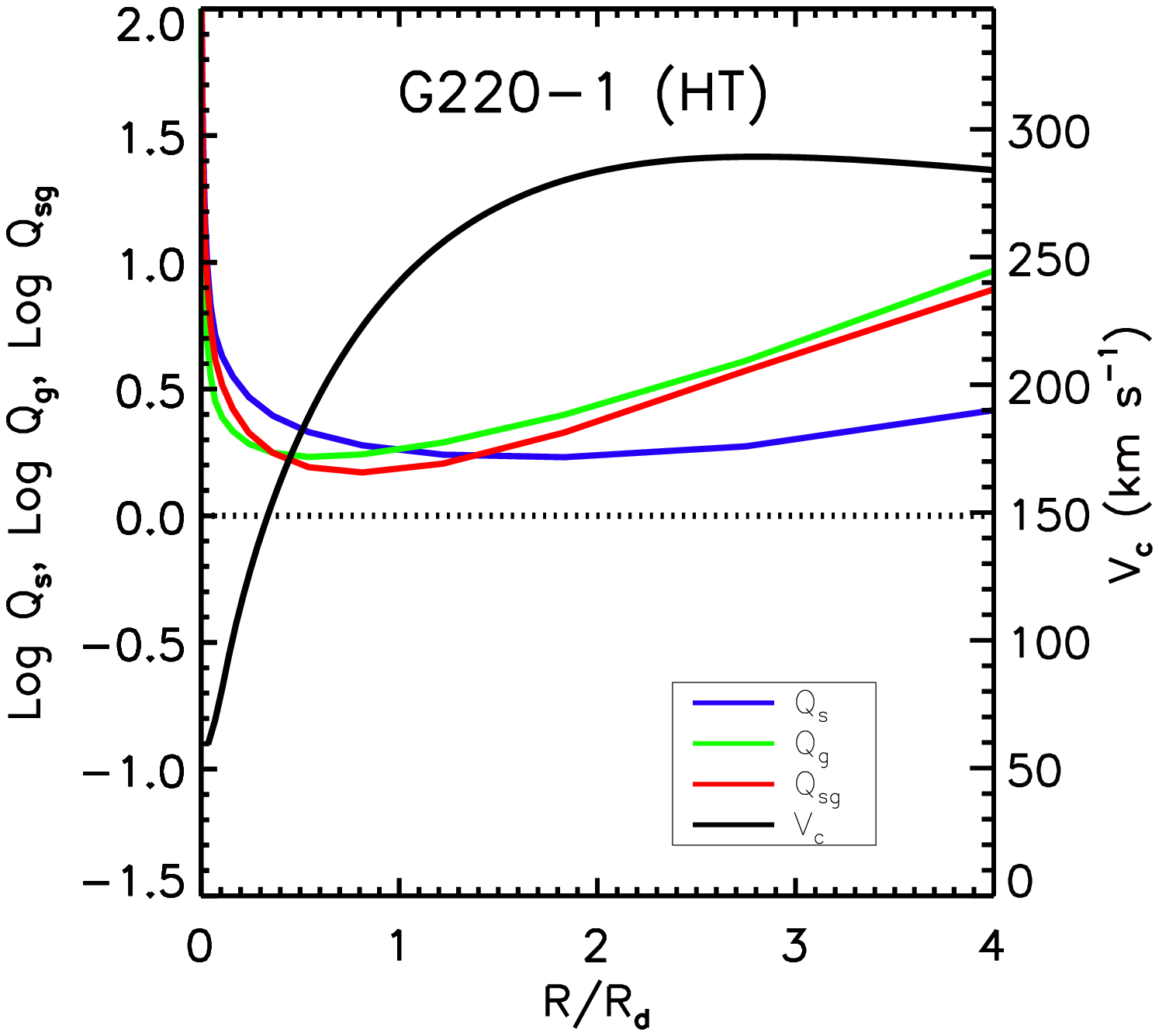}
\hspace{0.1in}
\includegraphics[width=3.0in]{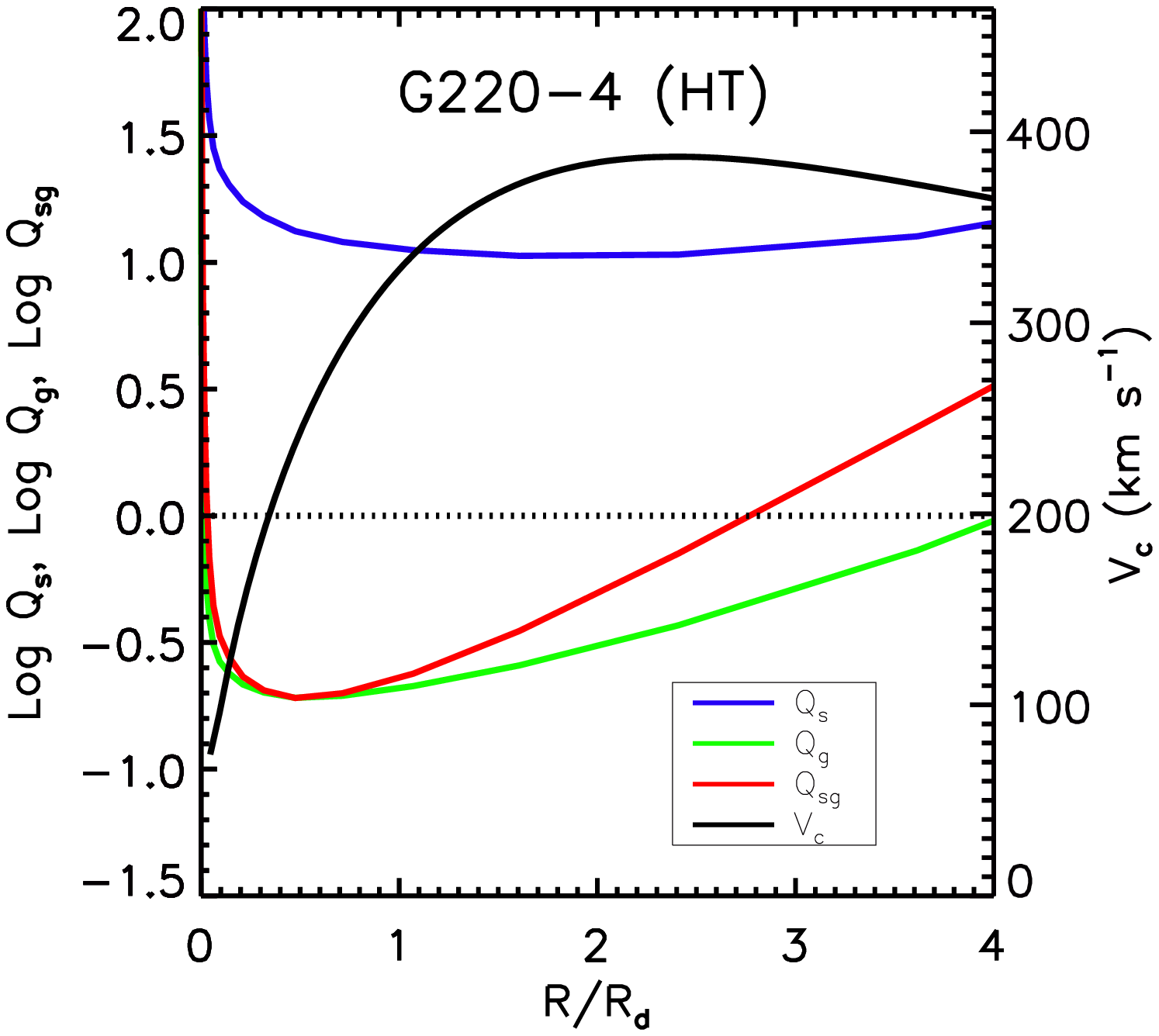}
\caption{\label{fig_3q} The initial gravitational instability and rotation
  curve of ({\em top panels}) low $T$ galaxy models G100-1 and G220-1 with
different rotation velocities, and ({\em bottom panels}) high $T$ models
G220-1 and G220-4 with different gas fractions. 
(Model parameters are given in Table~1.) 
Shown are the radial profiles
of the instability parameters, $Q_{\rm s}$ for stars ({\em blue}), $Q_{\rm g}$
for gas ({\em green}), and $Q_{\rm sg}$ for stars and gas together ({\em
red}), respectively. The black line is the curve of rotational velocity
$V_{\rm c}$, with scale given on the right.  $R_{\rm d}$ is the radial disk
scale length, given in Table 1.
The line of marginal stability $Q = 1$ is also shown. }
\end{center}
\end{figure}

The isolated galaxy models in our simulations cover a wide range of total mass 
and gas fraction. The models with disk fraction $m_{\rm d}$ = 0.05 have flat
rotation curves out to the virial radius, while in models with $m_{\rm d}$ =
0.10, the rotation curves drop slightly after a few disk scale lengths, as
shown in Figure~\ref{fig_3q}.  This Figure also shows the radial profiles of
the Toomre instability parameters for stars $Q_{\rm s}$, gas $Q_{\rm g}$, and
the combination $Q_{\rm sg}$ for several selected models with different mass,
sound speed, and disk gas fraction. They are calculated from
equations~(2)--(4) using the initial conditions, taking the most unstable
wavenumber $k$ at each radius. Note that $Q_{\rm s}$ is shown, rather than
Toomre's Q for the stars, so a factor $3.36/\pi =1.07$ should be divided out
when compared to commonly quoted $Q$ values \citep[e.g.\
in][]{kennicutt89,martin01}. Axisymmetric gravitational instability occurs 
when $Q < 1$ (or, in the stellar case $Q_{\rm s} <1.07$).

All three Q-curves have a similar dependence on galactic radius, dropping to a
minimum point at radius $R_{\rm min}$ and then rising. The stellar $Q_{\rm s}$
varies slowly, while in many galaxies $Q_{\rm g}$ and $Q_{\rm sg}$ vary more
dramatically with radius. Comparison of the models with the same gas fraction
but different total mass (top row) shows that the minimum value of $Q_{\rm
  g}$ and $Q_{\rm sg}$ decrease as galaxy mass increases, while the size of
the unstable region where $Q_{\rm sg} < 1$ increases with the mass of the
galaxy. Comparison of the models with the same total mass but different gas
fraction (bottom row) shows that the minimum values of $Q_{\rm g}$ and $Q_{\rm
sg}$, decrease as gas fraction increases, while the size of the unstable area
increases with gas fraction. Gas-rich, massive galaxies are more unstable than
gas-poor, low-mass galaxies. The more unstable the galaxy, the smaller $R_{\rm
min}/R_{\rm d}$ is. For example, G220-4 has $R_{\rm min} \approx$ 0.4 $R_{\rm
d}$, while G100-1 has $R_{\rm min} \approx$ 1.4 $R_{\rm d}$. The size of the
unstable regions range from $<2\ R_{\rm d}$ in small galaxies to $\approx
5\ R_{\rm d}$ in the most unstable galaxies. 

\begin{figure}
\begin{center}
\includegraphics[width=4in]{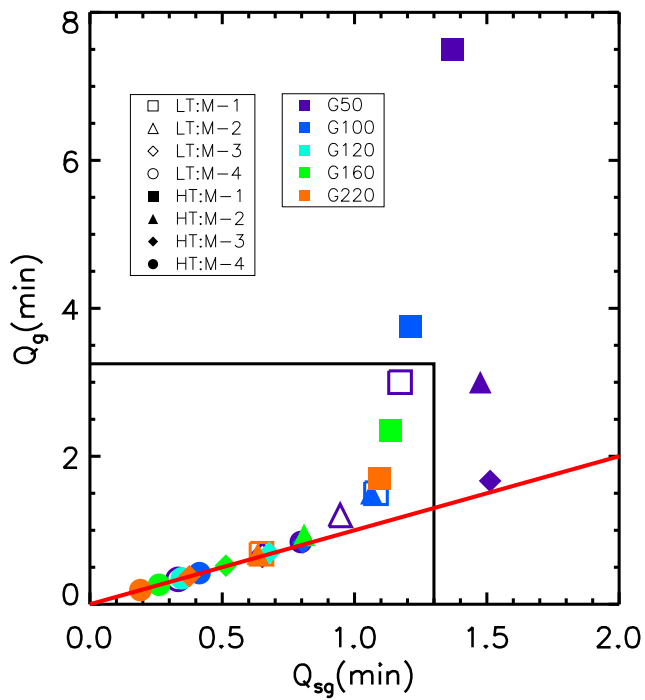}
\caption{\label{fig_qg_qsg} Comparison of minimum initial $Q_{\rm g}$ and
  $Q_{\rm sg}$ for both low $T$ ({\em open symbols}) and high $T$
  ({\em filled symbols}) models described in Tables 2 and 3, as shown in the
  legend. 
Labels from M-1 to M-4 are submodels that indicate increasing gas fraction,
as given in Table~1.
The color of the symbol indicates the rotational velocity for each
model.
The red line indicates $Q_{\rm g} = Q_{\rm sg}$. The
  rectangle encloses galaxies that form stars in the first 3 Gyrs in
  the simulations.}
\end{center}
\end{figure}

The Toomre parameter typically measured is $Q_{\rm g}$
\citep[e.g.,][]{kennicutt98a}, while instability is actually
determined by $Q_{\rm sg}$. A comparison of the minimum initial
$Q_{\rm g}(min)$ and $Q_{\rm sg}(min)$ for all our models is shown in
Figure~\ref{fig_qg_qsg}.  When less than unity, $Q_{\rm g}(min)
\approx Q_{\rm sg}(min)$, as shown also in Figure~\ref{fig_3q}. For values
above unity, though, $Q_{\rm g}(min) \gg Q_{\rm sg}(min)$.  Insofar as our
empirical initial conditions reproduce real galaxies, this suggests that
$Q_{\rm g}$ and $Q_{\rm sg}$ are equivalent in strongly unstable galaxies, but
in stable or marginally stable galaxies, $Q_{\rm g}$ alone may be insufficent
to describe the stability of the system, because it does not take into
account the destabilizing effect of the stellar disk mass. In small
galaxies such as dwarfs, or in gas-poor galaxies, stars tend to be
more important than gas in causing gravitational instability in disks.
Both stars and gas may need to be measured to fully characterize
gravitational instability in galactic disks.

Star formation is closely related to the value of $Q_{\rm sg}(min)$,
as shown by the rectangle in Figure~\ref{fig_qg_qsg}. The most
unstable model in the sample, G220-4 (high T) with $Q_{\rm sg}(min)
\approx 0.2$, has the most violent starburst, while the stable models
outside the rectangle do not form any stars in the first 3 billion
years. The critical value for star formation, $Q_{\rm sg} \approx 1.6$
will be derived with equation (7) in \S~\ref{sec_tsf} below. When
$Q_{\rm sg} > 1.6$, it is hard to form stars in a galaxy.
 
\section{STAR FORMATION MORPHOLOGY}
\label{sec_sf}

\begin{figure}
\begin{center}
  \includegraphics[width=3.0in]{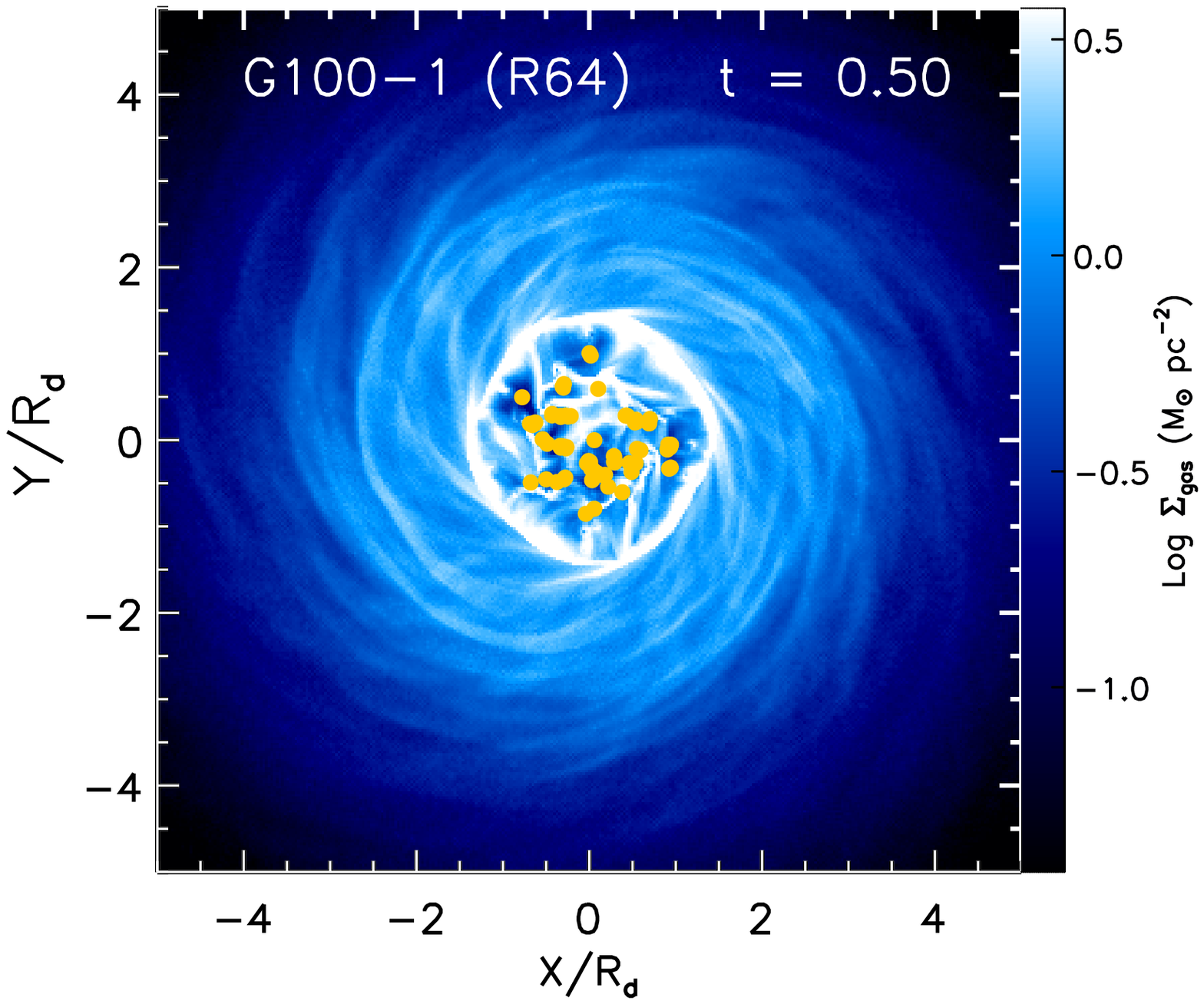}
  \hspace{0.3in}
  \includegraphics[width=3.0in]{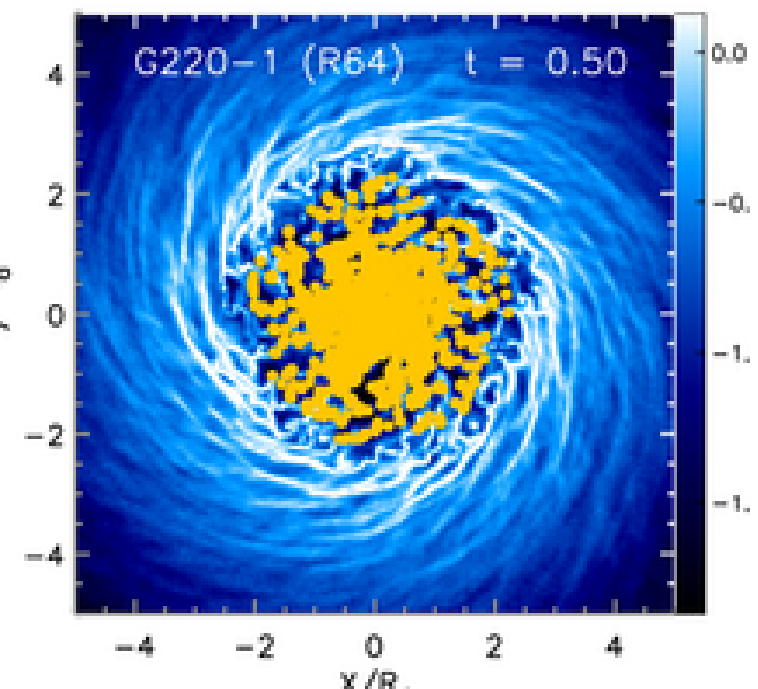}\\
  \includegraphics[width=3.0in]{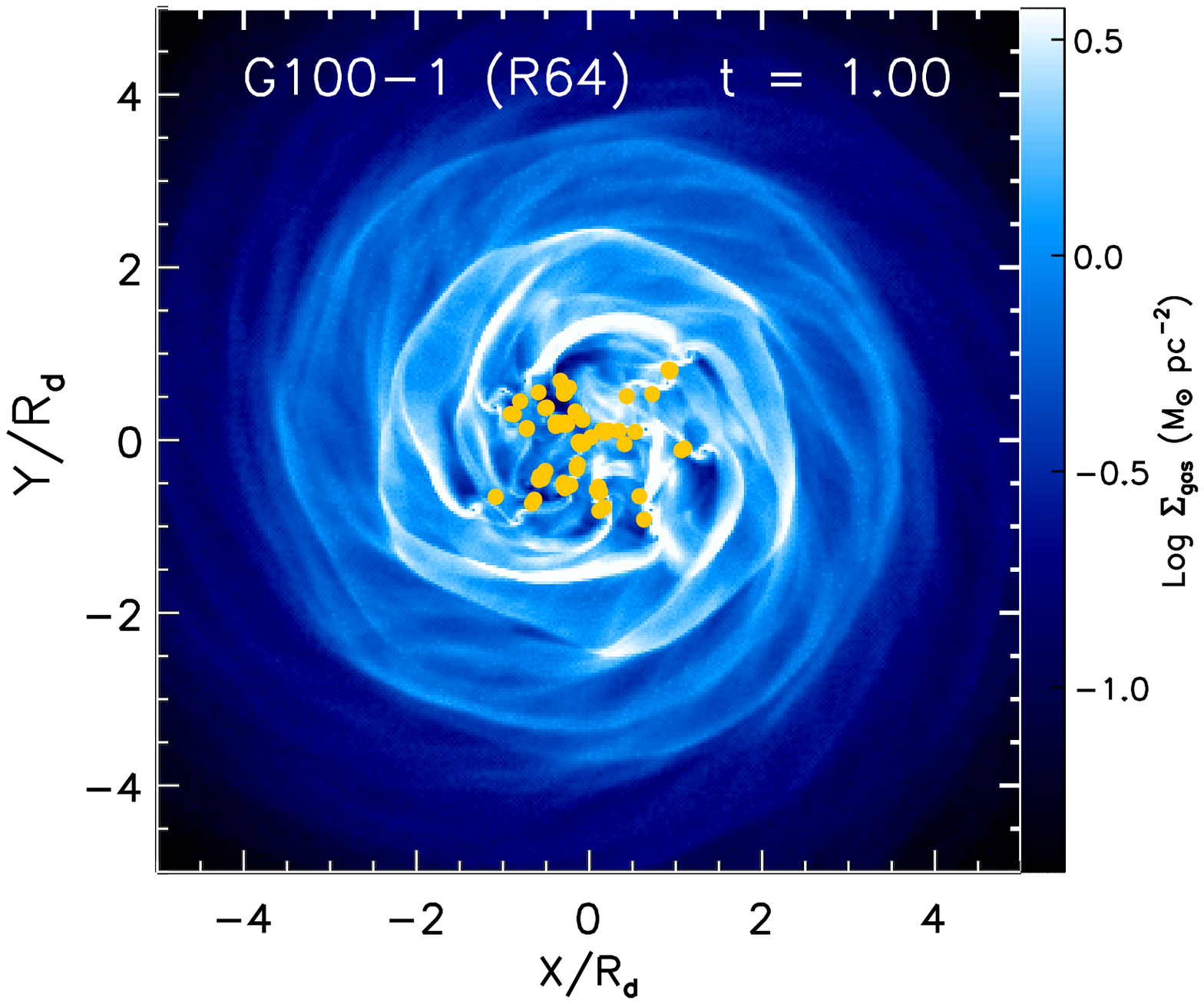}
  \hspace{0.3in}
  \includegraphics[width=3.0in]{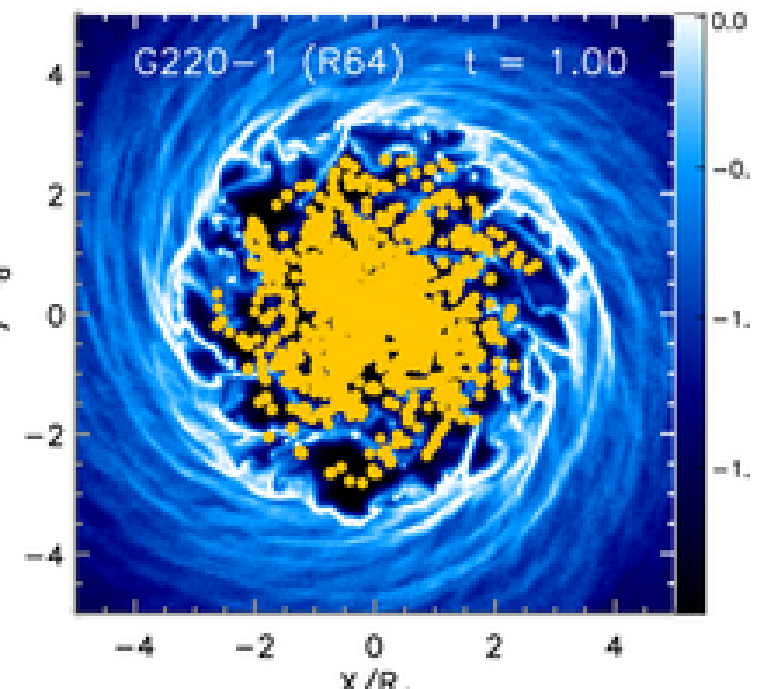}
\caption{\label{fig_map1} Star formation in the highest resolution (R64, with 
$N_{\rm tot} = 6.4 \times 10^6$) low $T$ models, with differing rotational
velocities and total mass G100-1 (\textit{left column}) and G220-1 
(\textit{right column}). The blue-white image shows atomic
gas surface density as indicated by the
color bars, while the yellow dots are sink particles representing 
both molecular gas and stars.
Time $t$ is given in units of 
gigayears.}
\end{center}
\end{figure}

Figure~\ref{fig_map1} shows sink particle locations in several of our low T
galaxy models, superposed on the atomic gas column density distribution. 
Again, keep in mind that sink particles represent regions of gravitational
collapse that include both molecular gas and stars. To produce the gas column
density image, the atomic gas density (SPH particles not accreted into sinks)
is projected to the x-y plane and then smoothed with an adaptive kernel that
always includes at least 32 gas particles. The columns in
Figure~\ref{fig_map1} show time evolution while the rows show different
models. 

Figure~\ref{fig_map1} shows that spiral structure develops in each
galaxy model, but it consists only of flocculent spiral arms (with
numerous short arms), not grand design (with two long symmetric arms,
such as M51), or multiple arm (with several long arms, such as our
Milky Way) patterns. This suggests that gravitational instability does
not cause isolated galaxies to form long spiral arms. We discuss this
further in \S~\ref{sec_dis}.  Gravitational collapse resulting in star
formation occurs as the gas density increases in the spiral arms. 
Collapse occurs first in the higher density regions near the galactic
center, and then extends over time to larger radii.

Star formation depends on the mass of the galaxy. 
Small galaxies such as G100-1 form stars slowly, while massive ones
such as G220-1 have vigorous starbursts at early time.  The size
of the star-forming region relative to the scale of the galaxy
increases with galaxy mass as well. For example, at $t = 1$~Gyr, star
clusters form only within $1\ R_{\rm d}$ in G100-1, but in G220-1,
they form out to $3\ R_{\rm d}$.

\begin{figure}
\begin{center}
\includegraphics[width=2.5in]{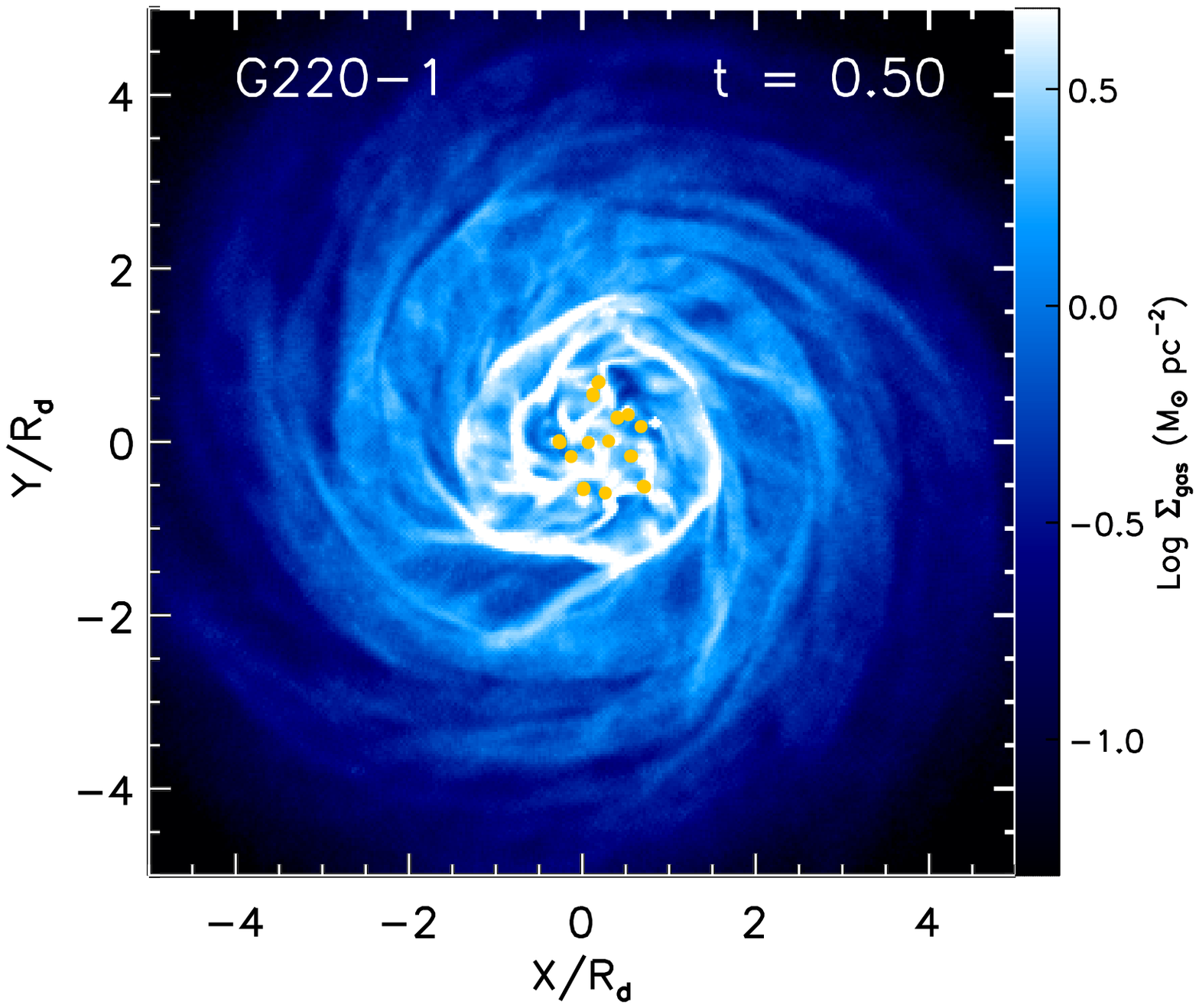}\\
\includegraphics[width=2.5in]{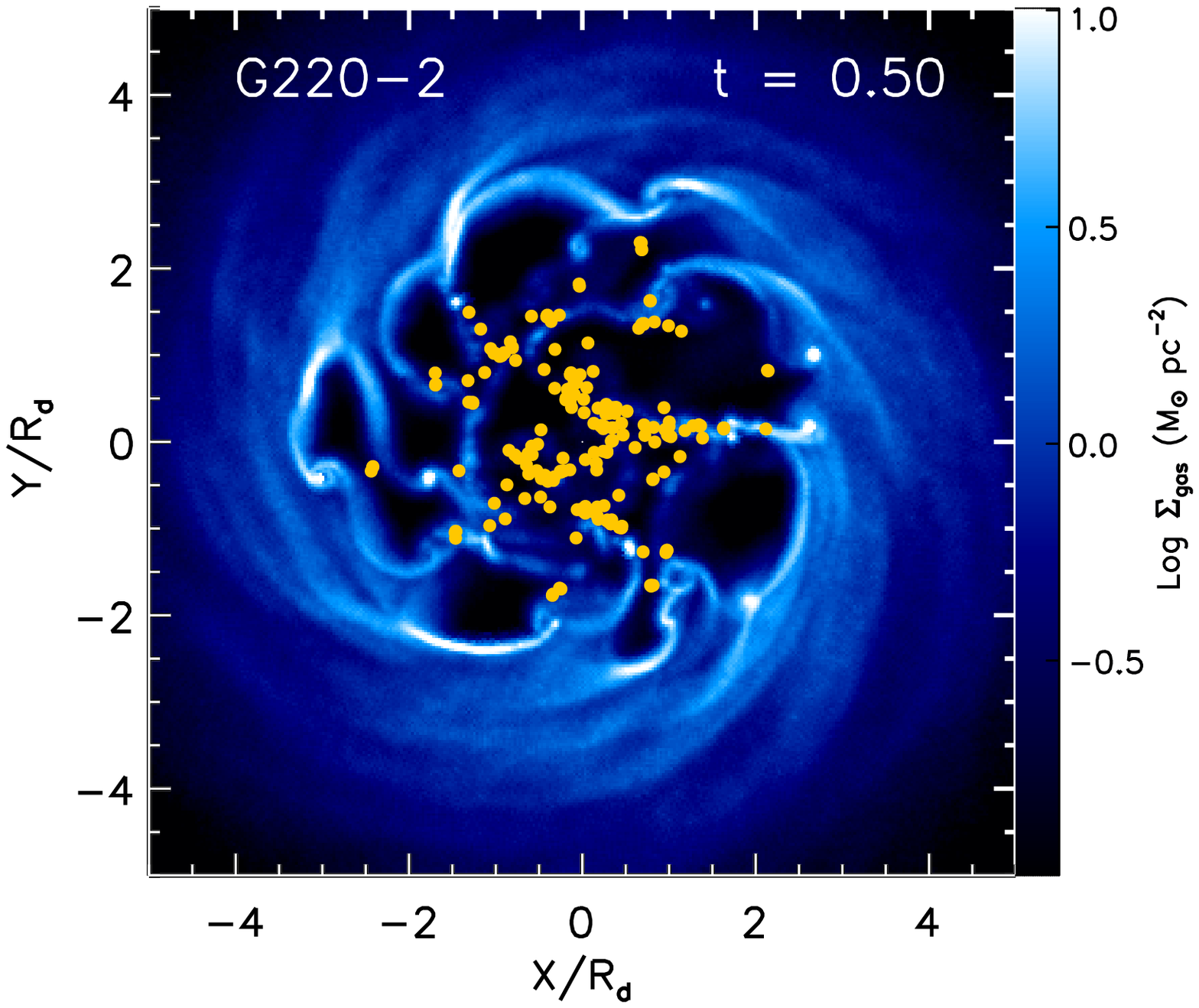}\\
\includegraphics[width=2.5in]{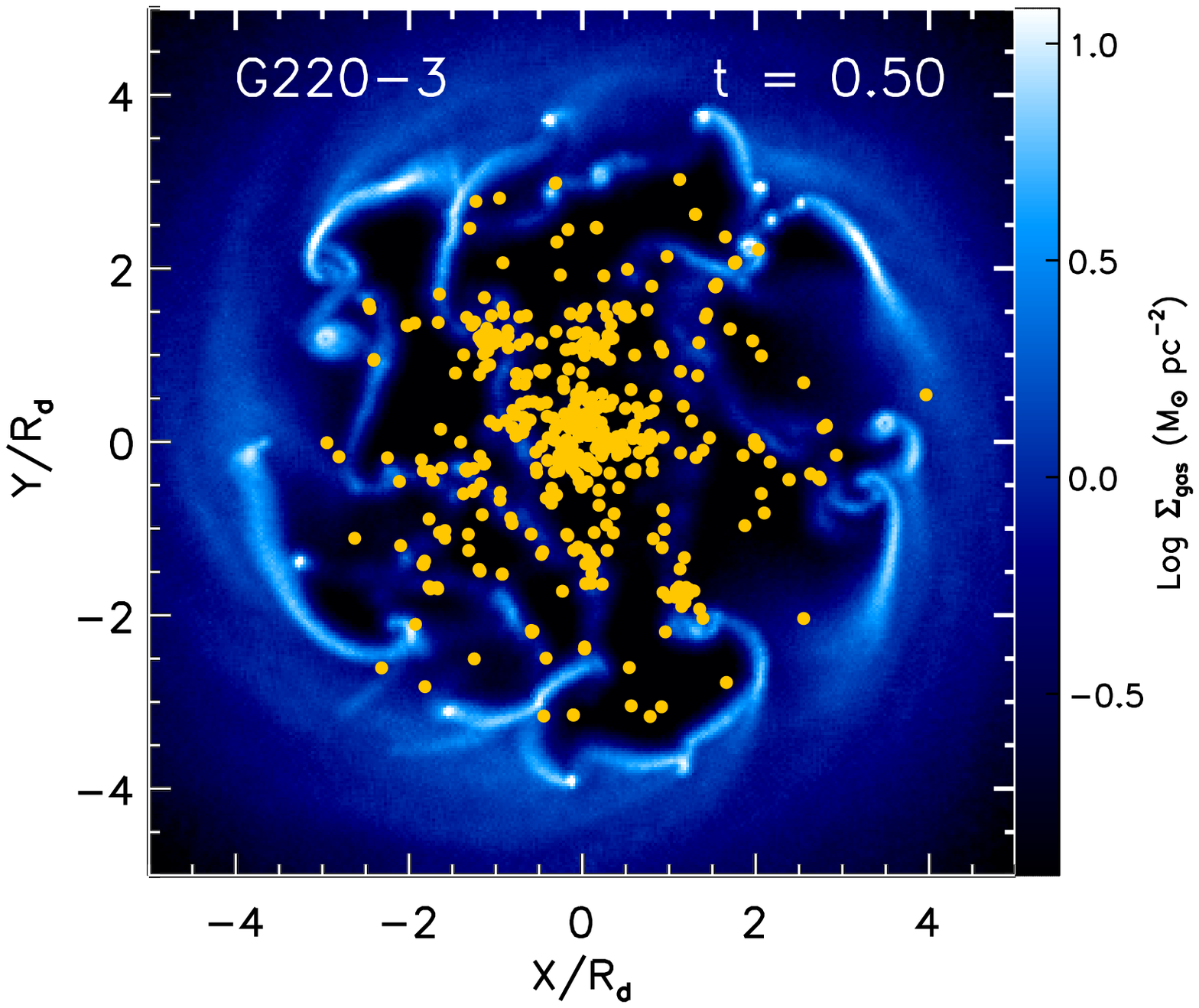}
\caption{\label{fig_map2} Same as Figure~\ref{fig_map1}, but for
  high T models with the same total mass but different gas fractions,
  $f_{\rm g}$ = 0.2, 0.5, 0.9 of the disk mass for G220-1, 2, 3,
  respectively, and disk fraction $m_{\rm d} = 0.05$ of the total
  mass.  }
\end{center}
\end{figure}

Star formation also depends on the gas fraction of the galaxy disk. 
Figure~\ref{fig_map2} shows models at one time with the same
total mass but different gas fractions. As the galaxy becomes more gas
rich, collapse becomes more vigorous, and the relative size of the
star formation region increases. The high density regions in the spiral arms
collapse most quickly, aligning star formation with the arms.  For example, at
$t = 0.5$~Gyr, clusters form within $1\ R_{\rm d}$ in G100-1, while in G100-4
they form out to 5 $R_{\rm d}$.

Taken together, Figures~\ref{fig_map1} and~\ref{fig_map2} illustrate
that rapid, widespread star formation takes place in massive or gas-rich,
unstable galaxies, while slow, centralized star formation occurs in more
stable galaxies. 

\begin{figure}
\begin{center}
\includegraphics[width=2.5in]{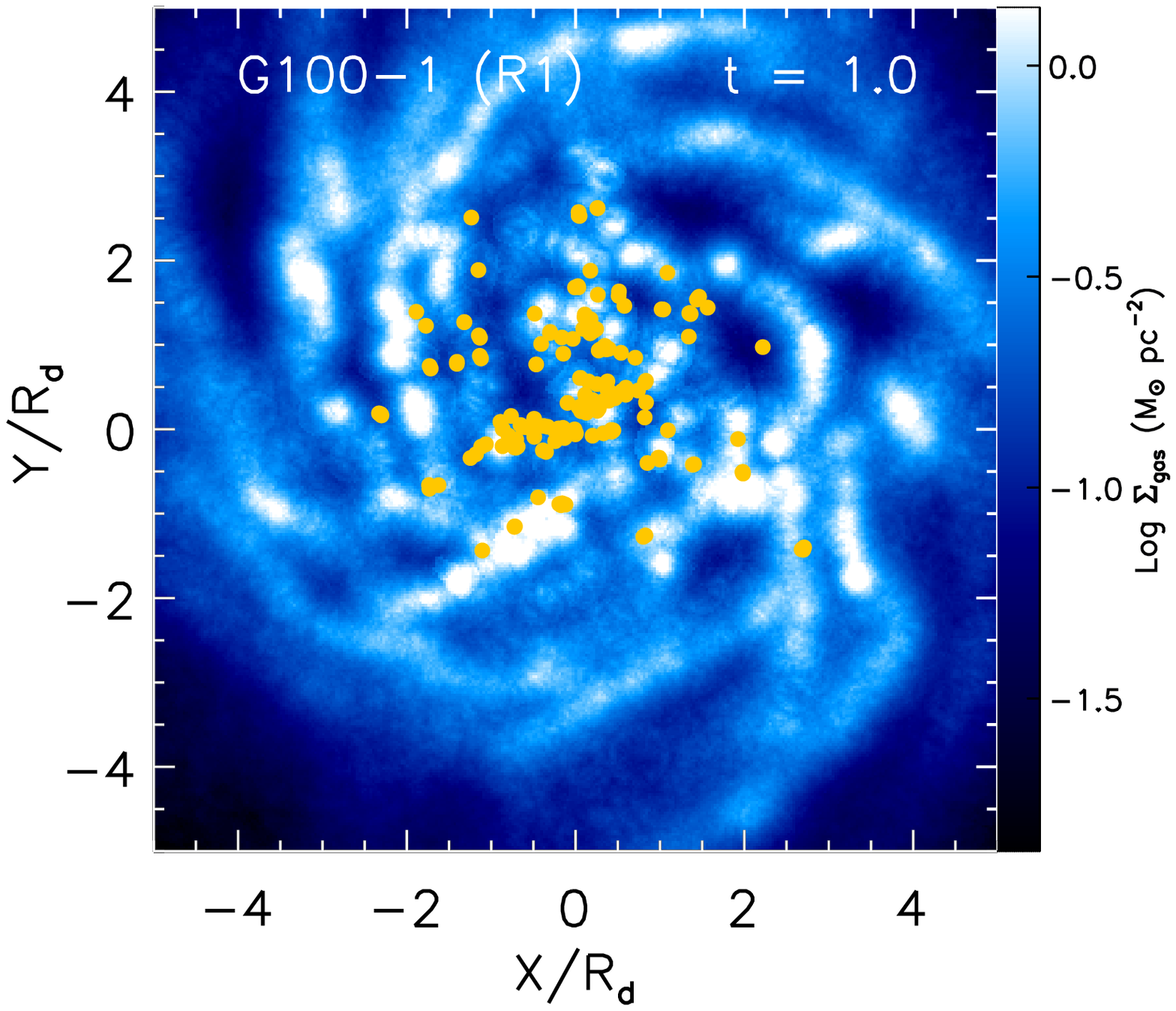}\\
\includegraphics[width=2.5in]{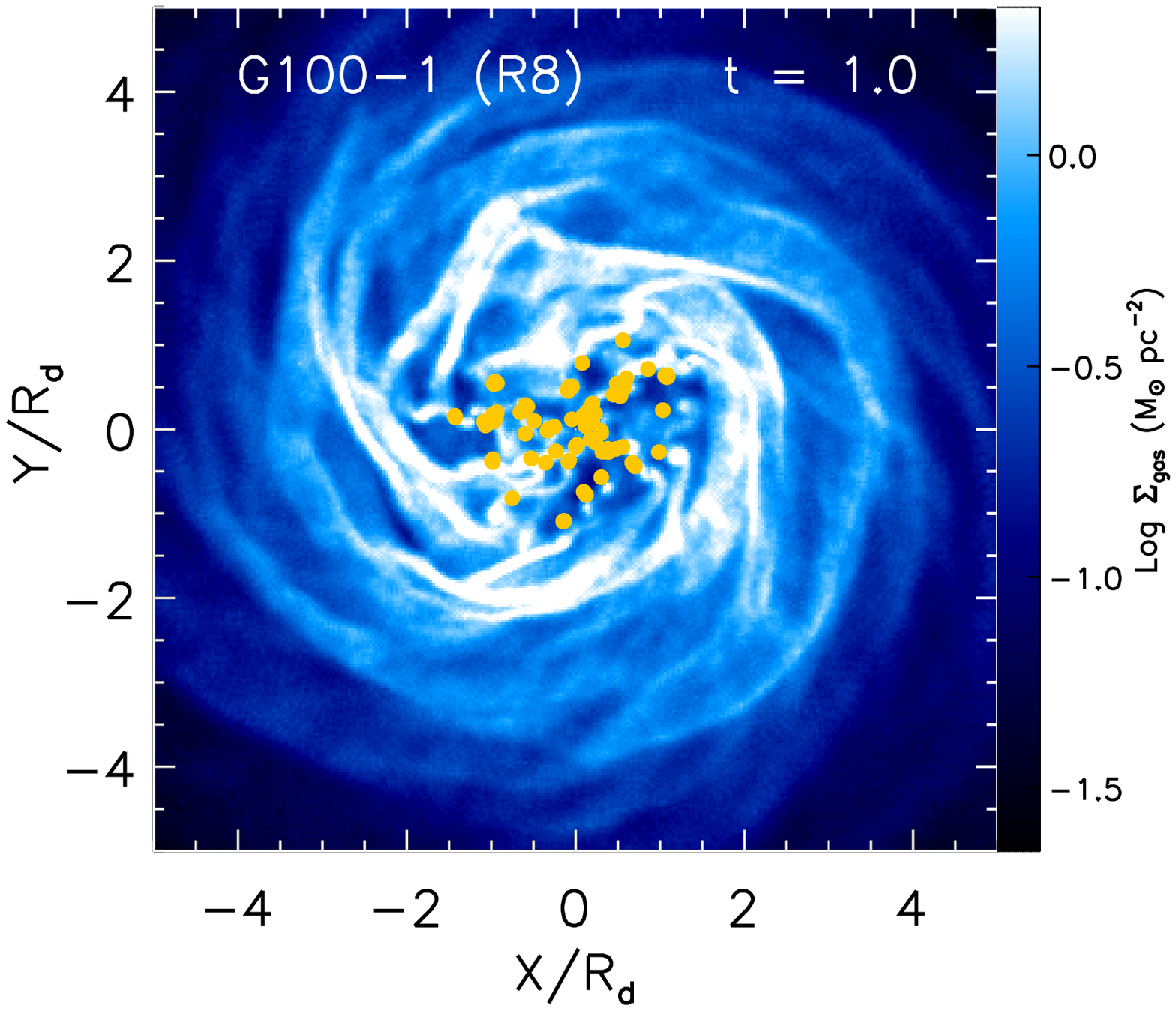}\\
\includegraphics[width=2.5in]{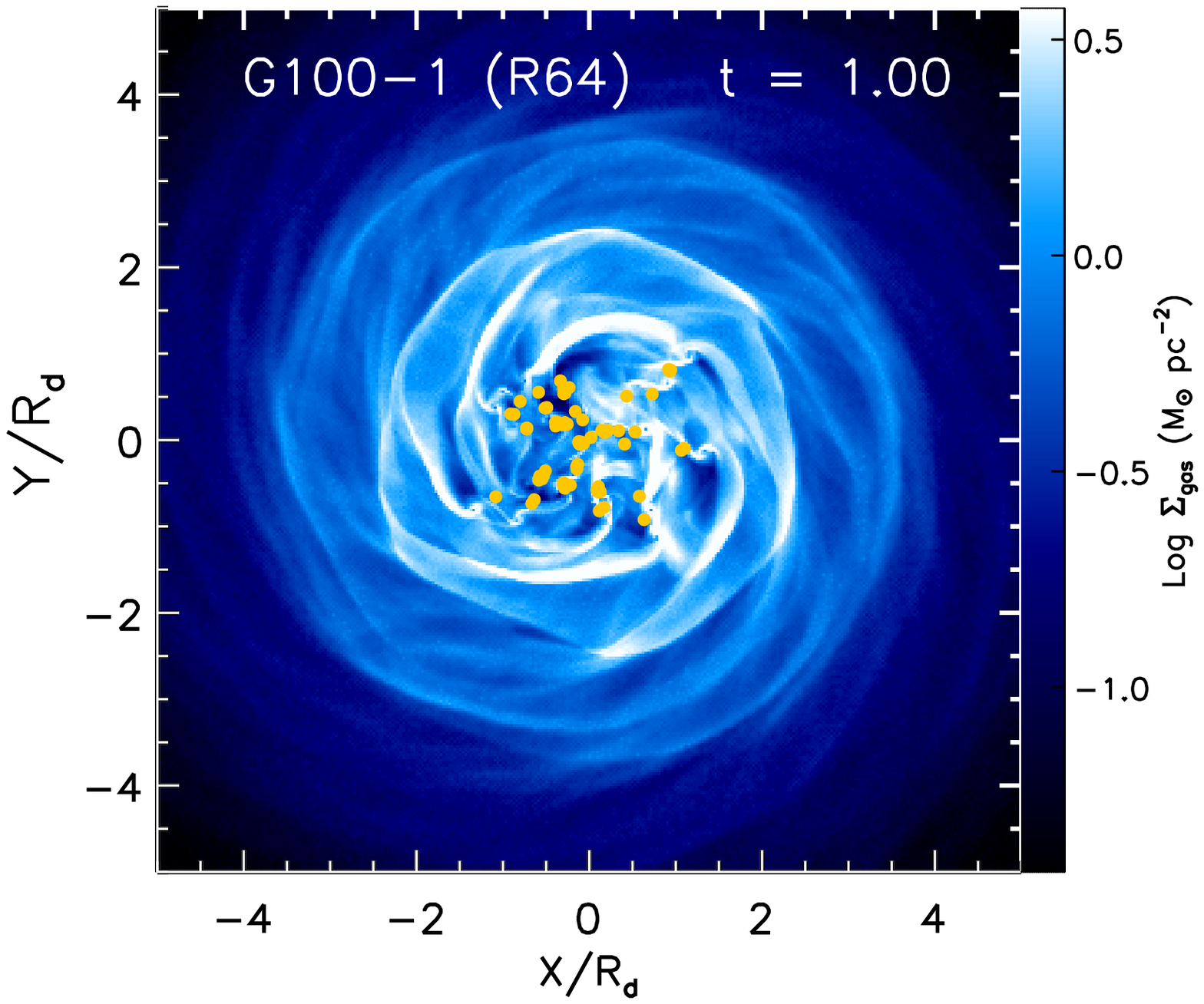}
\caption{\label{fig_rose} Gas surface density maps from the resolution
  study, showing versions of G100-1 (LT) with total particle number of $N_{\rm
tot} = 10^5$ (R1), $8\times 10^5$ (R8) and $6.4\times 10^6$ (R64). 
  Model R1, which is underresolved according to the criterion of
  \citet{bate97}, shows strong spurious fragmentation.  On the other
  hand model R8, which is marginally resolved, shows only minor
  differences with the highest-resolution
model R64, suggesting the criterion is adequate.  }
\end{center} 
\end{figure}

Figure~\ref{fig_rose} shows the maps of gas surface density from our
resolution study of model G100-1. In the low resolution run R1 where
the \citet{bate97} mass resolution criterion is not satisfied, strong
spurious fragmentation occurs. However, models R8 and R64, which comply with
the three criteria, look similar, and both look different from the
first one.  This suggests that the mass-resolution criterion of
\citet{bate97} is adequate for our problem. We will confirm this by
examination of resolution effects on several other diagnostics that we
study below. 

\begin{figure}
\begin{center}
\includegraphics[width=4.0in]{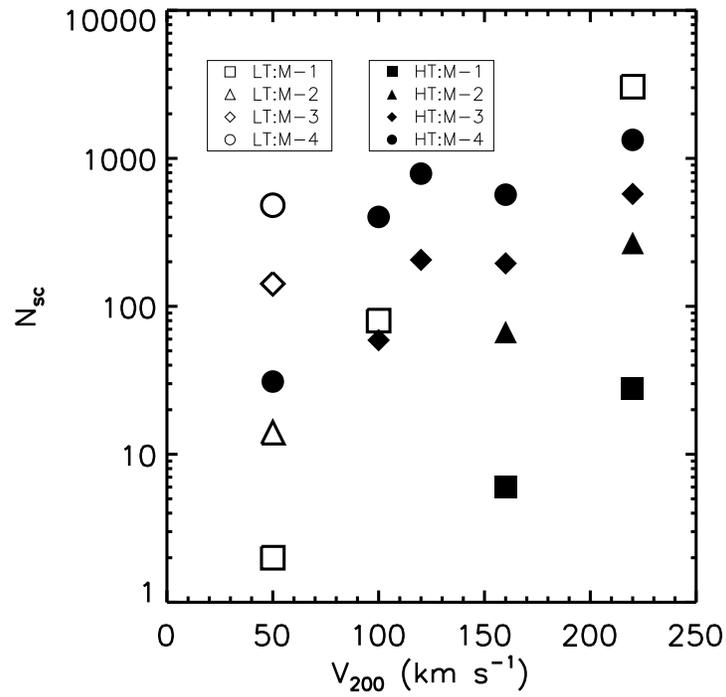}
\caption{\label{fig_nsink} 
  Number of stellar clusters 
  (sink particles)
  formed after 3~Gyr in both low $T$ ({\em open symbols}) and high $T$
  ({\em filled symbols}) models, with properties given in 
  Tables~1--3.
Again M-1 to 4 indicate submodels that reflect different gas fractions,
    as given in Table~1.
}
\end{center}
\end{figure}

As galaxies become more strongly gravitationally unstable the number
of clusters formed also increases (Figure~\ref{fig_nsink}). Unstable
galaxies, with higher rotational velocites $V_{\rm c}$ or higher gas
fractions (higher submodel numbers in most cases) produce more clusters than
their more stable, small or gas poor counterparts. The low $T$ models
are more unstable than their high $T$ counterparts, and they also
produce more clusters for the same galactic parameters.

\begin{figure}
\begin{center}   
\includegraphics[width=4in]{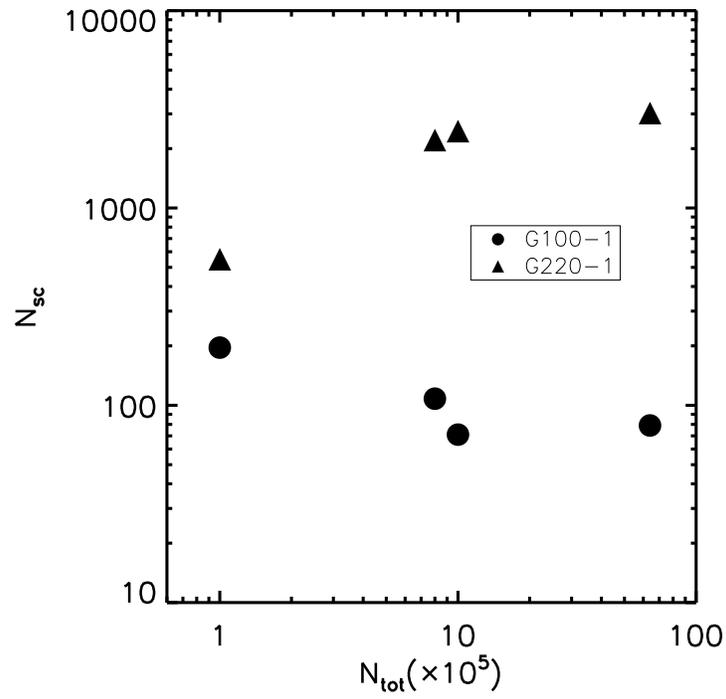}
\caption{\label{fig_nsink_res}
  Same as Figure~\ref{fig_nsink} but for models of
  G100-1 ({\em circles}) and
G220-1 ({\em triangles}) with different resolutions, as indicated by
the total number of particles $N_{\rm tot}$. }
\end{center}
\end{figure}

To check whether the cluster numbers shown in Figure~\ref{fig_nsink}
have numerically converged, we compare in Figure~\ref{fig_nsink_res}
the cluster number in the runs of the resolution study of models G100-1 and
G220-1. Moving from run R1 to run R8, increasing linear resolution by a factor
of two, shows a change by a factor of two in cluster number. However,
further increase in resolution shows rather smaller variations of
$\approx 30$\%, but in random directions, suggesting that convergence
effects are being overwhelmed by cosmic variance caused by chaotic
dynamics (see discussion of Figure~6 in \citealt*{klessen00}).

\begin{figure}
\begin{center}
\includegraphics[width=3in]{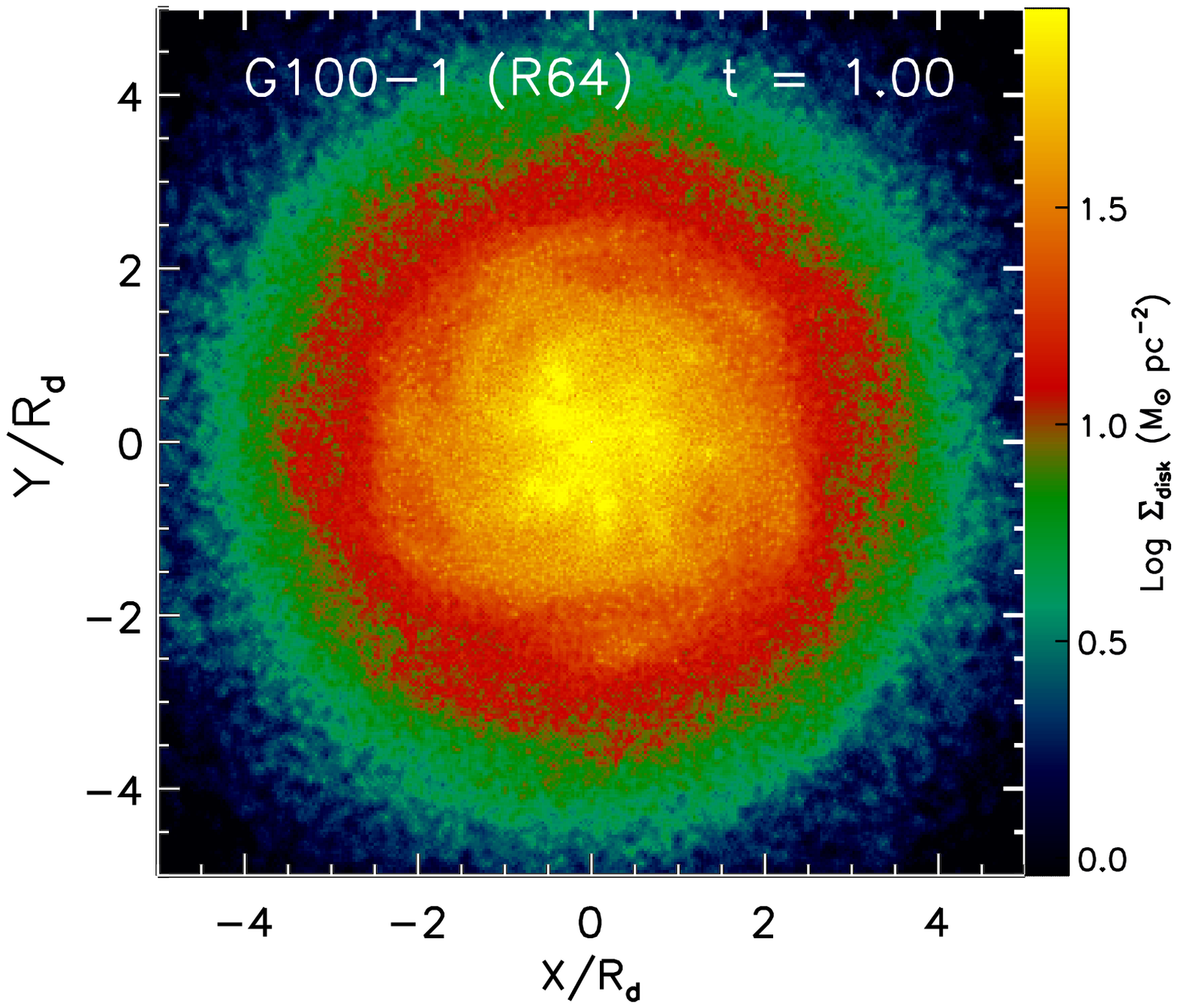}
\hspace{0.3in}
\includegraphics[width=3in]{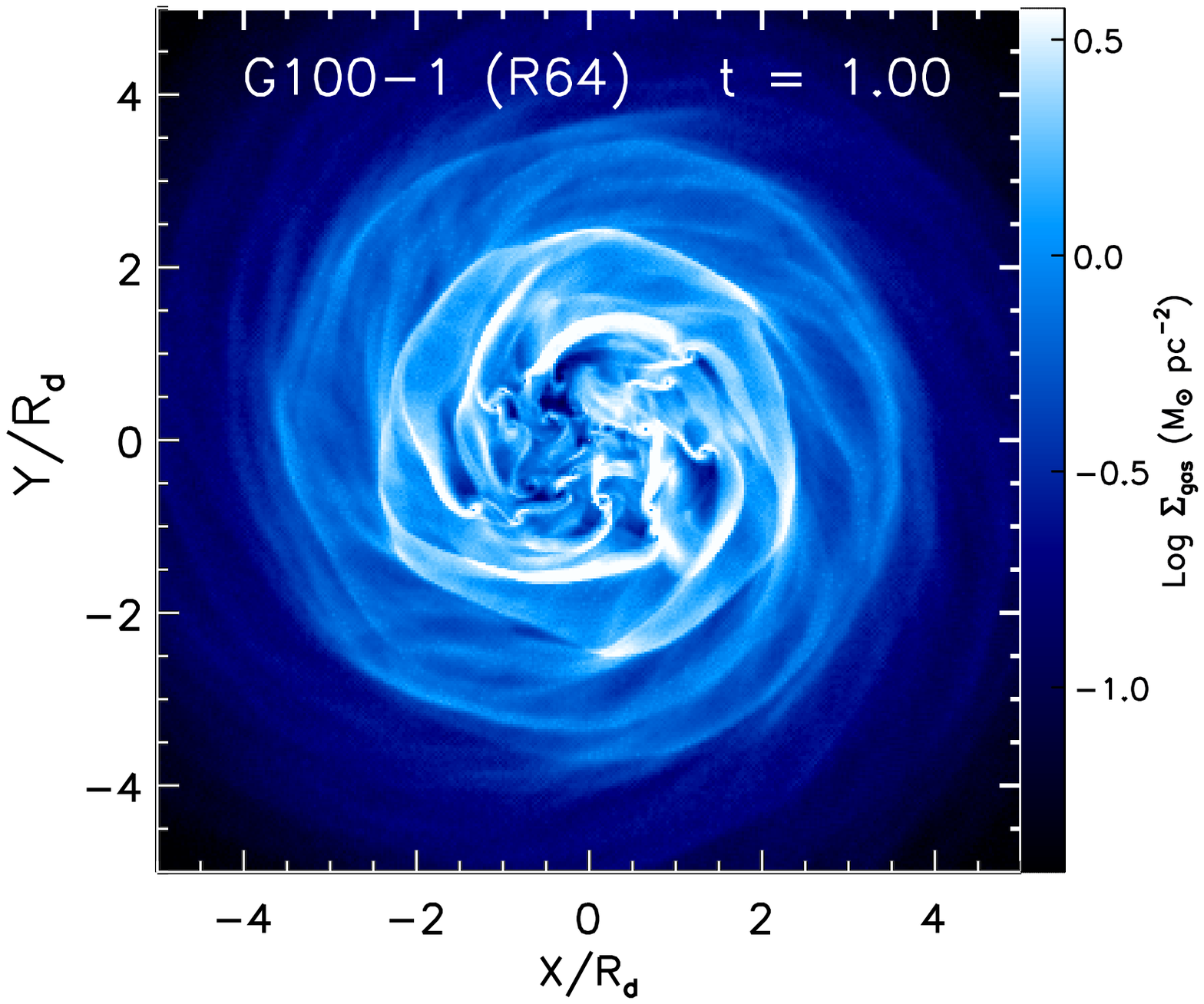}\\
\includegraphics[width=3in]{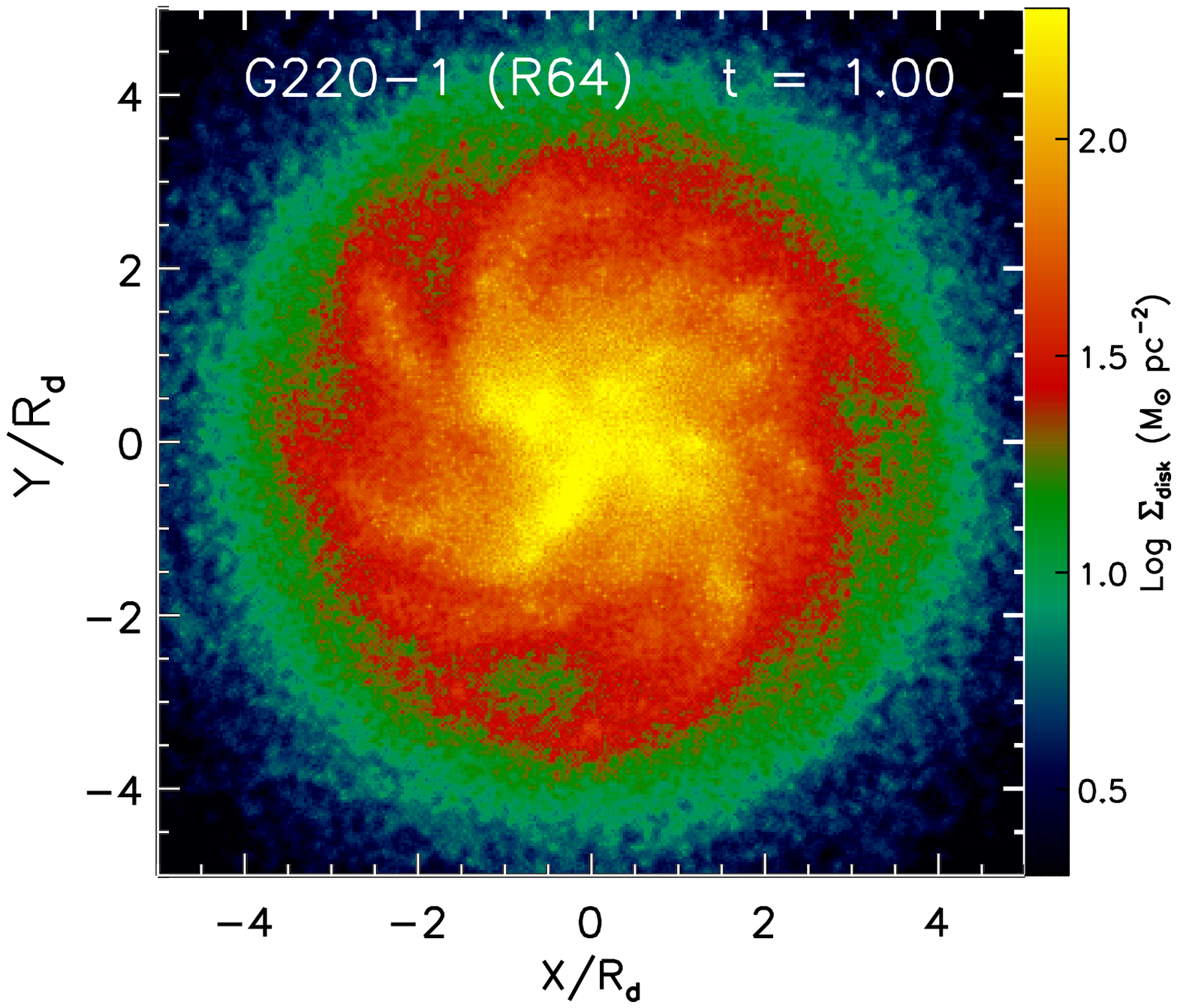}
\hspace{0.3in}
\includegraphics[width=3in]{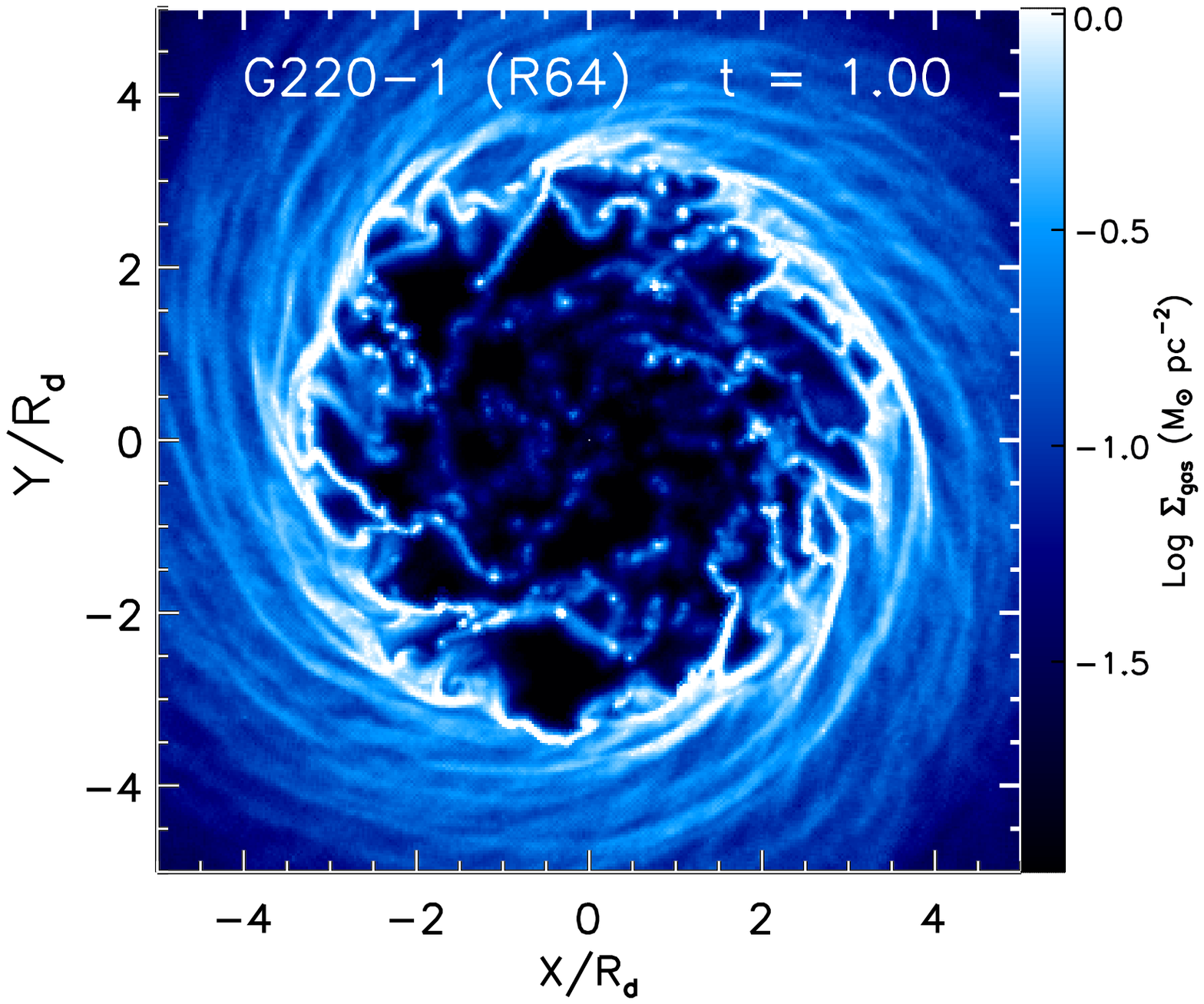}
\caption{\label{fig_map3} Comparison of the old stellar disk (\textit{left
    column}) and the 
atomic
gas disk (\textit{right column}) of LT models G100-1
  and G220-1 at $t = 1$~Gyr. Both models have $N_{\rm tot}=6.4\times10^6$. The
images are the surface densities of stars and gas, respectively, as shown by
the color bars. The 
sink particles, which
represent both molecular gas and stars,
are not included in these
images.} 
\end{center}
\end{figure}

The old stellar disk and the gas disk evolve differently. A comparison
of the two is shown in Figure~\ref{fig_map3}, showing their different
morphologies. The stellar disk does not evolve much over time in any of our
models, probably because of its lack of dissipation. In contrast, the gas disk
evolves quickly,  as seen in model G220-1. High density regions formed by
gravitational instability  collapse, turning atomic gas into molecular gas and
stars, as represented by the sink particles (not shown in Fig.\ \ref{fig_map3}). 
The resulting disk has a gas distribution that is mostly atomic in the
outer disk but molecule-dominated in the central region
(better seen in Fig.\ \ref{fig_map1}), agreeing with observations by
\citet{wong02}. 

\section{STAR FORMATION HISTORY}
\label{sec_tsf}

Star formation occurs at different times in different galaxies. Some
galaxies appear to have formed the bulk of their stars shortly after
formation, while some, even after billions of years, are still forming
stars \citep{kennicutt98a}. Stars form at different times and with
different global efficiencies in different model galaxies as well.
Because we neglect both gas recycling and galaxy interactions, our
models cannot represent the full star formation history of any real
galaxy.  Instead, they represent physics experiments that give the
response of gravitational instability to the chosen initial
conditions in each model.  The results of these experiments do,
however, appear very relevant to explaining the behavior of real
galaxies. 

\begin{figure}
\begin{center}
\includegraphics[width=3.0in]{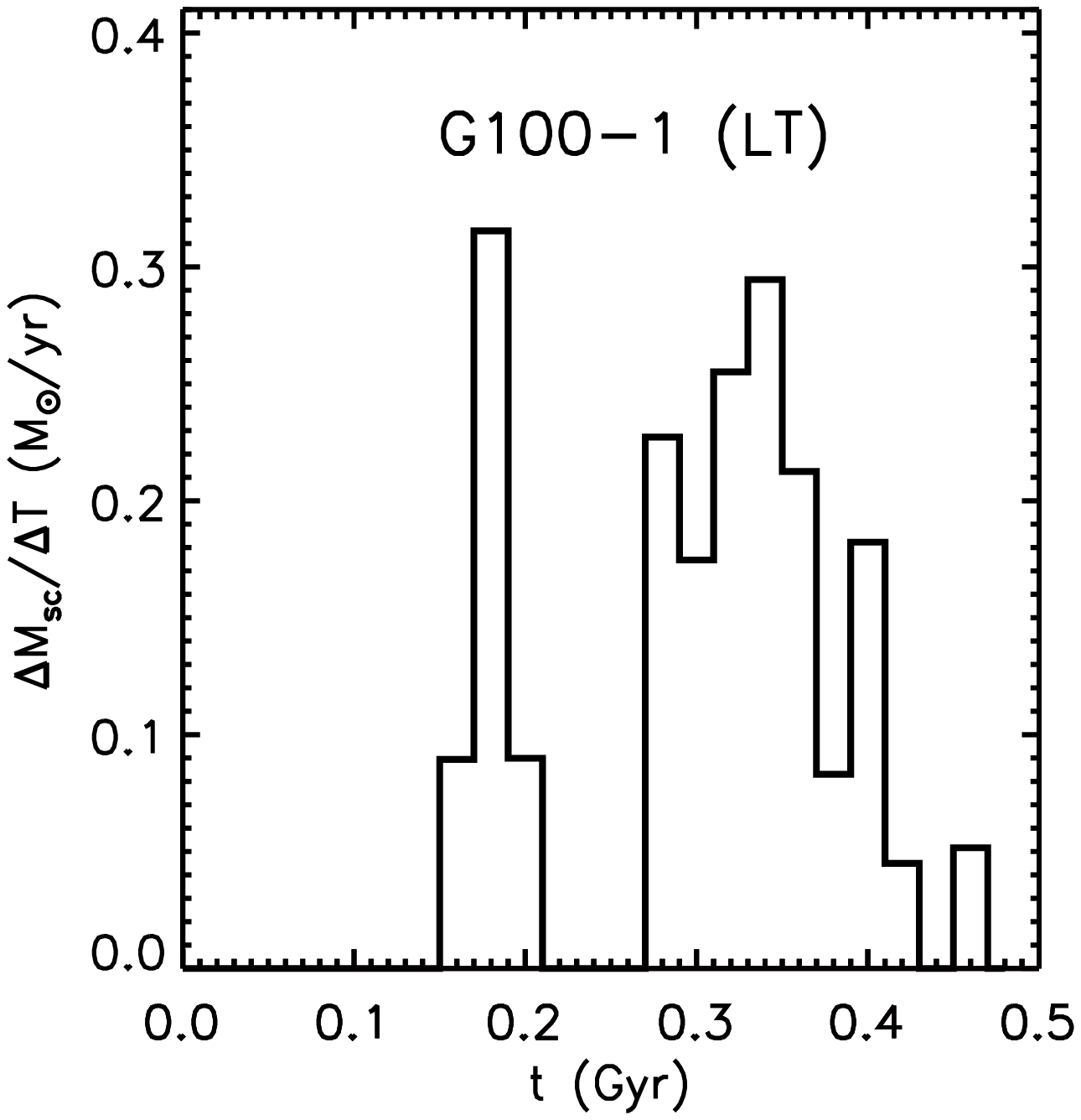}
\includegraphics[width=3.0in]{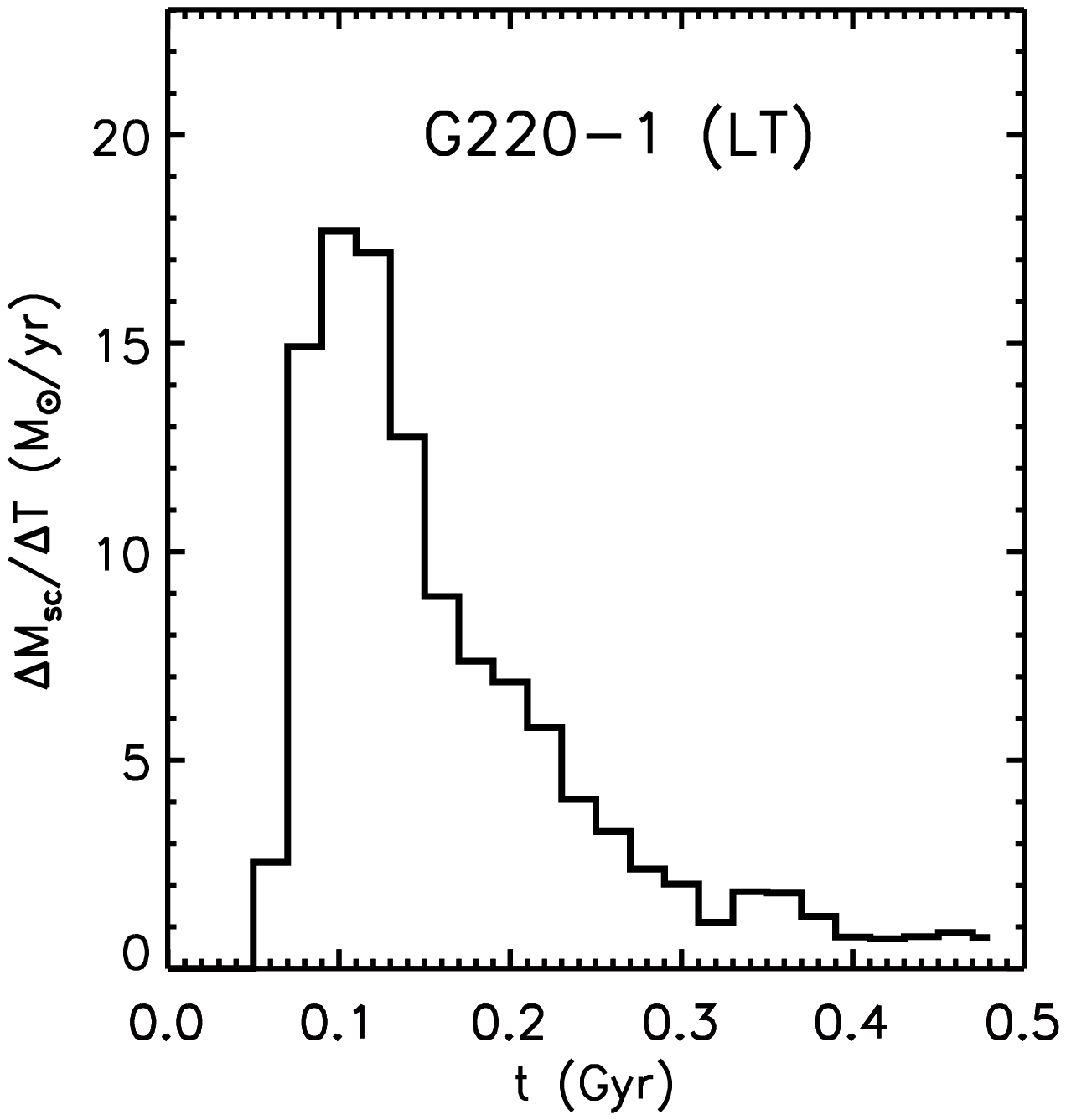}\\
\includegraphics[width=3.0in]{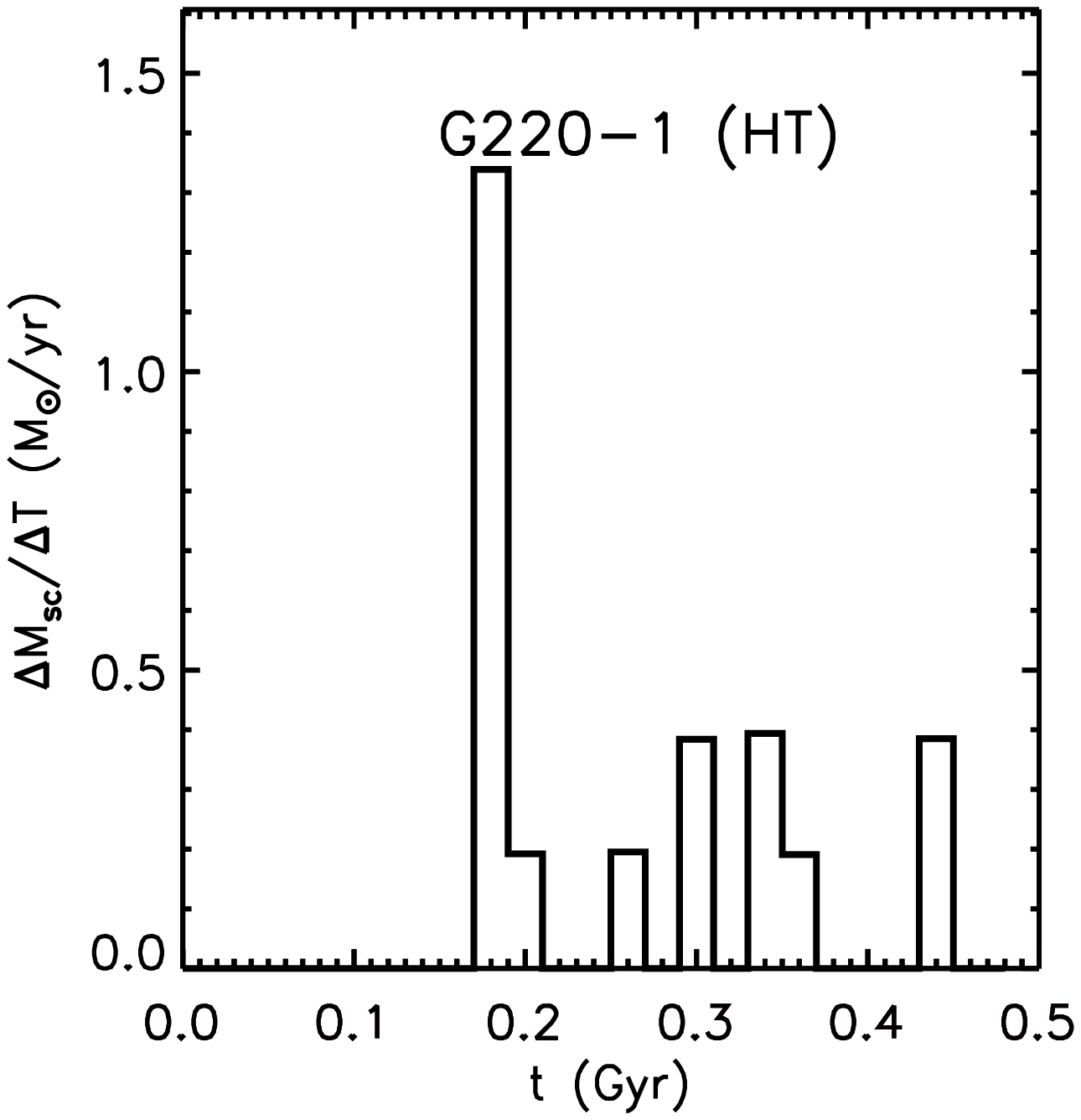}
\includegraphics[width=3.0in]{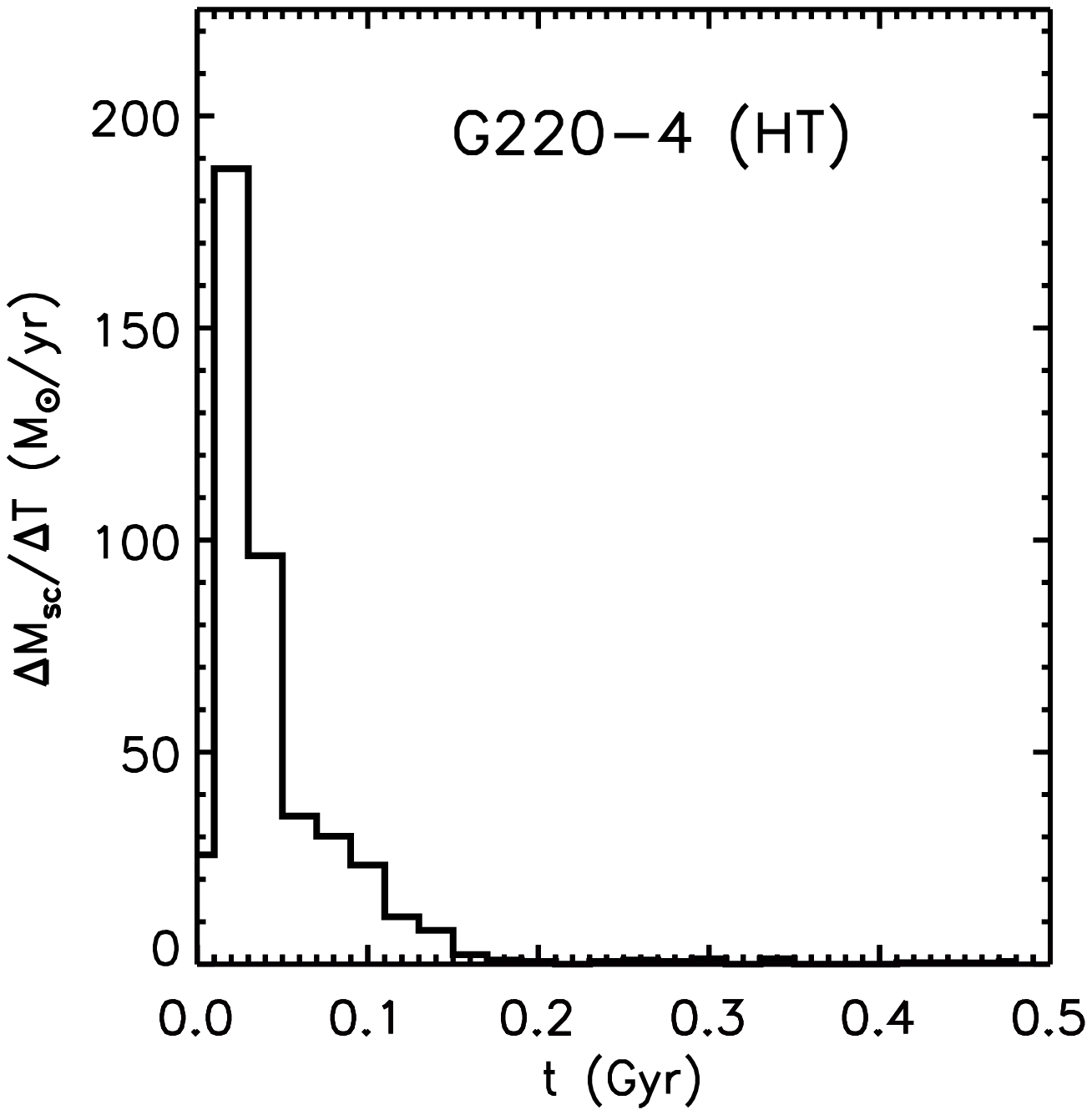}
\caption{\label{fig_when} Star formation histories for {\em top panels}) low
$T$ galaxy models G100-1 and G220-1 with different rotation velocities, and
({\em bottom panels}) high $T$ models G220-1 and G220-4 with different gas
fractions. 
$\Delta M_{\rm sc}$ is the mass collapsed into star clusters in a time
interval of $\Delta T=$ 20 Myrs (the bin size). The mass of star clusters is
taken as 30\% of the mass of the sink particles 
as explained in the text. } 
\end{center}
\end{figure}

\begin{figure}
\begin{center}
\includegraphics[width=4in]{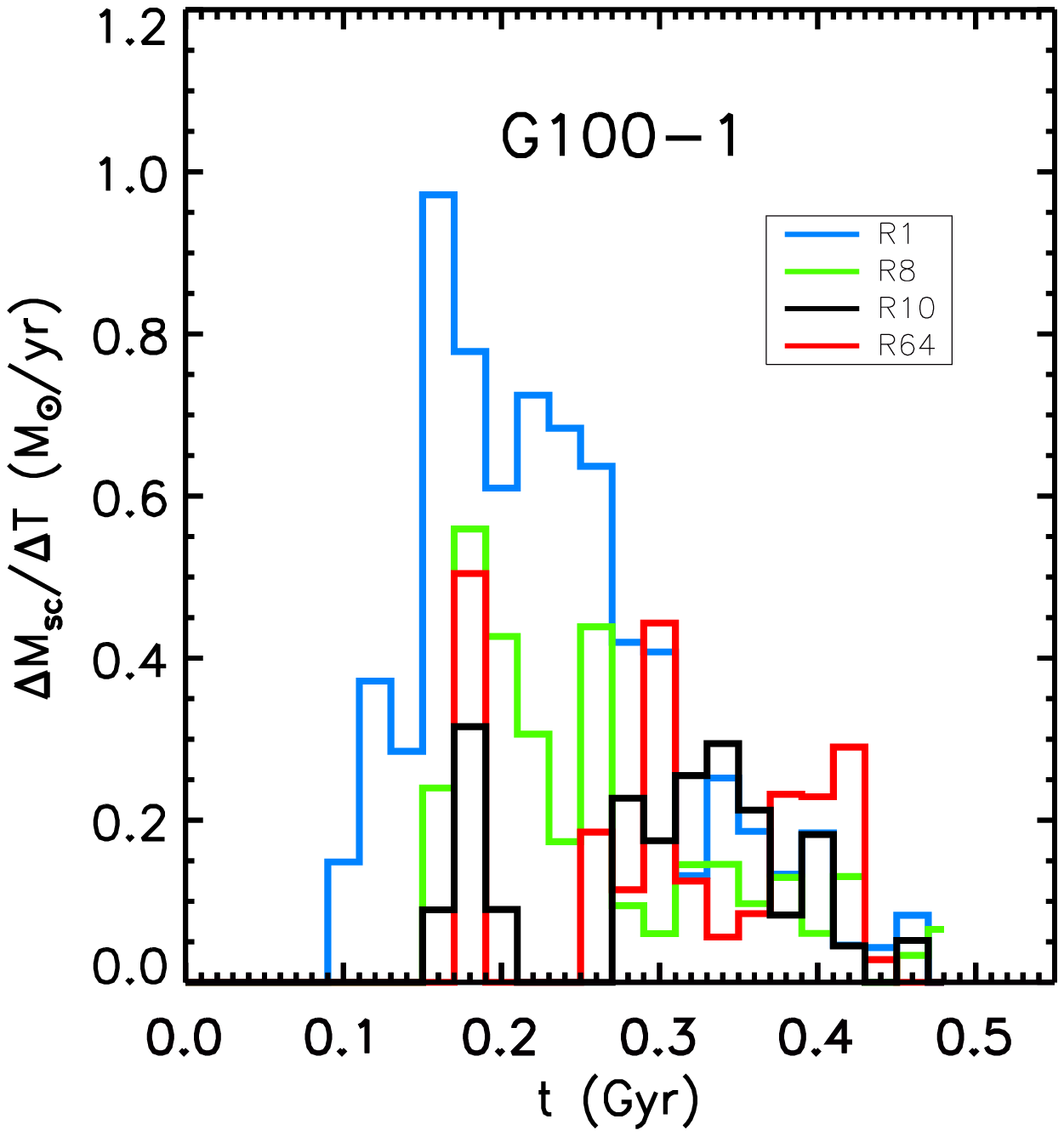}
\caption{\label{fig_when_res}
Same as Figure~\ref{fig_when}, but for resolution
  study of model G100-1 with R1 ({\em blue}), R8 ({\em green}), and
  R64 ({\em red}). The standard resolution of R10 is also shown ({\em
    black}). } 
\end{center}
\end{figure}

Figure~\ref{fig_when} shows the star formation history over the first
500 Myrs of representative models, measured by the mass of gas accreted into
sink particles over periods of 20 Myr, multiplied by our assumed local SFE of 30\%.
This figure has the same arrangement as Figure~\ref{fig_3q}. Comparison of
models with the same gas fraction but different total mass, and models with
the same total mass but different gas fraction, illustrates that the star
formation timescale and the mass of gas turned into stars depend strongly on
the total mass and the gas fraction of the galaxy. Stars form early and
efficiently in unstable galaxies, in contrast to late and slow formation in
more stable ones. Figure~\ref{fig_when_res} demonstrates that
failing to resolve the Jeans length can lead to a factor of two
overestimate in the star formation rate, as can be seen by comparison
of R1 to the higher resolutions. Once the Jeans length is resolved,
the time history is generally well converged, although details vary
between models with different resolution.

\begin{figure}
\begin{center}
\includegraphics[width=3.5in]{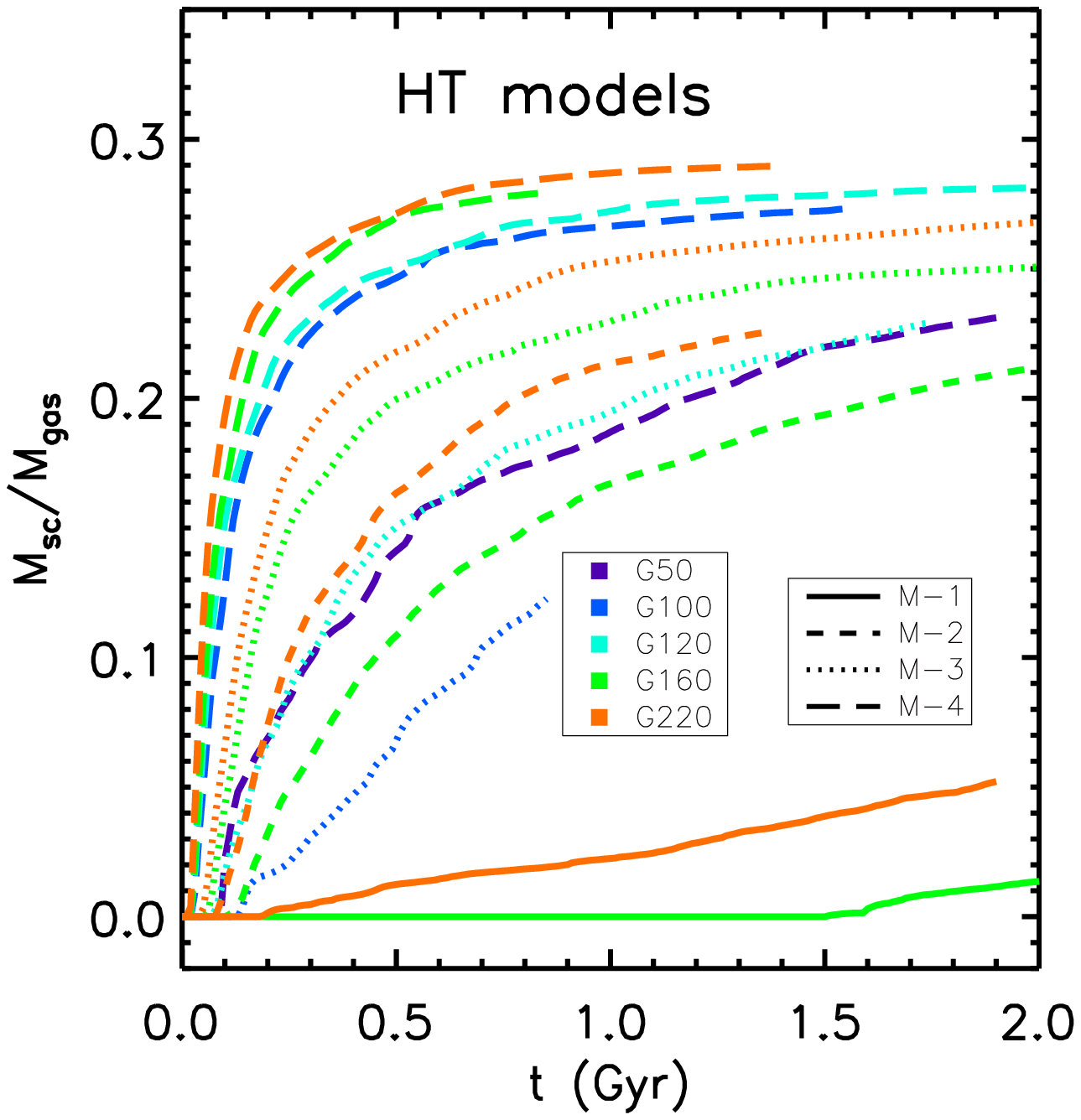}\\
\includegraphics[width=3.5in]{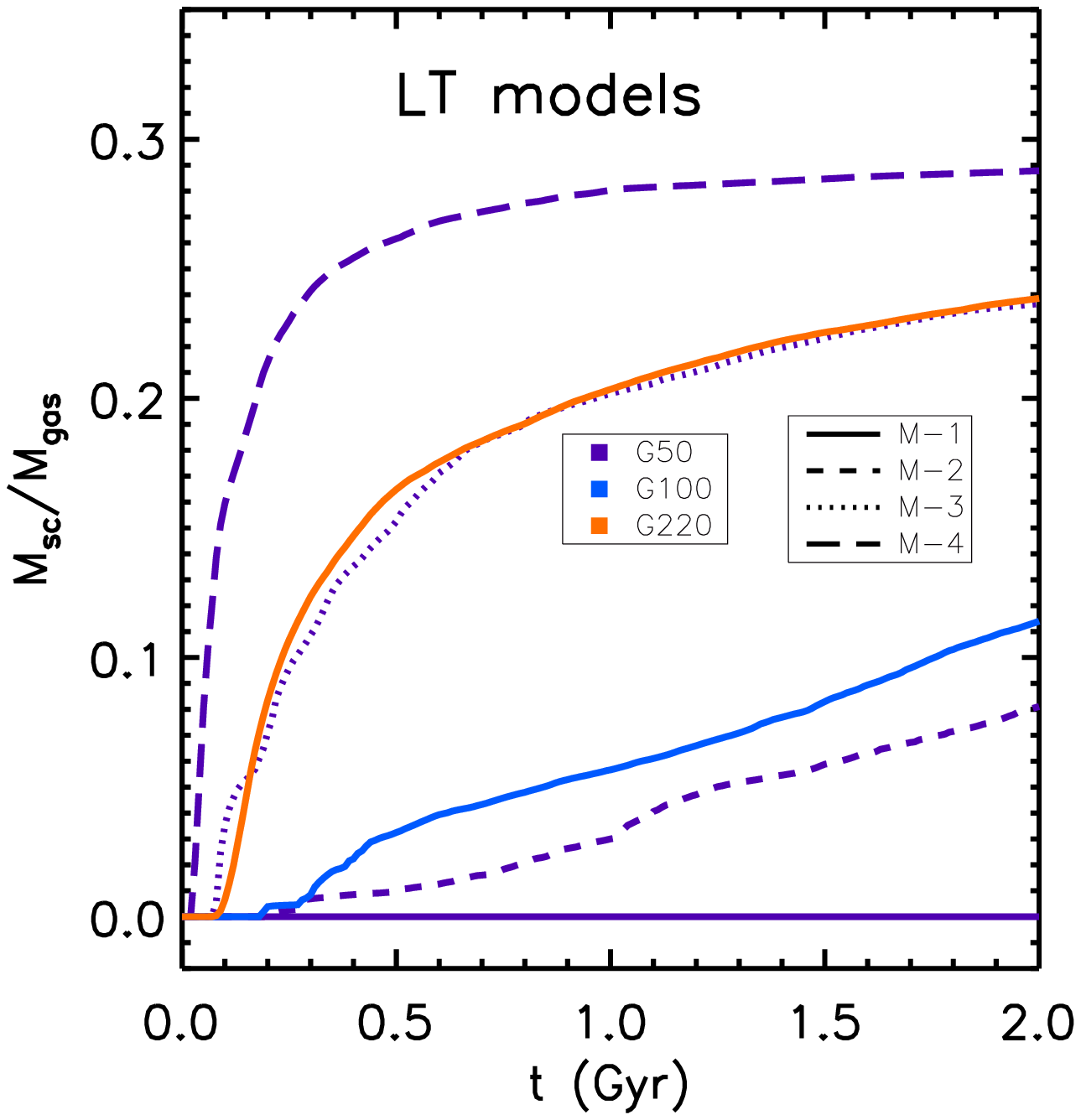}
\caption{\label{fig_sink} 
Time history of the mass in star clusters $M_{\rm sc}$
in all ({\em a}) high $T$ and ({\em
    b}) low $T$ galaxy models in Table~2 and 3, scaled by initial
  total gas mass $M_{\rm gas}$. Again $M_{\rm sc}$ is taken as 30\% of the
    mass of the sink particles.
M-1 to 4 indicate submodels 
with increasing
gas fractions, as 
given in Table~1.
}
\end{center}
\end{figure}

Figure~\ref{fig_sink} shows integrated time histories of star
formation normalized by the initial total gas mass $M_{\rm
  sc}(t)/M_{\rm gas}$. (Note that the global SFE can never exceed our
assumed local SFE of 30\%.) These histories reflect the same dependence on
stability as the differential histories shown in Figure~\ref{fig_when}. Fast
star formation occurs early and goes on until all the gas has collapsed into
sink particles after a few hundred megayears, while slow star formation occurs
late and tends to last over many gigayears. 

\begin{figure}
\begin{center}
\includegraphics[width=4in]{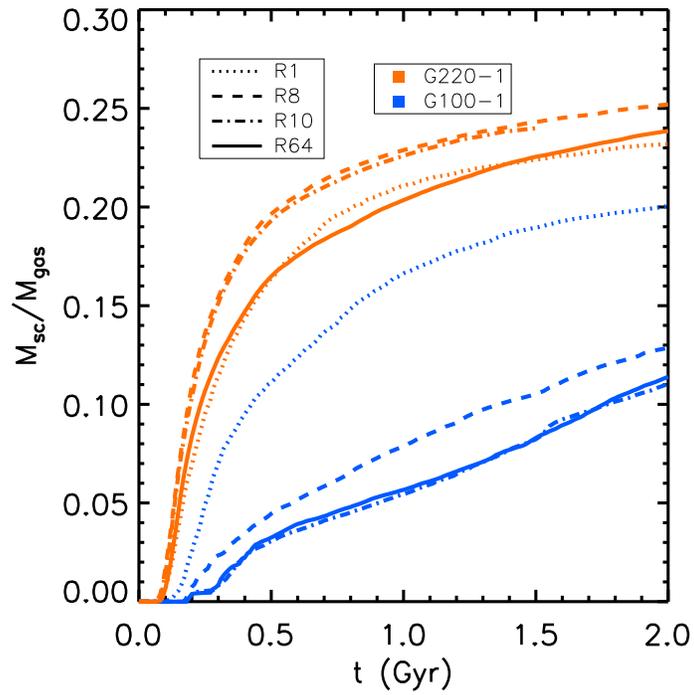}
\caption{\label{fig_sink_res}
 Same as Figure~\ref{fig_sink} but for resolution study of models G100-1 and
G220-1, with resolution levels of R1
  ({\em dotted line}), R8 ({\em dashed line}), and R64 ({\em solid
line}). The standard resolution R10 model is also shown ({\em dash-dotted
 line}).} 
\end{center}
\end{figure}

In agreement with Figure~\ref{fig_when}, Figure~\ref{fig_sink_res} shows that
spurious fragmentation can dominate the accreted mass (multiplied in the
Figures by our assumed SFE) in our low-resolution model R1, but that the
accreted mass has converged to within 10\% in model R8 as compared to model
R64. The integrated mass converges better than the differential mass shown in
Figure~\ref{fig_when_res}, suggesting that the variations seen there reveal
fluctuations in a process that produces a more stable average behavior. Both
models of G100-1 and G220-1 show convergence of the integrated 
mass accreted in runs that satisfy the three numerical criteria, 
suggesting that our standard criteria are sufficient for the problem studied
here.

The mass curves in Figure~\ref{fig_sink} appear to have a functional
behavior of the form
\begin{equation}
\label{eq_mas}
\frac{M_{\rm sc}}{M_{\rm gas}} = M_0\left[1 - \exp(-t/\tau_{\rm SF})\right] \,,
\end{equation}
where $M_0$ is the maximum fraction of initial total gas turned into star clusters,
and $\tau_{\rm SF}$ is the star formation timescale. Again, bear in mind that
this is a description of the physical response of a galactic disk to the initial
conditions for each model, not a description of the star formation
history of a real galaxy interacting with its cosmic environment.

\begin{figure}
\begin{center}
\centerline{a}
\includegraphics[width=3.5in]{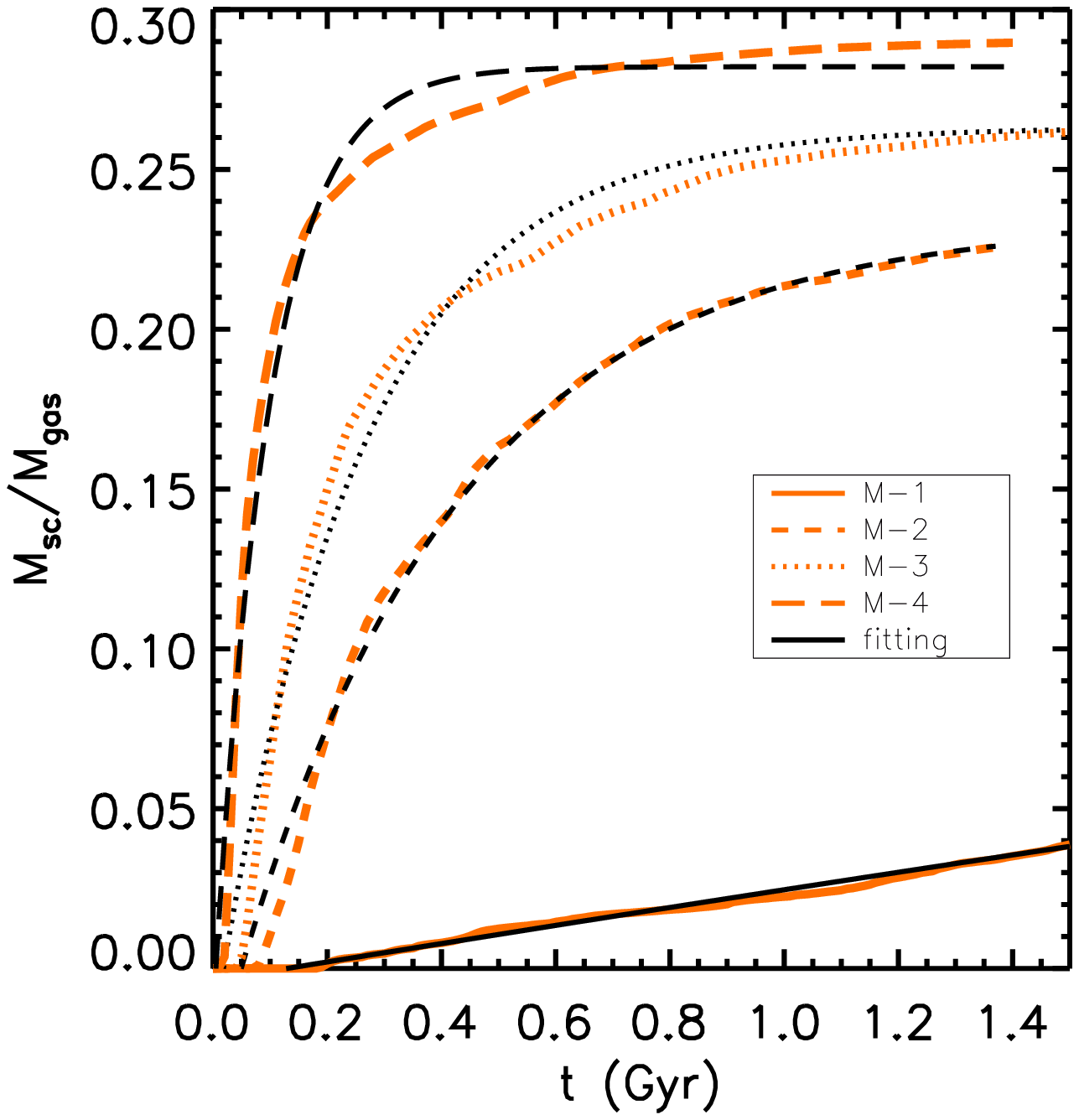}
\centerline{b}
\includegraphics[width=3.5in]{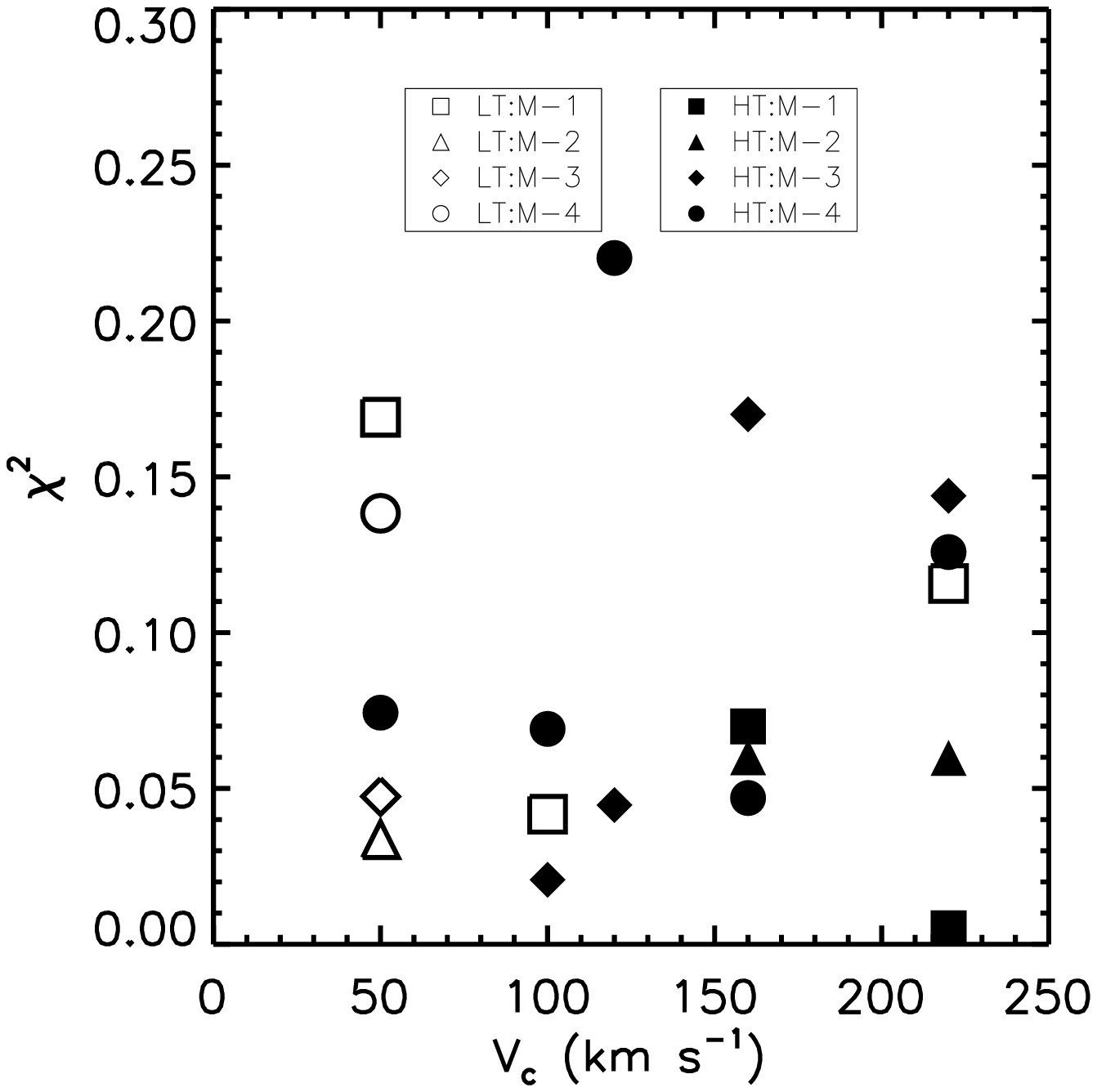}
\caption{\label{fig_fit} Same as Figure~\ref{fig_sink} but with fits to
  equation~(\ref{eq_mas}). \textit{a:} Examples of fits to high $T$
  model G220.  \textit{b:} The relative goodness of the fits indicated
  by $\chi^2$, as defined in the text. }
\end{center}
\end{figure}

Figure~\ref{fig_fit}{\em (a)} demonstrates that equation~\ref{eq_mas} 
fits models of G220 rather well. Figure~\ref{fig_fit}{\em (b)} shows the
statistical relative goodness of the fits, indicated by the parameter
$\chi^2 = \sum (y_i - y_f)^2$, where $y_i$ are points drawn from the
simulation history, while $y_f$ are analytic values from the best-fit
parameters in equation~(\ref{eq_mas}).

\begin{figure}
\begin{center}
\centerline{a}
\includegraphics[width=3.5in]{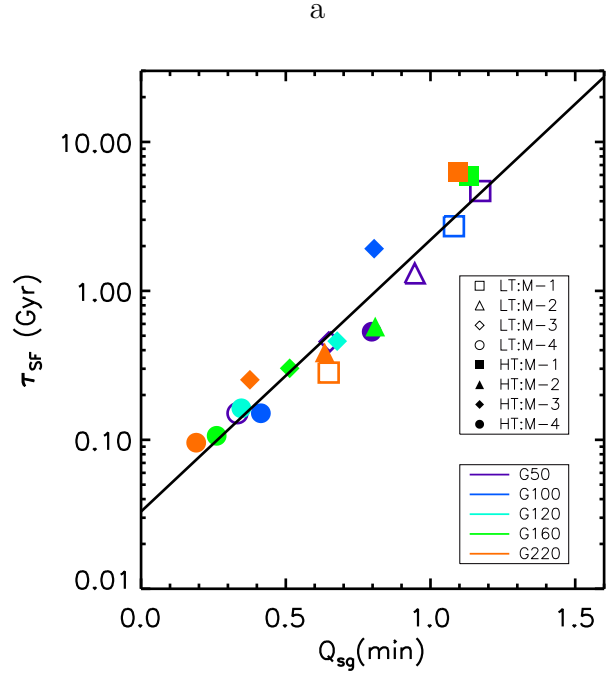}
\centerline{b}
\includegraphics[width=3.5in]{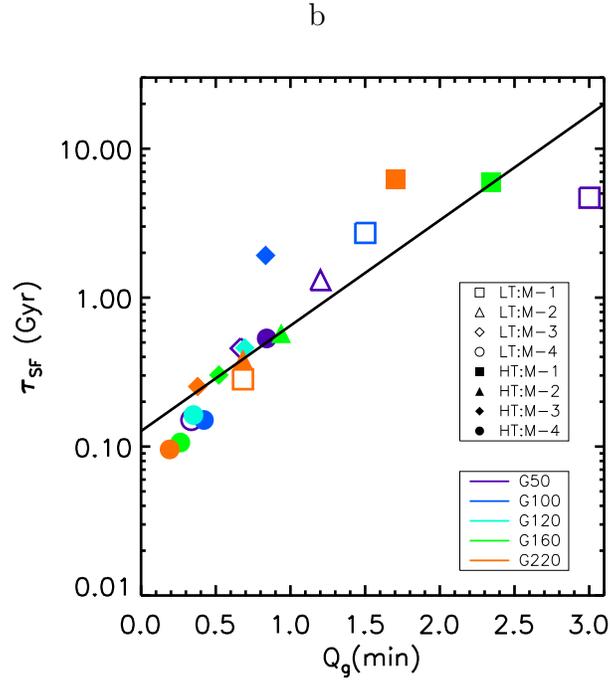}
\caption{\label{fig_tsf} Star formation timescale $\tau_{\rm SF}$ correlates
  with the initial disk instability (\textit{a}) $Q_{\rm sg}$ (from
\citealt{li05}) and (\textit{b}) $Q_{\rm g}$ for both low $T$ (open symbols) and
high $T$ (filled symbols) models. The solid lines are least-squares fits to
the data.} 
\end{center} 
\end{figure}

The star formation timescale $\tau_{\rm SF}$ appears to be closely
correlated with the minimum initial values of the gravitational
instability parameters $Q_{\rm g}(min)$ and $Q_{\rm sg}(min)$
\citep{li05}. Figure~\ref{fig_tsf} shows the relations between $\tau_{\rm
SF}$ and $Q_{\rm sg}(min)$ and $Q_{\rm g}(min)$ for all high-resolution
models. An exponential function describes both relations rather well, with the
correlation being rather better to $Q_{\rm sg}(min)$ than to $Q_{\rm g}(min)$.
The best fits are  
\begin{eqnarray}
\tau_{\rm SF} & = & \left(34 \pm 7 \mbox{Myr}\right) \exp\left[Q_{\rm sg}(min)/0.24\right], \\ 
\tau_{\rm SF} & = & \left(127 \pm 9 \mbox{Myr}\right) \exp\left[Q_{\rm g}(min)/0.61\right].
\end{eqnarray}

In stable galaxies with large $Q_{\rm sg}$, $\tau_{\rm SF}$ is large,
and stars form slowly over a long time. They resemble normal disk galaxies.
In unstable galaxies with small $Q_{\rm sg}$, $\tau_{\rm SF}$ is
small, and vigorous star formation occurs until the gas is used up or
the galaxy is stabilized again. They resemble starburst galaxies.
Typical observed starburst times of order $10^8$ yr are consistent with our
fits for $\tau_{\rm SF}$ in unstable galaxies \citep{kennicutt98a}. 

For a Milky Way-size disk model, $M_{\rm 200} \approx 1.3\times 10^{12}\
M_{\odot}$, $m_{\rm d} \approx 6\%$ , and $f_{\rm g} \approx 7\%$  
(\citealt{dubinski96, dehnen98, dubinski05}), the calculated minimum
initial Toomre Q parameter of $Q_{\rm sg}(min) \approx 1.2$ with
effective sound speed $c_{\rm s} = 6$ km s$^{-1}$.  The predicted star
formation timescale according to Figure~\ref{fig_tsf} is around 5.2
Gyrs. If a bulge with mass of $\sim 10^{10} M_{\odot}$ and size of
$\sim 2$ kpc is included in the model, the resulting
minimum initial $Q_{\rm sg}(min) \approx 1.23$, which gives $\tau_{\rm
  SF} \simeq 6$ Gyrs, agreeing well with observations.

\begin{figure}
\begin{center}
\centerline{a}
\includegraphics[width=3.5in]{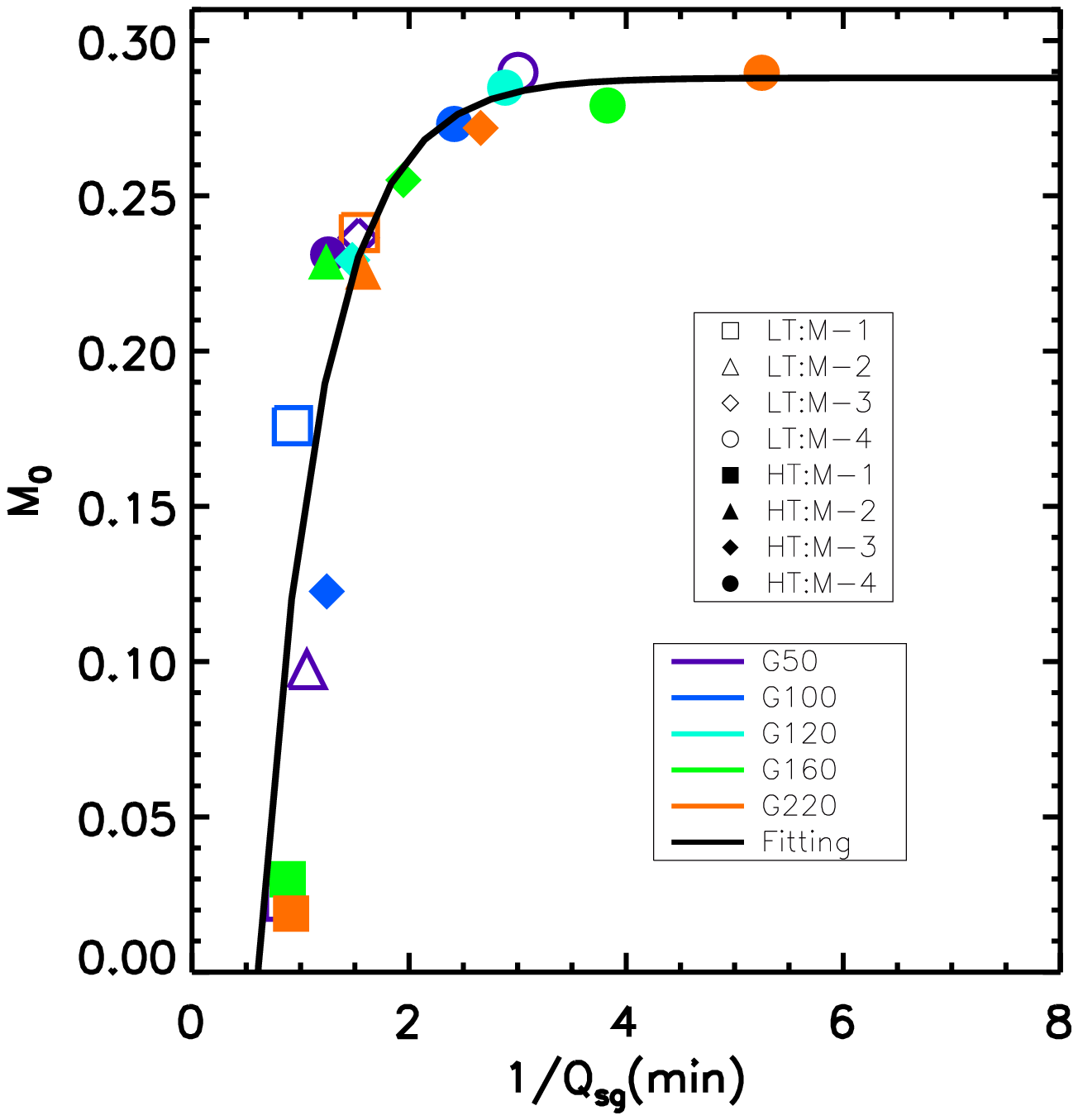}
\centerline{b}
\includegraphics[width=3.5in]{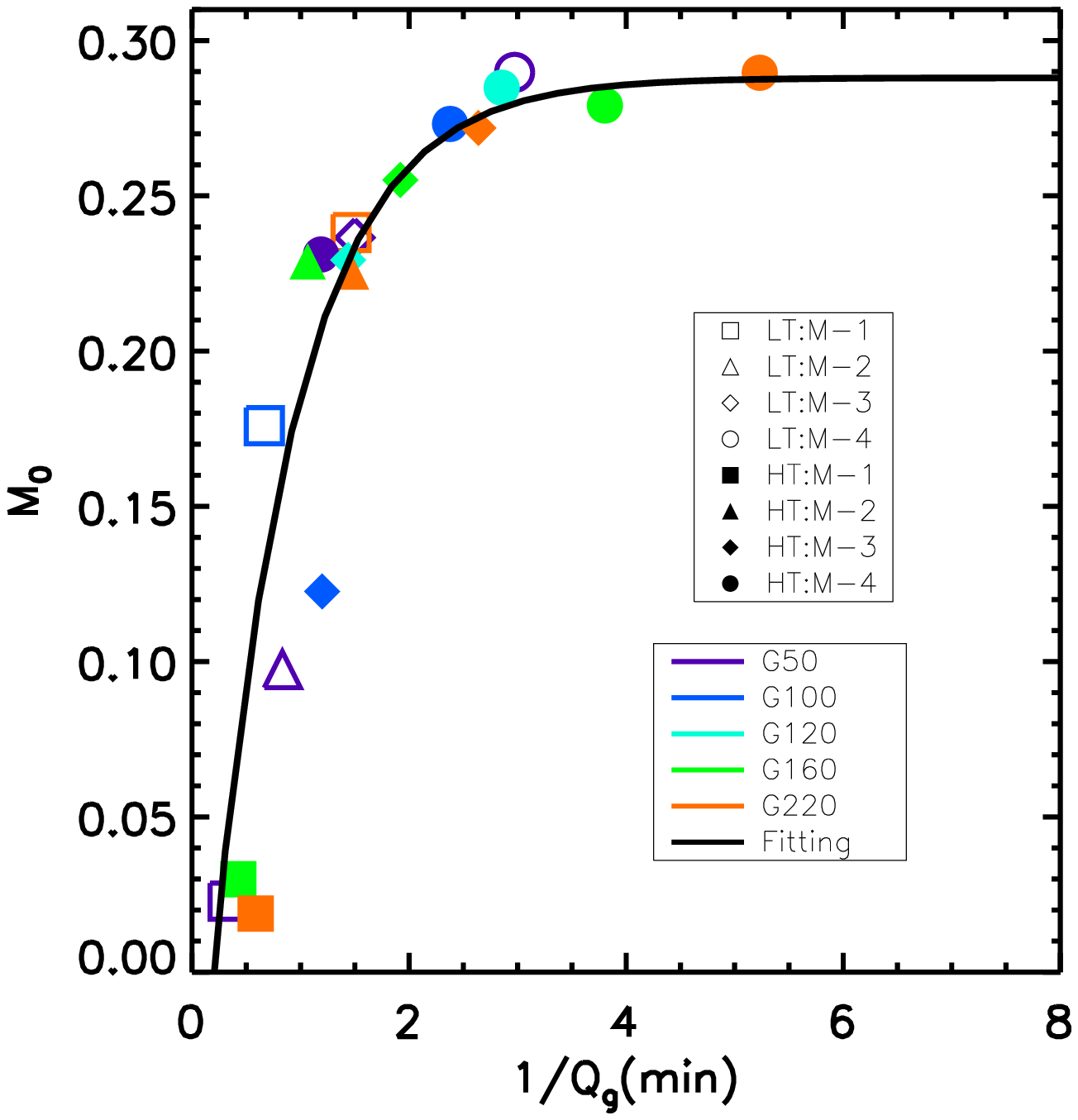}
\caption{\label{fig_m0} The maximum fraction $M_0$ of gas accreted
  onto clusters as a function of  (\textit{a}) $Q_{\rm sg}(min)$ and
  (\textit{b}) $Q_{\rm g}(min)$. Solid lines indicate best fits of the form
  given in equation (\ref{eq_m0}).} 
\end{center}
\end{figure}

Figure~\ref{fig_m0} shows the final fraction of total gas formed into stars 
$M_0$ as a function of the minimum initial instability parameters $Q_{\rm sg}(min)$ and
$Q_{\rm g}(min)$. The solid line is the best fit to both the low $T$
and high $T$ models, which gives: 
\begin{eqnarray}
\label{eq_m0}
M_0  & = & 0.29\{1 - 2.88\exp \left[-1.74/Q_{\rm sg}(min)\right]\}, \\
M_0  & = & 0.29\{1 - 1.28\exp \left[-1.28/Q_{\rm g}(min)\right]\} \,. 
\end{eqnarray}
(Note that our assumption of a local SFE of 30\% gives an upper limit
to $M_0$ of 0.3.) In order for a galaxy to form stars, $M_0$ should be
positive. This gives the empirical constraints $Q_{\rm sg} < 1.6$ and $Q_{\rm
g} < 5.2$.  

\section{RADIAL AND VERTICAL DISTRIBUTIONS} 
\label{sec_dist}

\begin{figure}
\begin{center}
\includegraphics[width=3in]{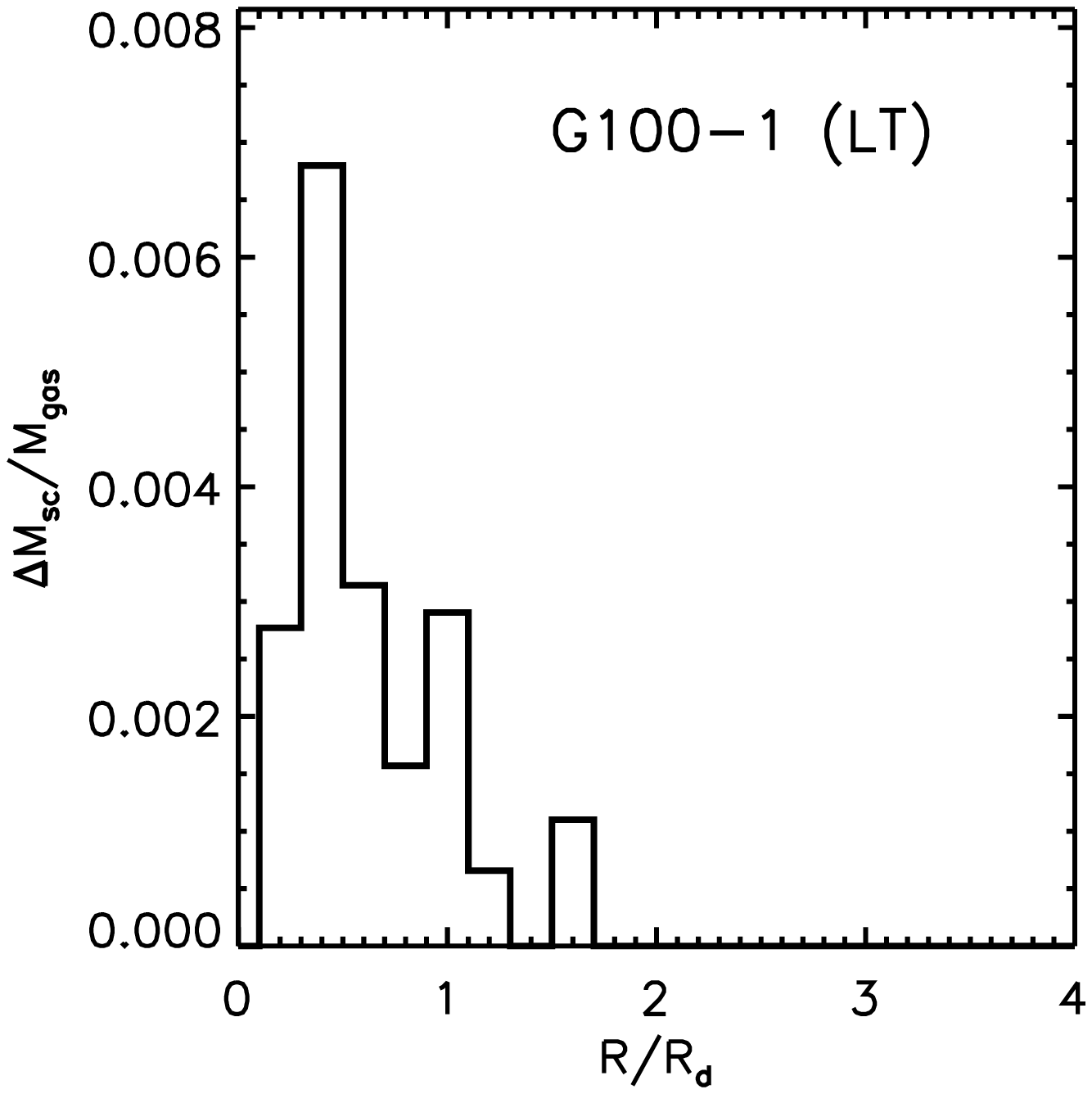}
\includegraphics[width=3in]{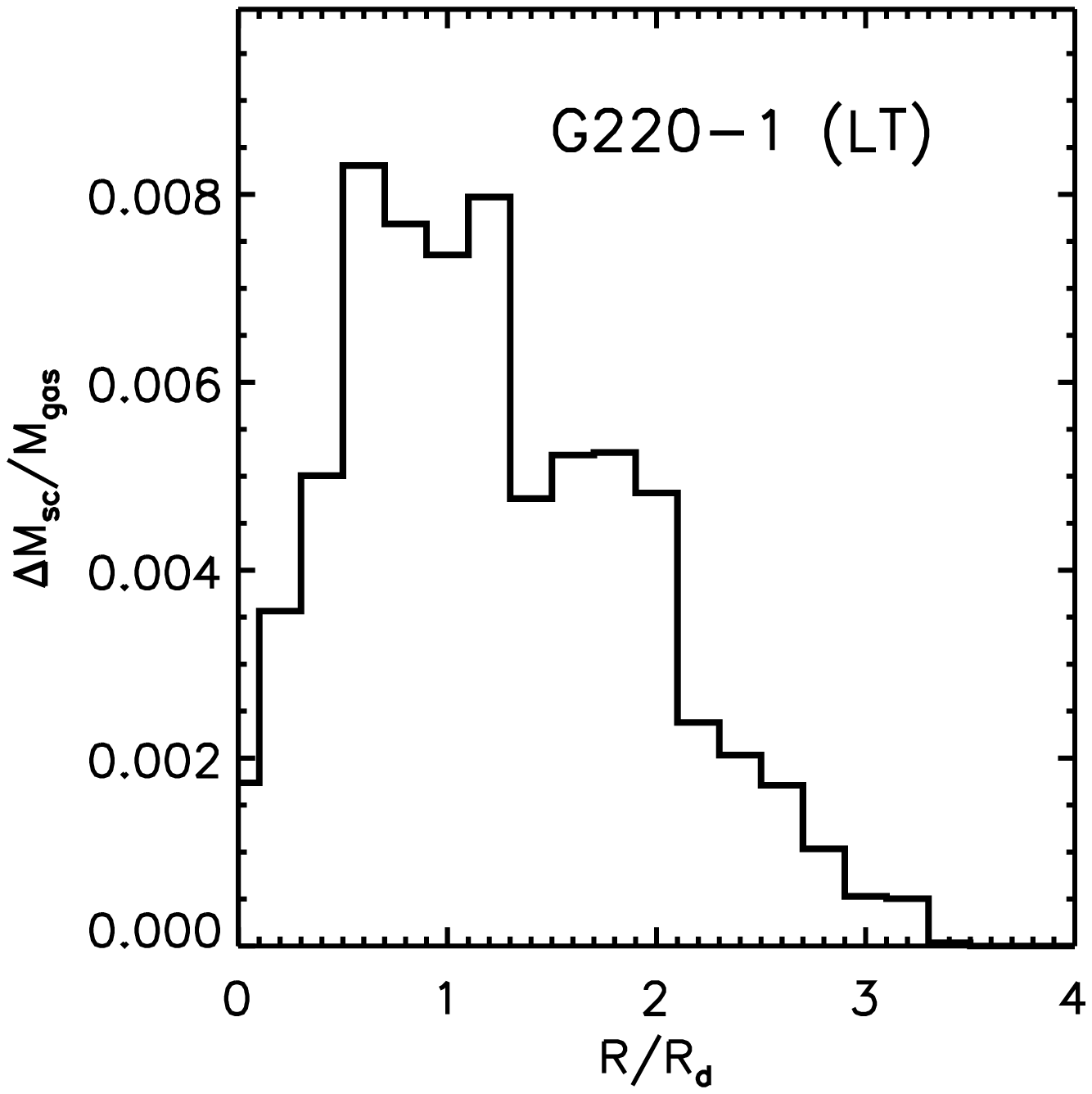}\\
\includegraphics[width=3in]{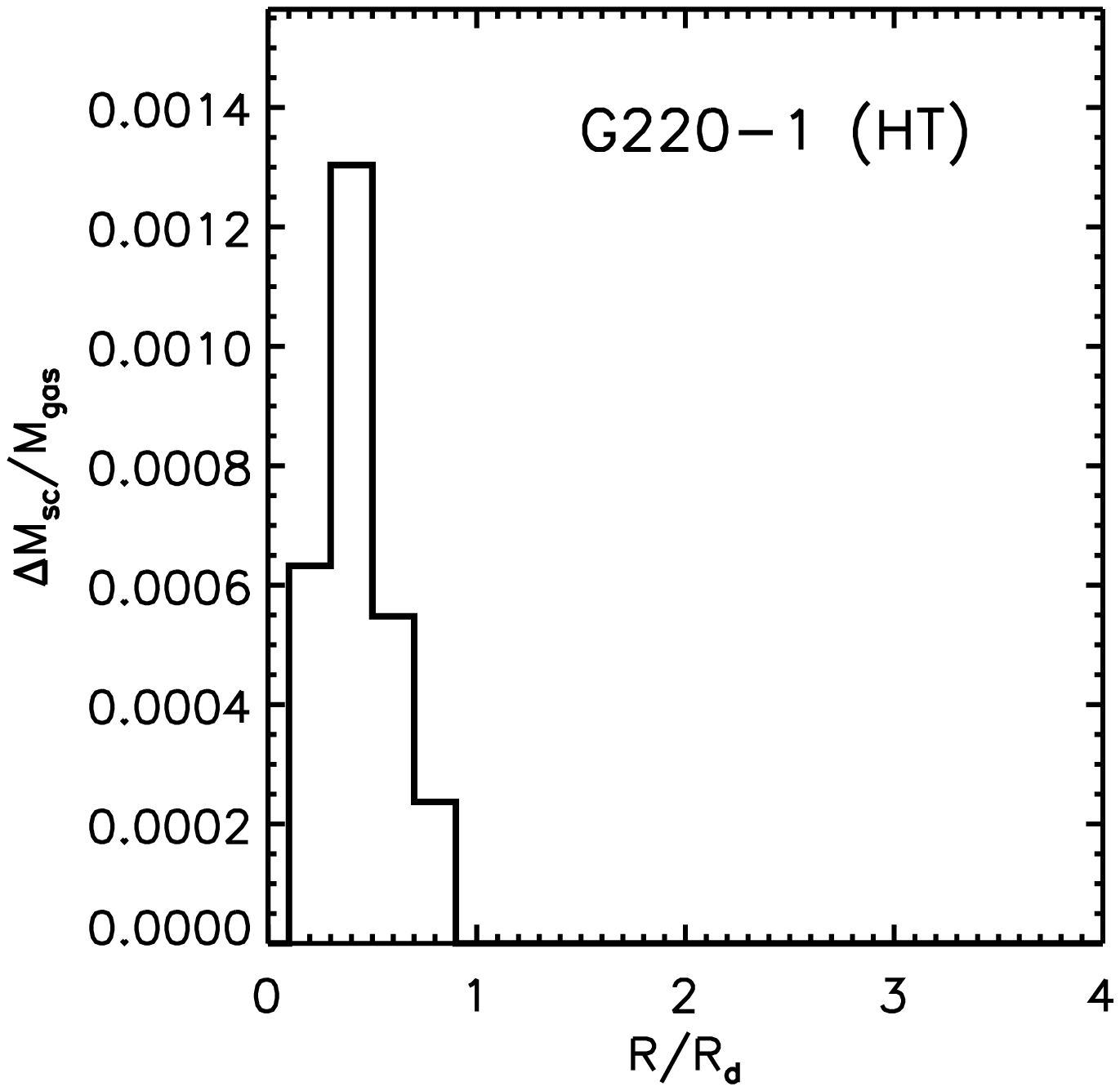}
\includegraphics[width=3in]{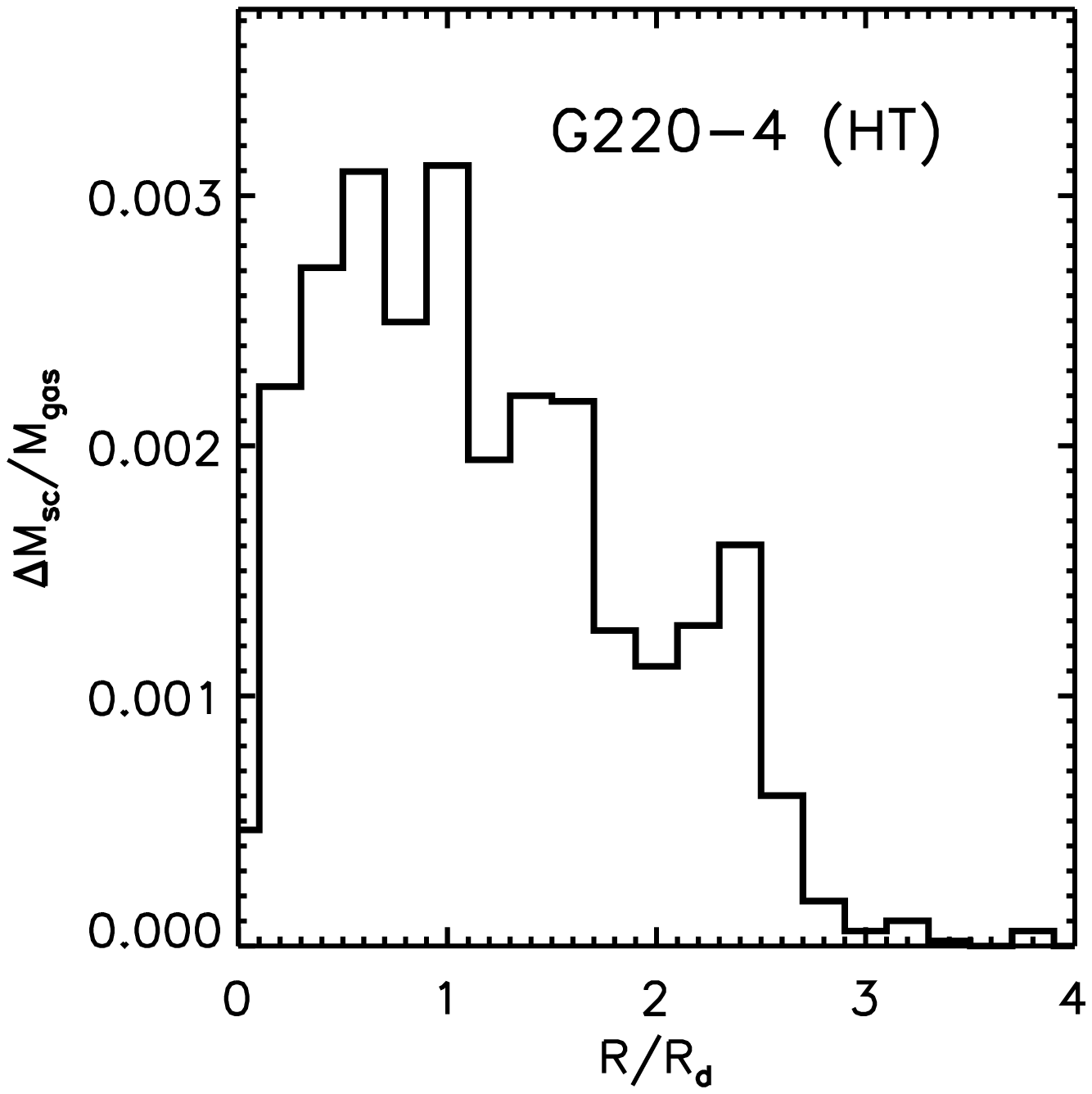}
\caption{\label{fig_where_r} Radial distribution of mass in star clusters
$\Delta M_{\rm sc}$ in a bin size of 0.2 $R_{\rm d}$, normalized to initial
gas mass $M_{\rm gas}$ after 3 Gyrs for ({\em top panels}) low $T$ galaxy
models G100-1 and G220-1 with different rotation velocities, and ({\em bottom
panels}) high $T$  models G220-1 and G220-4 with different gas fractions.}
\end{center}
\end{figure}

Figure~\ref{fig_where_r} shows the radial distribution of 
gravitational collapse and star formation in different low $T$ models. The
arrangement of the panels is the same as in Figure~\ref{fig_3q} for direct
comparison. Star formation takes place in the unstable regions
outlined by the Q-curves in Figure~\ref{fig_3q}. 

\begin{figure}
\begin{center}
\includegraphics[width=4in]{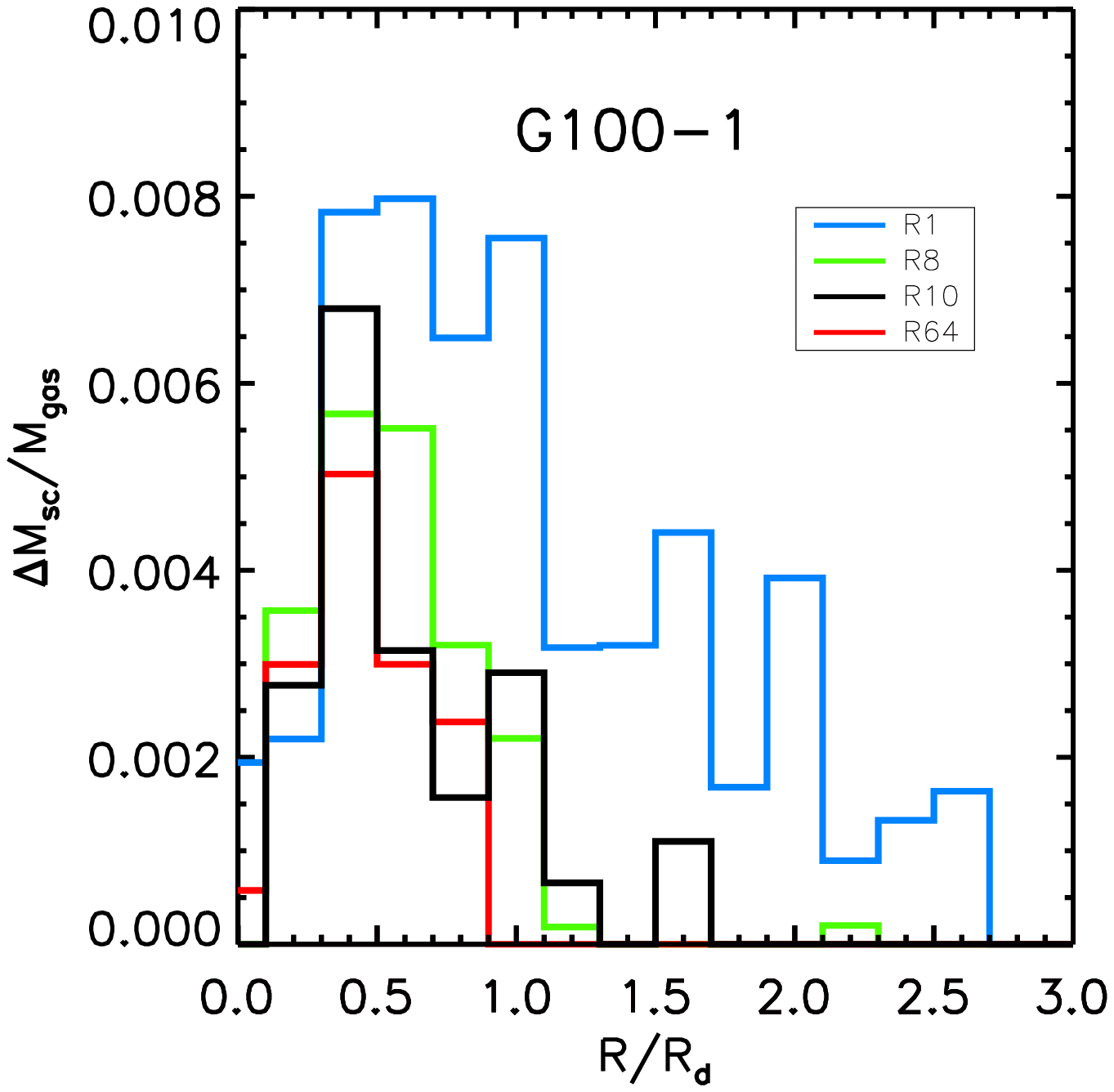}
\caption{\label{fig_res_where_r} 
  Same as Figure~\ref{fig_where_r} but for resolution study of model G100-1,
with R1 ({\em blue}), R8 ({\em green}), and R64 ({\em red}).  The
standard resolution R10 is also shown ({\em black}).} 
\end{center}
\end{figure}

The most unstable radius, where $Q_{\rm sg}$ reaches a minimum,
coincides with the radial peak of star formation in most models. 
The normalized size of the star formation region $R_{\rm SF}$
increases with the mass, or the gas fraction of the galaxy. For example,
$R_{\rm SF} \approx$ 2 $R_{\rm d}$ in G100-1, but expands to $\approx$ 5
$R_{\rm d}$ in G220-4.  Figure~\ref{fig_res_where_r} shows that 
the underresolved R1 model produces too much star formation at most
radii, but that for the standard resolution R10 model
the radial distribution of clusters has converged quite well.

\begin{figure}
\begin{center}
\includegraphics[width=3in]{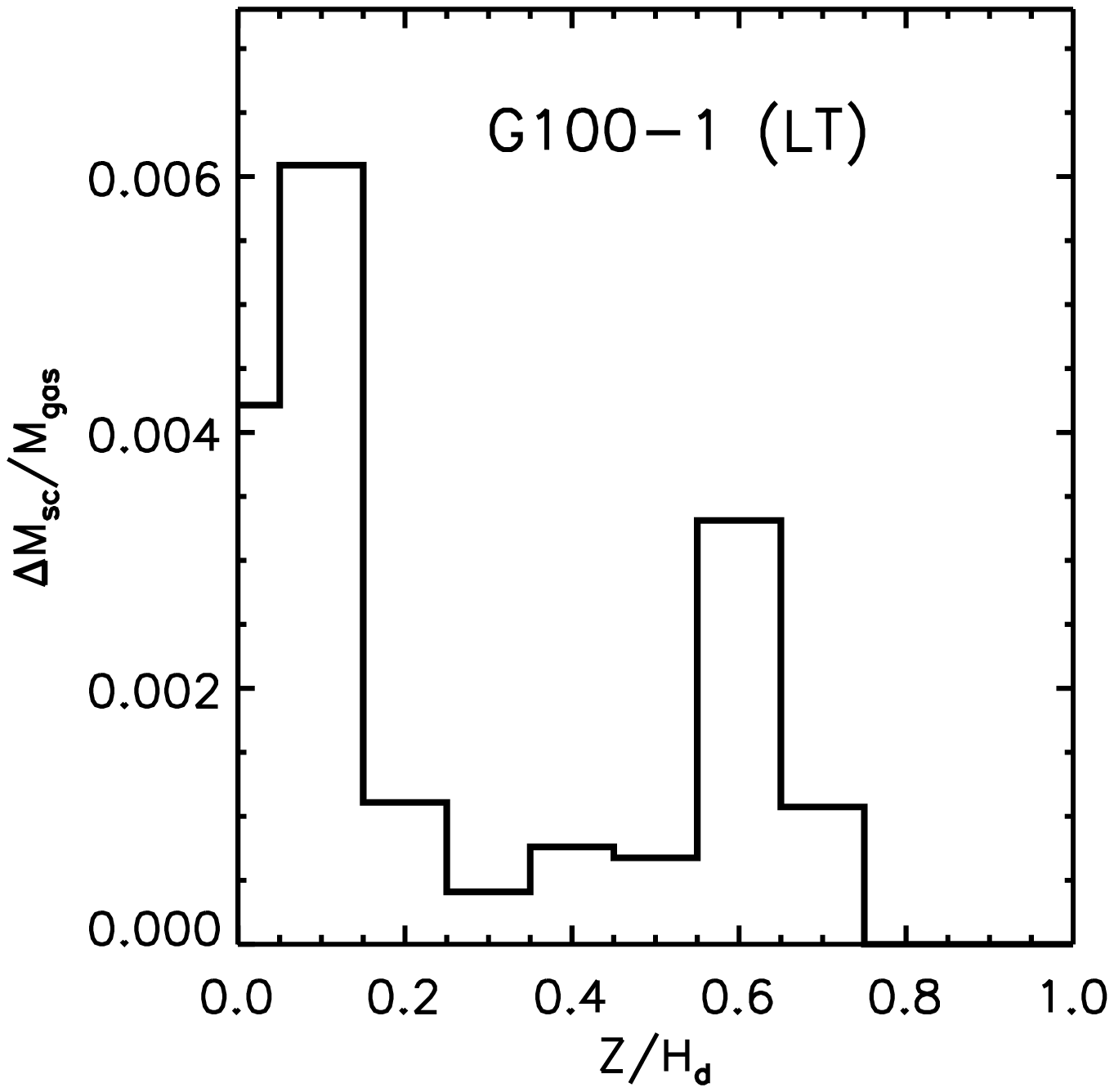}
\includegraphics[width=3in]{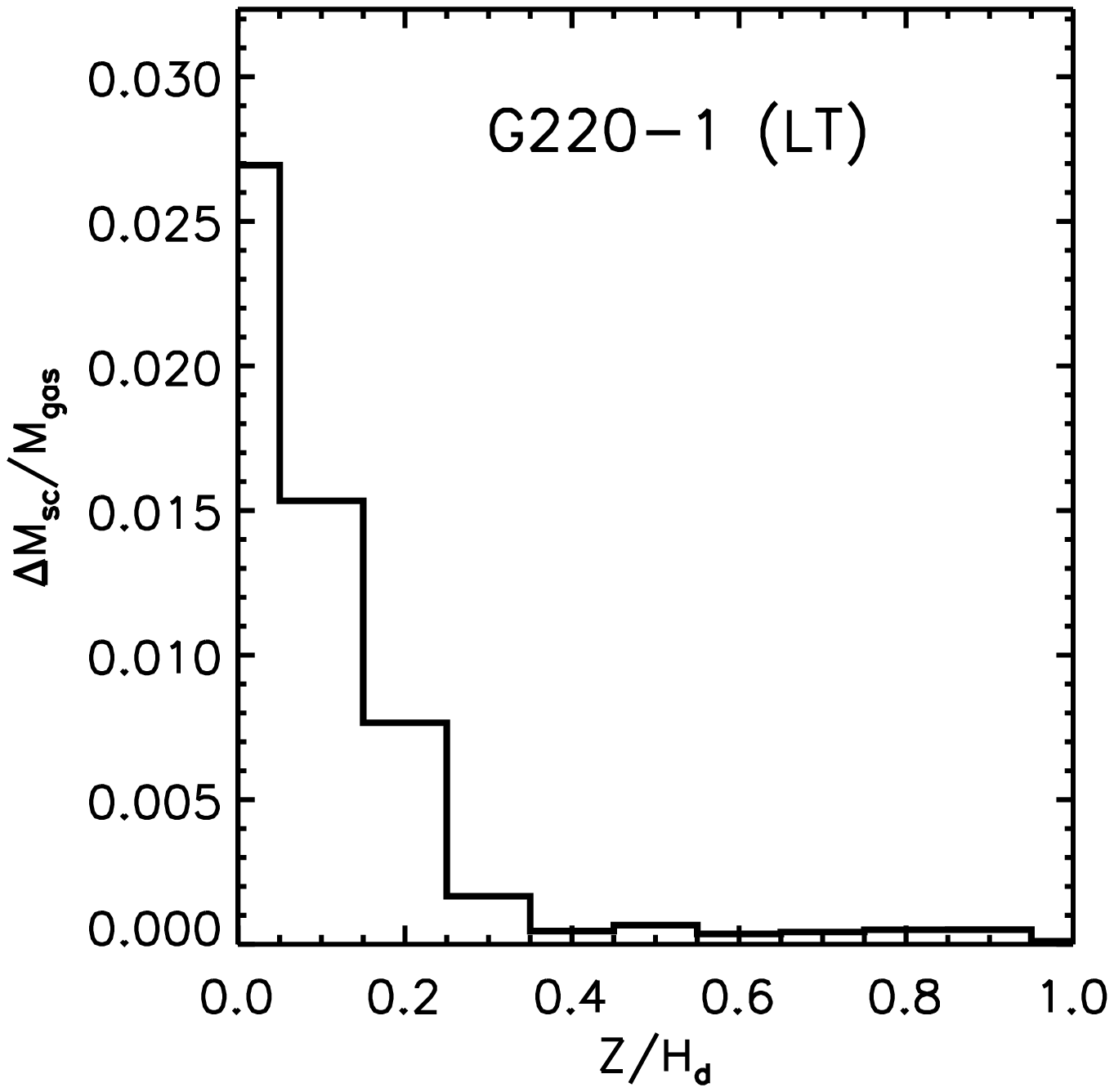}\\
\includegraphics[width=3in]{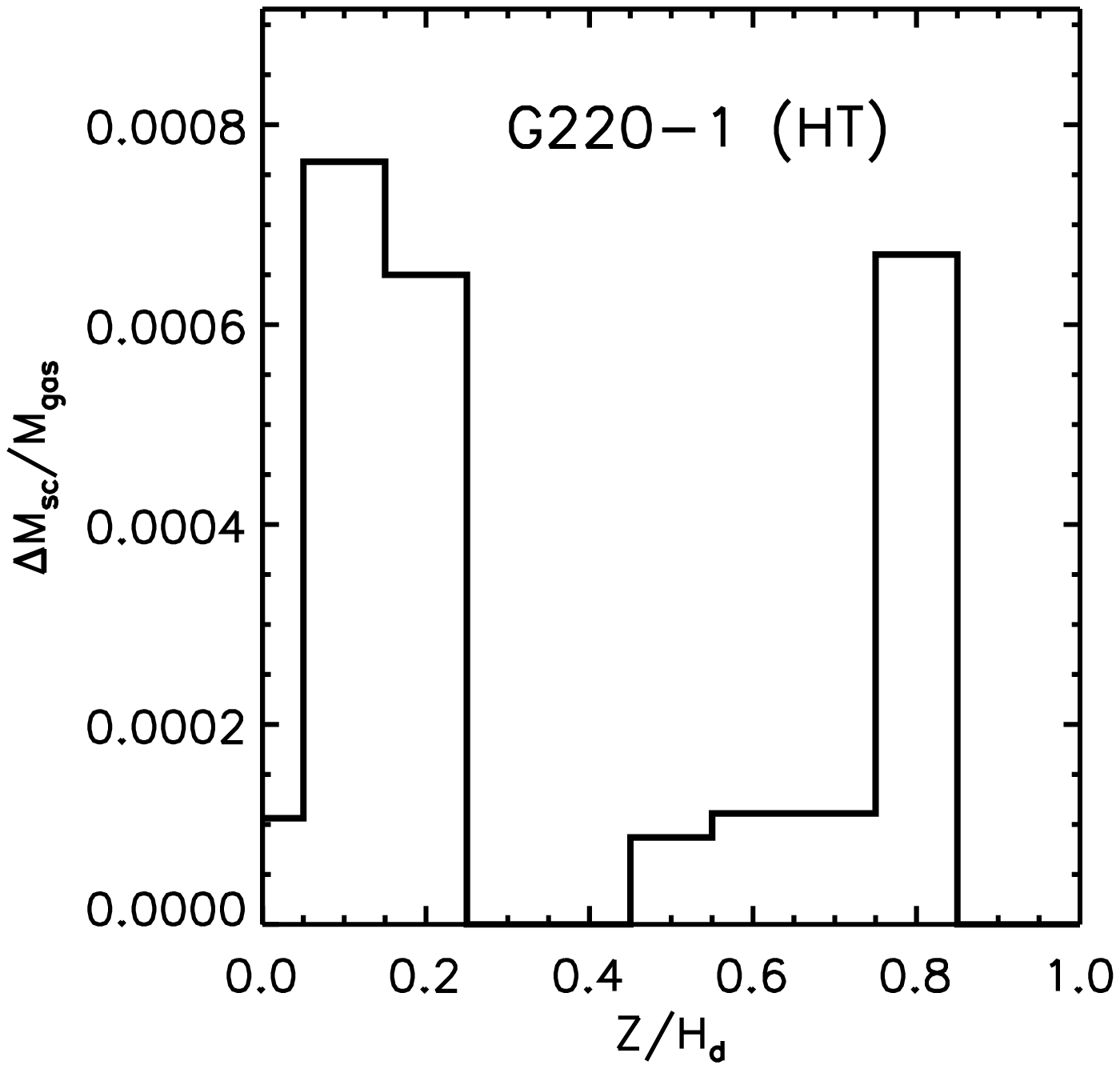}
\includegraphics[width=3in]{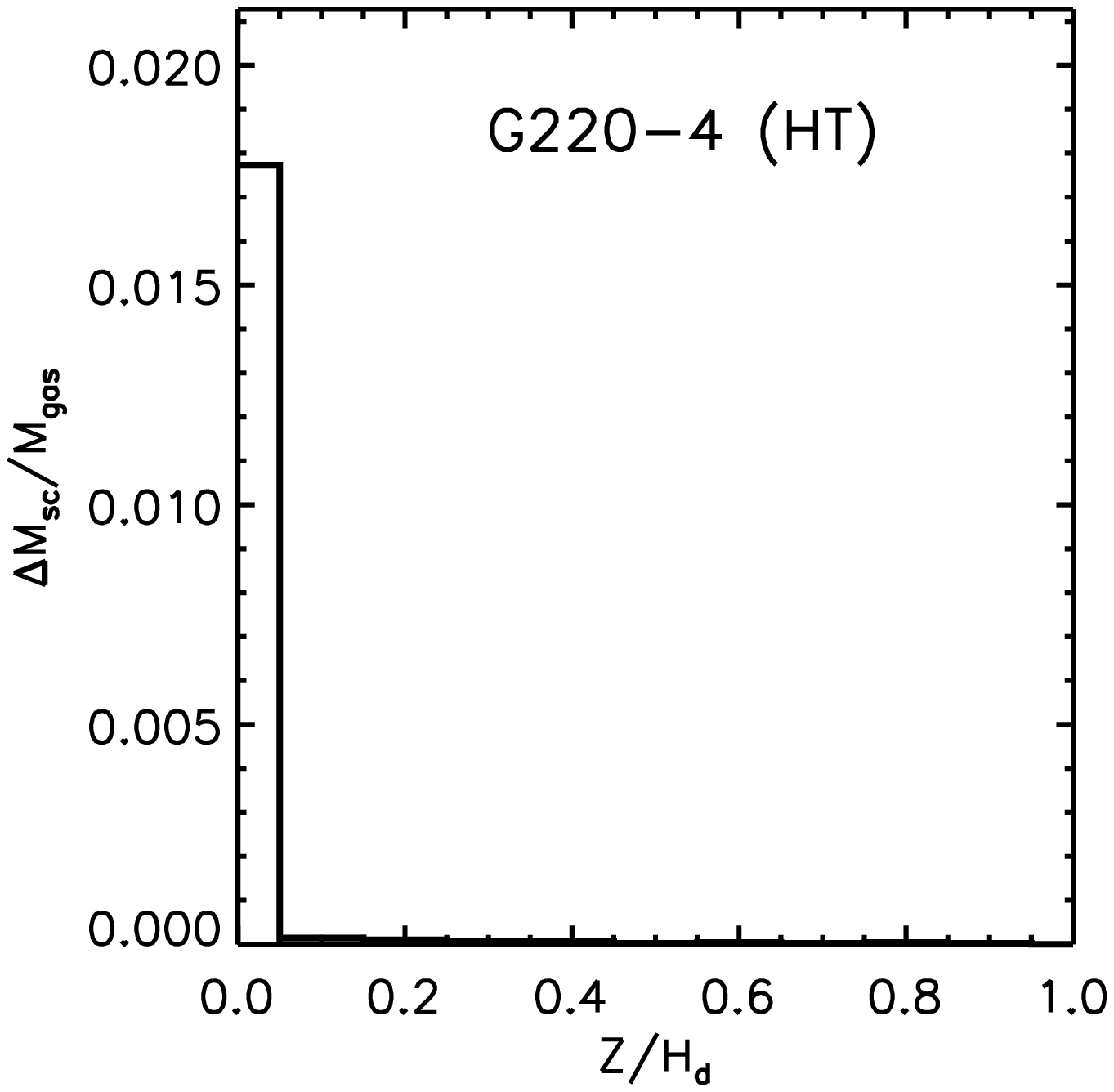}
\caption{\label{fig_where_z} Vertical distribution of mass in star clusters
$\Delta M_{\rm sc}$ in a bin size of $0.1 H_{\rm d}$, normalized to initial
gas mass $M_{\rm gas}$ after 3 Gyrs for ({\em top panels}) low $T$ galaxy
models G100-1 and G220-1 with different rotation velocities, and ({\em bottom
panels}) high $T$  models G220-1 and G220-4 with different gas fractions. 
The distance from the plane $Z$ is normalized by the vertical disk scale length
  $H_{\rm d}$.} 
\end{center}
\end{figure}

Figure~\ref{fig_where_z} shows the vertical distribution of collapse and
star formation in the same models as in Figure~\ref{fig_where_r}. The
bin size of the histogram is $0.1 H_{\rm d}$, where the vertical disk scale
length of the stars $H_{\rm d} = 0.2\ R_{\rm d}$ as set up in the models. 
The vertical thickness of the star forming region also depends on the mass and
gas fraction of the galaxy. In small, or gas-poor galaxies, stars form in a
wide spatial distribution along the vertical direction ($Z \approx 1\ H_{\rm
d}$). As the galaxy becomes more unstable, star formation becomes more
vigorous and is more concentrated in a narrow range in the galactic plane ($Z
< 0.3\ H_{\rm d}$).  

Both Figures~\ref{fig_where_r} and~\ref{fig_where_z} illustrate that the
spatial distribution of star formation in isolated disk galaxies depends on
the gravitational stability of the galaxy. In unstable galaxies, most of the
gas turns into stars quickly. The star formation regions reach several disk
scale lengths, but vertically they concentrate in the galactic plane. In more
stable galaxies, only a small fraction of the gas forms stars, over a long 
period. The star formation region is radially concentrated in the central
region but vertically extended out of the galactic plane.

\begin{figure}
\begin{center}
\includegraphics[width=4.0in]{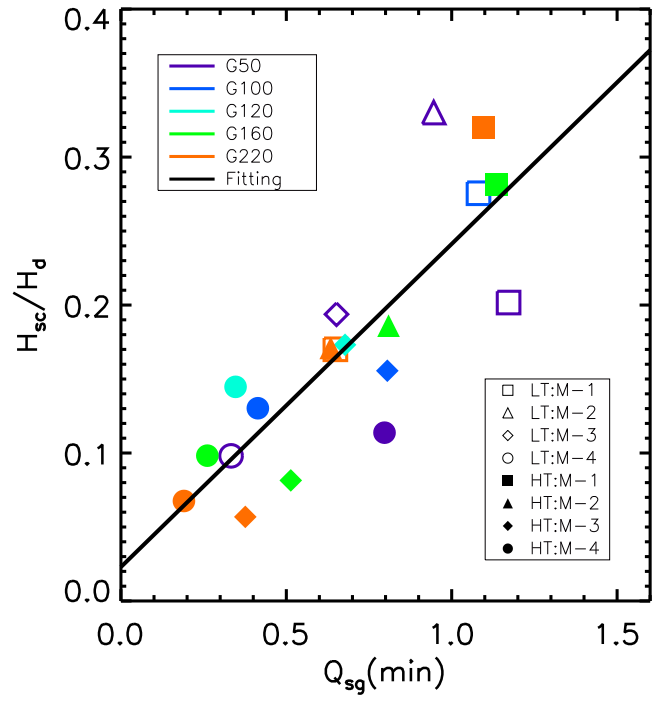}
\caption{\label{fig_zsink} 
  Vertical scale height of star clusters formed within 3 Gyrs normalized by
the vertical disk scale length, $H_{\rm sc}/H_{\rm d}$, correlates with the
initial disk instability $Q_{\rm sg}(min)$.}  
\end{center}
\end{figure}

The correlation is quantified in Figure~\ref{fig_zsink}, which shows that the
normalized vertical scale height of star clusters, $H_{\rm sc}/H_{\rm d}$
increases linearly with the minimum initial disk instability $Q_{\rm
sg}(min)$. $H_{\rm sc}$ is calculated from the variance of the vertical
distribution of the star clusters. Regions of gravitational collapse will form
molecular clouds with high extinction in addition to stellar clusters.  The
distribution of sink particles thus also roughly traces the distribution of
molecular gas, as briefly discussed by \citet{li05}. Our result that unstable
galaxies have smaller $H_{\rm sc}$ than stable galaxies thus implies
that they also have thinner dust lanes. This result agrees well with the
observations by \citet{dalcanton04}, who find that thin dust lanes only form
in nearby galaxies with $V_{\rm c} \ge 120$~km~s$^{-1}$. They find that
gas-poor modern galaxies are typically Toomre unstable using the
\citet{rafikov01} criterion when $V_{\rm c} \ge$ 120 km~s$^{-1}$. 

\section{CORRELATED STAR FORMATION}
\label{sec_corr}
Observations of star clusters in the Large Magellanic Cloud have shown
that there is a correlation between spatial separation $\Delta S$ and age
difference $\Delta T$. \citet{elmegreen98} show that $\log \Delta T \sim 0.35
\log \Delta S$, and argue that this correlation is due to the hierarchical
structure of turbulence (see also \citealt{nomura01}). 

\begin{figure}
\begin{center}
\includegraphics[width=2.5in]{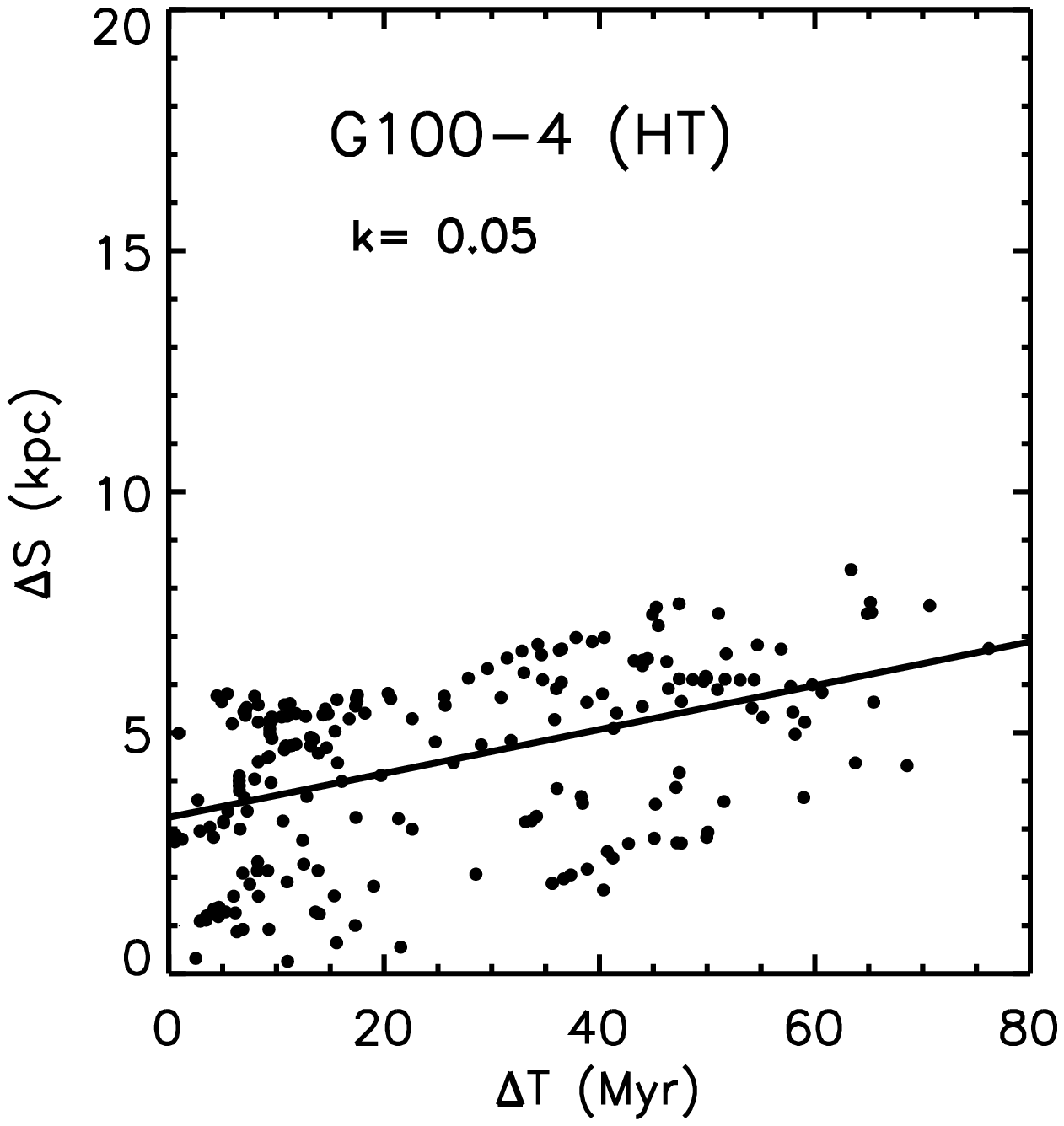}\\
\includegraphics[width=2.5in]{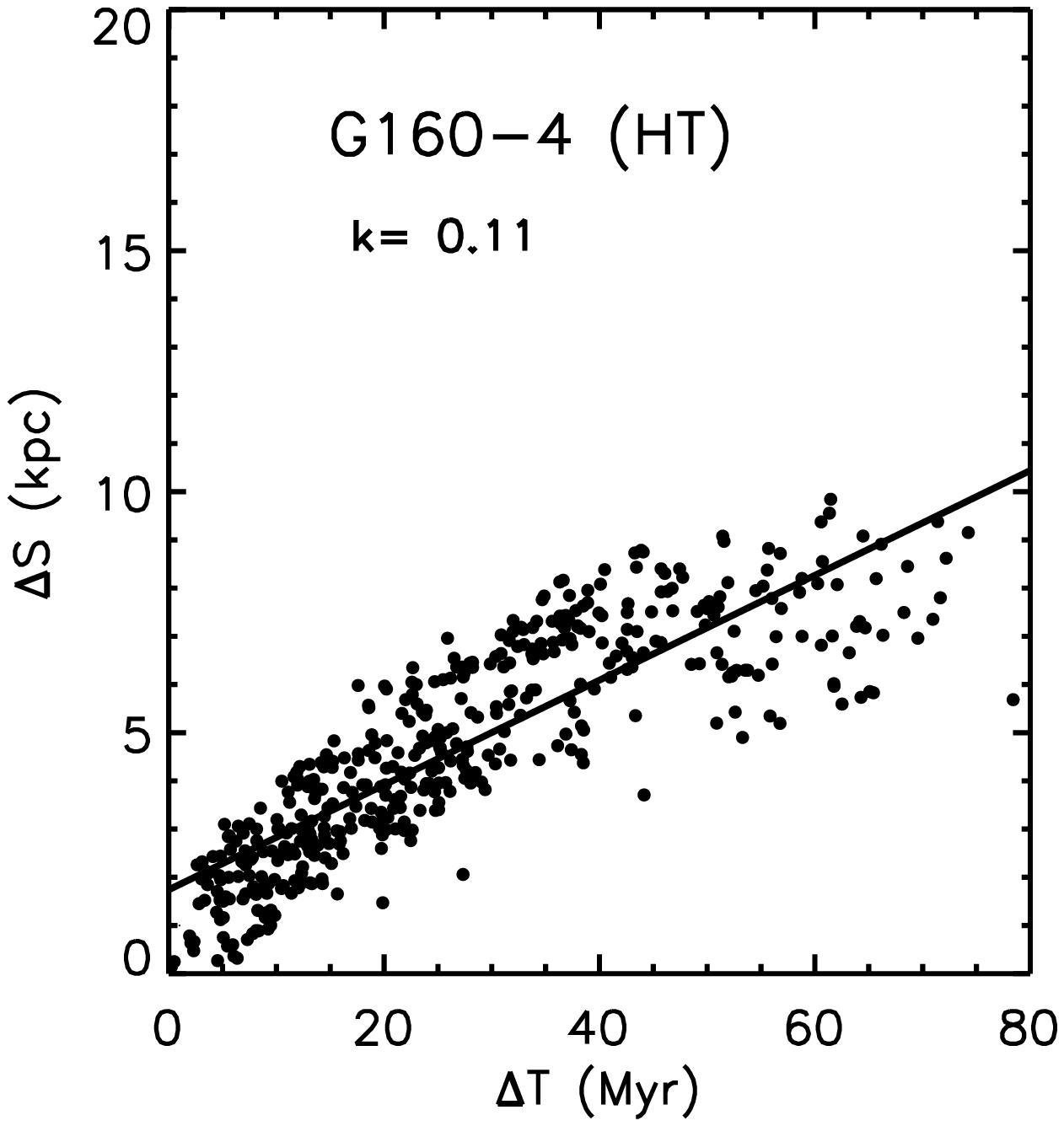}\\
\includegraphics[width=2.5in]{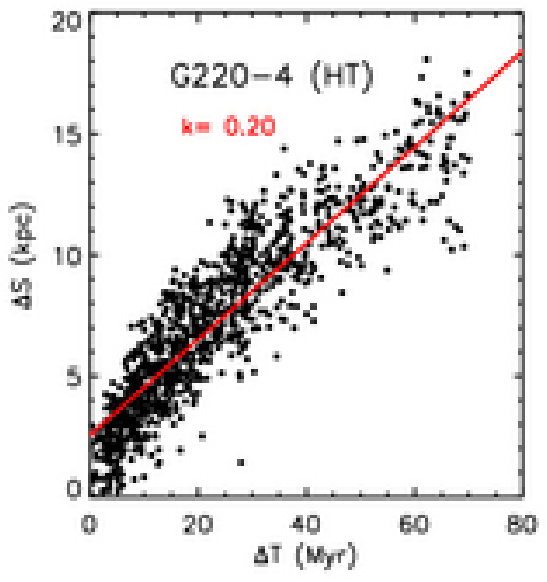}
\caption{\label{fig_clustering1} Linear correlation between separations of
  cluster pairs $\Delta S$ and their age differences $\Delta T$ in models
  G100-4, G160-4, and G220-4 at model times equal to their characteristic star
  formation timescales $\tau_{\rm SF}$.  The clusters shown here have age
  10~Myr~$< T < 100$~Myr. The red lines represent least squares fits,
  with slopes $k$ given in each panel.}
\end{center}
\end{figure}

Figure~\ref{fig_map1} shows that the distribution of star clusters in
our models also appears structured rather than
smooth. Figure~\ref{fig_clustering1} shows that for any random reference
cluster, the separation $\Delta S$ of the cluster pairs is linearly
correlated to their relative age $\Delta T$. Neighboring clusters tend to form
coevally, while distant pairs tend to form at different times. This behavior
is seen in all of our models.  

\begin{figure}
\begin{center}
\includegraphics[width=2.5in]{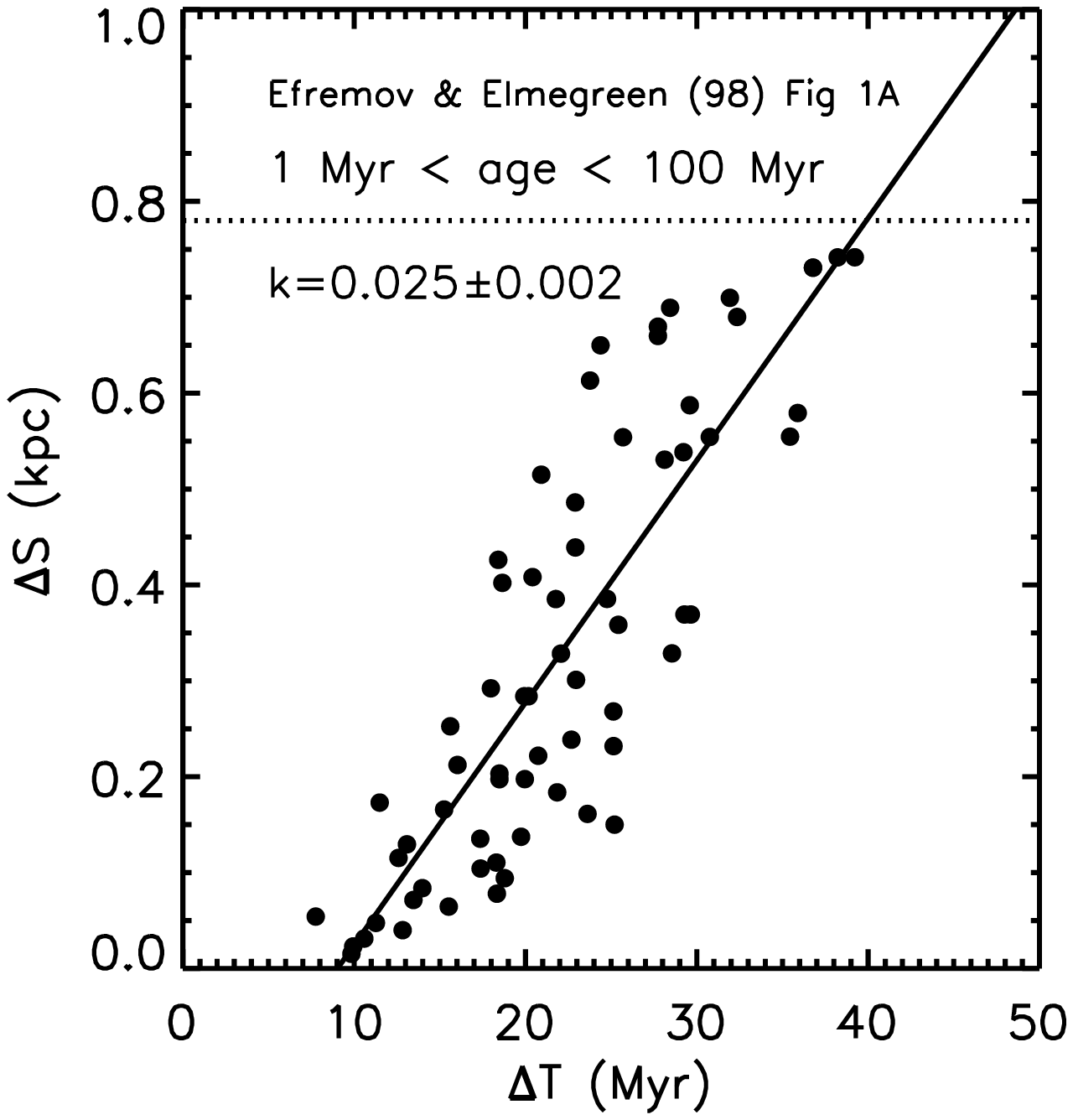}\\
\includegraphics[width=2.5in]{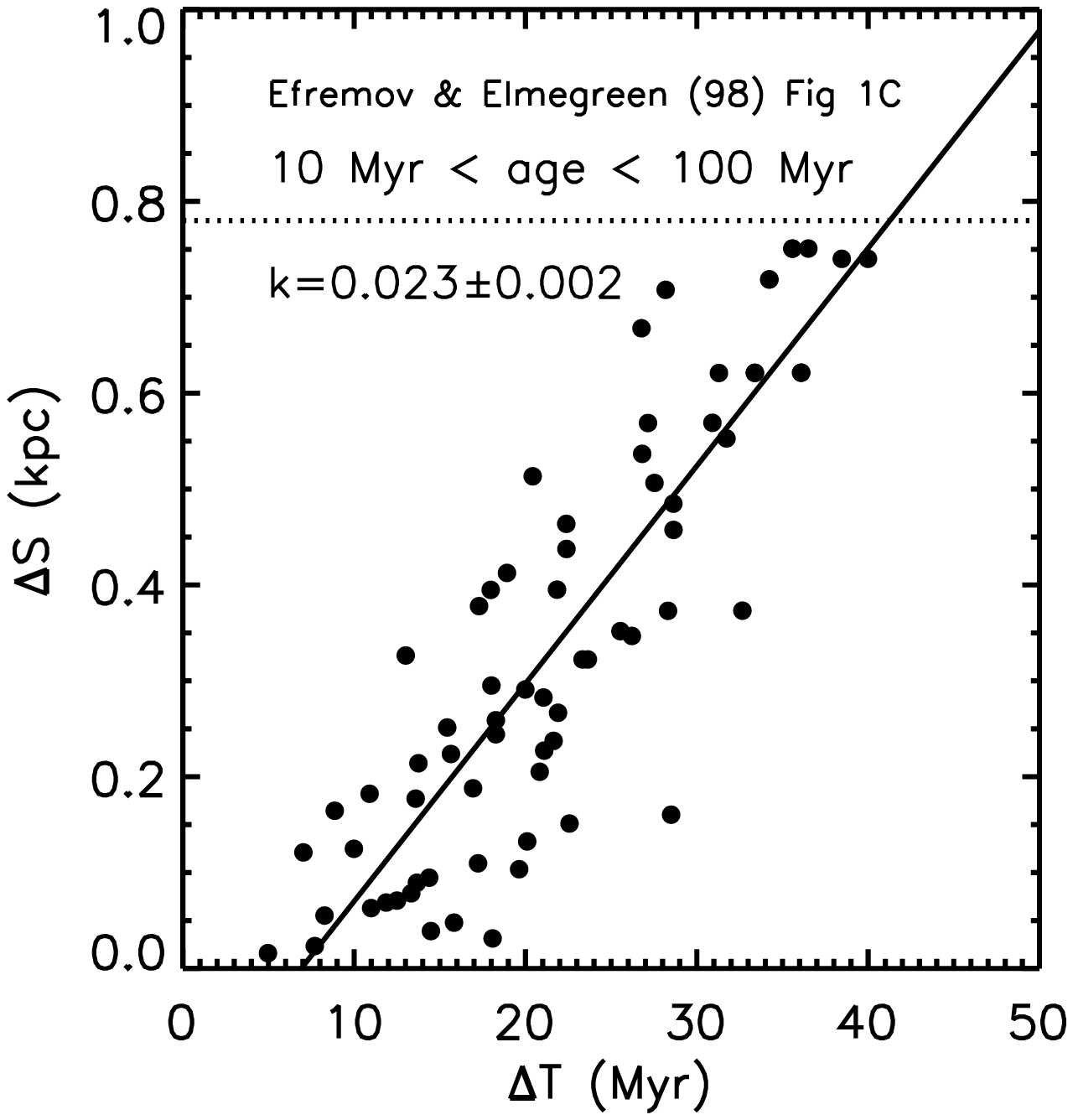}\\
\includegraphics[width=2.5in]{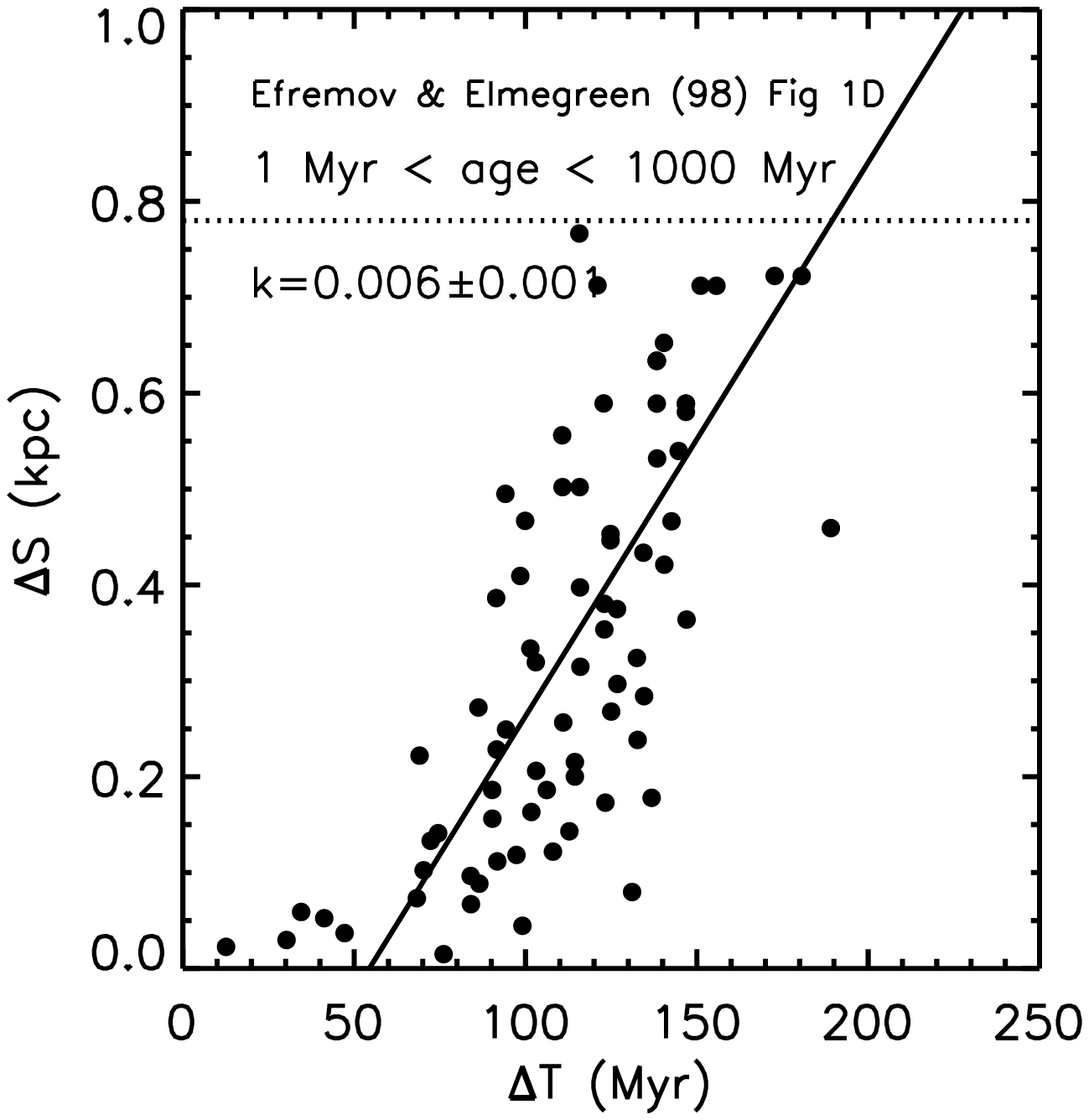}
\caption{\label{fig_lmc} Linear correlation between separations of
  cluster pairs $\Delta S$ in the Large Magellanic Cloud and their age
  differences $\Delta T$, reproduced from \citet{elmegreen98}. {\em a:}
  For 1~Myr~$< T < 100$~Myr, {\em b:} for 10~Myr~$< T < 100$~Myr, and
  {\em c:} for 1~Myr~$< T < 1000$~Myr. Only data with $\Delta S \le
  0.78$~kpc, corresponding to~$1^{\circ}$ (and indicated by the dotted
  line), is plotted here to maintain consistency with the original
  fits. }
\end{center}
\end{figure}

For comparison with our models, the data used by \citet{elmegreen98}
is replotted on a {\em linear} scale in Figure~\ref{fig_lmc}.  Their
$\Delta S$ is converted from degrees to kpc in the original plots
(Figure 1 of \citealt{elmegreen98}), by using the mean distance to the Large
Magellanic Cloud of 45 kpc given in the same paper. Only data with $\Delta S
\le 1^{\circ}$ (corresponding to 0.78 kpc) is plotted here to maintain
consistency with the original fits. We see that this data can be well fit by a
linear correlation, similarly to our model results.

\begin{figure}
\begin{center}
\includegraphics[width=2.5in]{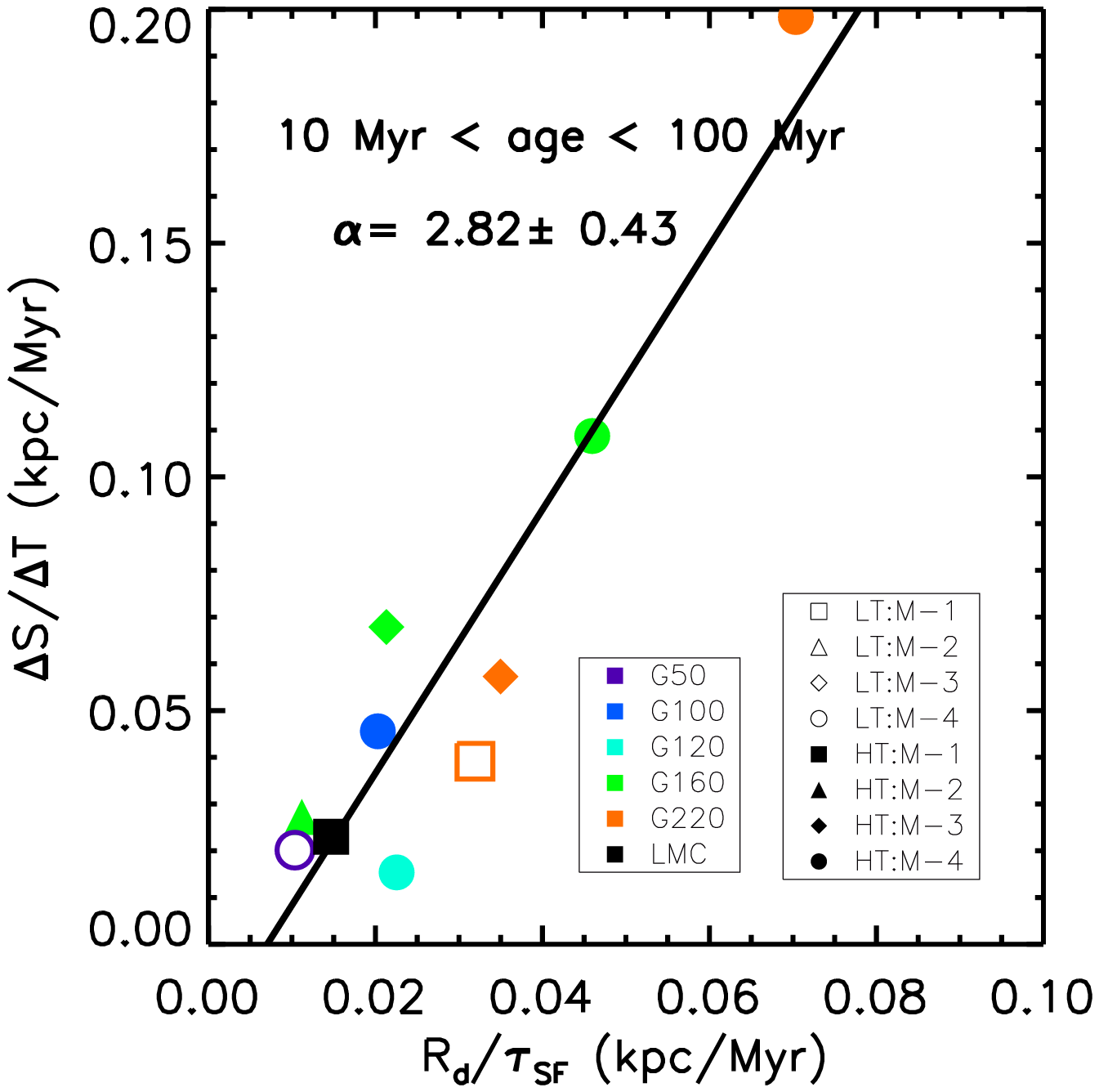}\\
\includegraphics[width=2.5in]{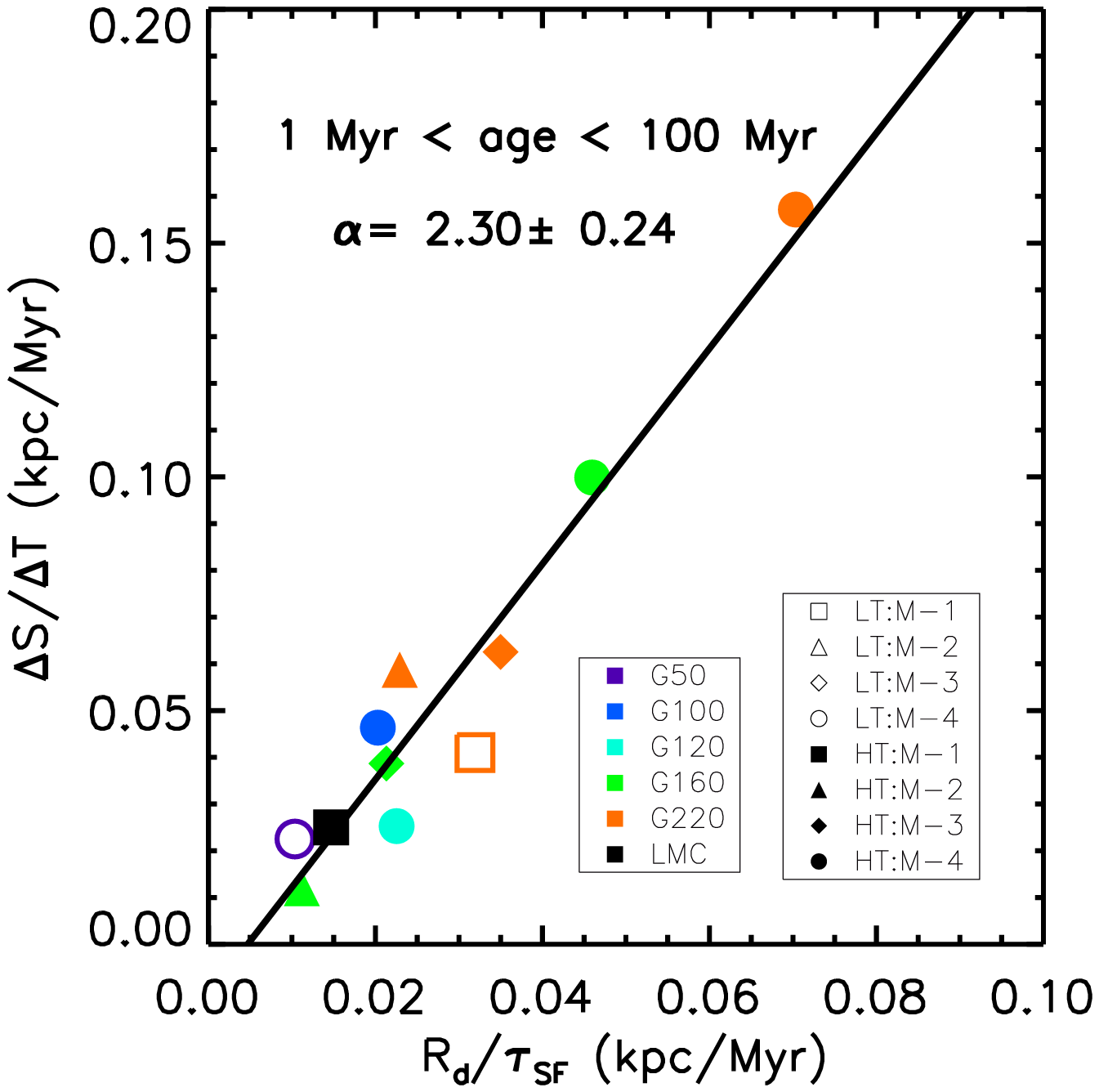}\\
\includegraphics[width=2.5in]{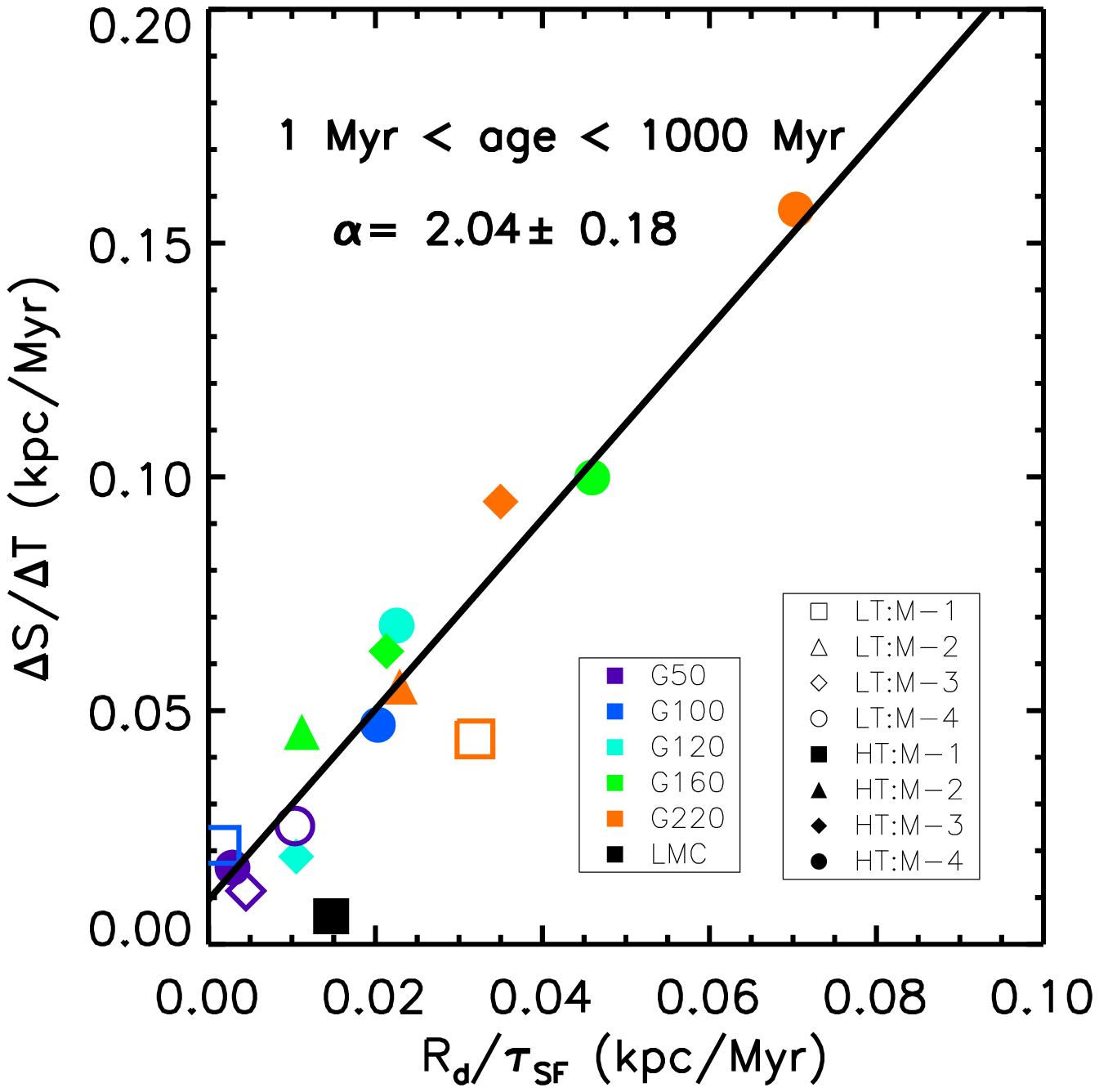}
\caption{\label{fig_clustering2} The correlation between $\Delta S/\Delta T$
  and the galaxy properties $R_{\rm d}/\tau_{\rm SF}$, for clusters with
different ages $T$ in all models. {\em a:} For 10~Myr~$< T < 100$~Myr, 
{\em b:} for 1~Myr~$< T  < 100$~Myr, and {\em c:} for 1~Myr~$< T <
1000$~Myr. The black line is the least-square fit to the data, while the
black square represents the fits shown in Figure~\ref{fig_lmc} to the 
observed star clusters in the Large Magellanic Cloud from
\citet{elmegreen98}. } 
\end{center}
\end{figure}

Our models suggest that a linear $\Delta S$--$\Delta T$ correlation is
the natural result of gravitational collapse in a differentially rotating disk. 
It occurs for all ages of clusters in all of our models, as is demonstrated in
Figure~\ref{fig_clustering2}. This Figure further shows that the linear slope
of the $\Delta S$--$\Delta T$ relation is directly proportional to the star
formation rate $\tau_{\rm SF}^{-1}$ derived in \S\ref{sec_tsf} multiplied by
the disk scale length $R_{\rm d}/\tau_{\rm SF}$. The stronger the instability,
the more frequently clusters form, and so the greater distance $\Delta S$ one
must traverse to find clusters with a given age difference $\Delta T$.

The Large Magellanic Cloud has a halo mass of $M_{\rm
  halo} \approx 1.6 \times 10^{10} M_{\odot}$, gas mass  $M_{\rm 
gas} \approx 1 \times 10^9 M_{\odot}$, and radial disk scale length $R_{\rm d}
= 1.47$ kpc \citep{marel01, alves04}. It is close to our model G50-4,
so we estimate it to have a comparable star formation timescale
$\tau_{\rm SF}$ for the current episode of $\sim 100$ Myr.  With this
estimate, our fits to the data drawn from \citet{elmegreen98} fall on the
correlation derived from our models strikingly well, as shown in
Figure~\ref{fig_clustering2}. (Note that even a substantially larger value 
of $\tau_{\rm SF}$ would not markedly change this conclusion.) 

These correlations survive among clusters as old as 1~Gyr, as shown in
Figure~\ref{fig_clustering2}{\em (c)}.  This suggests that the observed
$\Delta S$--$\Delta T$ correlation is simply the result of gravitational
interaction. Interstellar turbulence in the galactic disk cannot
determine cluster positions as long as 1~Gyr after cluster formation.
The mixing time---the generalization to supersonic flow of the
eddy turn-over time---is an order of magnitude shorter
(\citealt{avillez02,klessen03}), so turbulent structures will be
uncorrelated on those timescales.

Unstable galaxies have a high slope of $\Delta S$ versus $\Delta T$,
which appears to imply fast star formation, while more stable galaxies
have slower star formation and smaller slopes.

\section{DISCUSSION}
\label{sec_dis}

The simulations presented here model the nonlinear development of
gravitational instability in isolated galaxies.  They represent the
response of the galactic disks to the initial conditions imposed in
each case, but do not directly represent the star-formation history
of real galaxies interacting with their environment. Nevertheless, we
think our models yield useful insights into the behavior of real
galactic disks. Furthermore, in the simulations presented here, we do not
include explicit feedback, magnetic fields, or gas recycling from star-forming
regions, all of which will affect the detailed star formation history and gas
evolution, so they will eventually have to be considered. However, we
believe each will have minor effects on the questions considered here,
for reasons that we now describe.

The assumption of an isothermal equation of state for the gas actually implies
substantial feedback to maintain the effective temperature of the gas against
radiative cooling and turbulent dissipation. Real interstellar gas has a wide
range of temperatures. However, the rms velocity dispersion generally falls
within the range 6--12 km s$^{-1}$ (e.g., \citealt{elmegreen04}). Direct
feedback from the starburst may play only a minor role in quenching subsequent
star formation \citep[e.g.][]{kravtsov03,monaco04}, perhaps because most
energy is deposited not in the disk but above it as superbubbles blow out
\citep[e.g.][]{fujita03,avillez04}. At least three effects may conspire to
maintain the velocity dispersion in the observed narrow range. First,
radiative cooling drops precipitously in gas with sound speed $\lesssim$ 10 km
s$^{-1}$ as it becomes increasingly difficult to excite the Lyman $\alpha$
line of hydrogen. Second, the energy input $\dot{E}$ from supernovae at the
observed Galactic rate drives a flow with rms velocity dispersion of $\sim$
9.5 km s$^{-1}$ (\citealt{avillez00}), and the rms velocity dispersion depends
only on $\dot{E}^{1/3}$ (\citealt{maclow99, mk04}), so a wide range of star
formation rates leads to a narrow range of velocity dispersions.  Third,
magnetorotational instabilities may maintain a velocity dispersion of order
6~km~s$^{-1}$ even in the absence of any other feedback (\citealt*{sellwood99,
  dzi04}).  

\citet{kim01} have demonstrated that swing and magneto-Jeans
instabilities operating in a gaseous disk occur at a Toomre instability
criterion of $Q_{\rm g} \approx 1.4$, suggesting that magnetostatic support is
unimportant. Another longstanding question is how fragmentation can proceed in
the presence of a magnetic field. Material collects from considerable distance
in our models. If it is constrained to flow along field lines, collapse can
still occur in so long as the mass-to-flux ratio remains supercritical
along each field line. So long as the local mass-to-flux ratio remains
supercritical, the magnatic field can retard the contraction but not
prevent it (\citealt{heitsch01, vazquez05}).

In our simulations, we do not include any gas recycling so that sink particles
form until the gas is consumed. However, in reality there are at least three
sources of gas that ultimately need to be taken into account. First, in our
picture, the bulk of the cold, molecular gas, which is primarily contained in
giant molecular clouds, is already gravitationally unstable and either forming
stars or being dispersed on a free-fall time \citep*{hartmann01}. Some 
fraction of that dispersed mass may rejoin the diffuse, gravitationally stable
gas, while the rest will participate within tens of megayears in further star
formation not taken into account by our assumed SFE. Second, the massive stars
in newly-formed clusters evolve quickly, and at the end of their lives eject
part of their mass back into the interstellar gas, supplying a gas reservoir
for the next cycle of star formation until all the mass is locked up in low
mass stars or compact objects. Finally, galaxies are not isolated objects, but
accrete further mass from interactions and the intergalactic medium.  Our
models do not take this into account, but form the basis for models that do.
For a first example, see \citet*{li04b}.

There have been many models of the formation of spiral structure
(e.g., \citealt*{lin66, roberts69, toomre77, sellwood84, binney87,
  elmegreen03}) that offer a general picture that spiral structure is
a combination of sheared features from star formation and density
waves in the gas and stars. The wide variety of spiral structure
observed in different galaxies is due to the wide range of the swing
amplifier that depends on the response of the galactic disk to the
perturbation from gravitational instability, tidal forces, or bar
formation. Grand-design spiral structures are usually thought to be
produced by perturbations or interactions (e.g.,
\citealt{kormendy79}), as seen also in our simulations of galaxy
mergers using models similar to the ones described here \citep{li04b}.

\citet{elmegreen93} showed grand design spiral arms forming in
isolated galaxies in 2D simulations with resolutions from 42 $\times$
64 to 84 $\times$ 128 grid cells. However, their models assume a very
large disk mass fraction $m_{\rm d}$ = 0.2, so that the rotation curve
falls sharply at large radii, rather than remaining constant as
is typically observed. Moreover, their model uses a very cold disk with
a hot stellar population in the central part to produce a Q-barrier.
This produces a highly unstable disk that forms stars in a violent starburst. 
Their simulations using galaxy models having parameters more typical of normal 
spirals, and without the Q-barrier, show flocculent features only.
Furthermore, their inability to resolve the Jeans length probably induces
spurious fragmentation and artificial spiral arms \citep{truelove97}, as we
demonstrated in Figure~\ref{fig_rose}. Our 3D simulations with high
resolution, and more realistic galaxy models with $m_{\rm d} \le 0.1$
and flat rotation curves (shown in Figure~\ref{fig_3q}) show clear
flocculent arms.  \citet{elmegreen03} proposed a turbulent origin for
flocculent spiral arms, but our results suggest that gravitational
instability in the absence of external perturbations can fully explain them.

Our resolution study demonstrates that the gas density distribution, the
number of fragments and the collapsed mass converge sufficiently well when
$N_{\rm k} \ge 2$, supporting the validity of the numerical criteria used in 
our simulations. Simulations of isolated, isothermal disks by
\citet{robertson04} show large-scale collapse in the galactic center leading
to disks far smaller than observed, which they argued was caused by
the isothermal equation of state. Our model G100-1 has parameters close to
theirs but with an order of magnitude higher particle number that resolves the
Jeans length as suggested by the \citet{bate97} criterion.  It does not show
this behavior.  Similarly, \citet{governato04} argue that several
long-standing problems in galaxy simulations, such as compact disks, overly
efficient star formation, and lack of angular momentum may well be caused by
inadequate resolution, or violation of the other numerical criteria.

\section{SUMMARY}
\label{sec_sum}

To summarize our work, we have simulated star formation with
sufficient resolution to resolve gravitational collapse in
models ofa wide  range of disk galaxies with different total mass, gas
fraction, and initial gravitational instability. 

The gravitational collapse or star formation timescale $\tau_{\rm SF}$ depends
exponentially on the initial Toomre instability parameter for the combination
of stars and gas in the disk $Q_{\rm sg}$ given by \citet{rafikov01}.
The tight correlation between $\tau_{\rm SF}$ and $Q_{\rm sg}(min)$
suggests that the global star formation rate in isolated disk galaxies is
controlled by the nonlinear development of gravitational
instability. Quiescent star formation occurs where $Q_{\rm sg}$ is large, 
while vigorous starbursts occur where $Q_{\rm sg}$ is small. Star 
formation begins when the gas become unstable, with a rate controlled by $Q_{\rm
sg}(min)$. Galaxies with high initial mass or gas fraction are the most
unstable, forming stars quickly. The number of star clusters formed also
appears to depend on the strength of instability. 

Our models show that stars form in gravitationally unstable regions as defined
by $Q_{\rm sg}$. The star formation region in strongly unstable galaxies
extends multiple disk scale lengths in radius, but is concentrated to 
less than a tenth of a scale height vertically, while in marginally unstable
galaxies it concentrates radially in the central scale length, but extends for
more than half a scale height vertically. These results directly agree with
the observations of \citet{dalcanton04} that unstable galaxies show thin dust
lanes, while more stable galaxies show thick dust distributions. 

Star clusters do not form uniformly but rather are distributed so that their
spatial separation $\Delta S$ and age difference $\Delta T$ correlate linearly
with each other. The slope of the correlation depends directly on the star
formation rate multiplied by the disk scale length $R_{\rm d}/\tau_{\rm
  SF}$. Observations of Large Magellanic Cloud clusters by \citet{elmegreen98} 
quantitatively agree with this dependence. 

In future work we will study global and local Schmidt laws, star
formation thresholds, and the mass distribution of the clusters that
form in our models.

\acknowledgments We thank V. Springel for making both GADGET and his
galaxy initial condition generator available, as well as for useful
discussions, A.-K. Jappsen for participating in the implementation of
sink particles in GADGET, and F. Adams, J. Dalcanton, B. Elmegreen,
R. Kennicutt, J. Lee, C. Martin, R. McCray, T. Quinn, M. Shara, and J. van
Gorkom for useful discussions. We also thank the referee for valuable comments
that have helped to improve the manuscript.
This work was supported by the NSF under grants
AST99-85392 and AST03-07854, by NASA under grant NAG5-13028, and by the
Emmy Noether Program of the DFG under grant KL1358/1.  Computations were
performed at the Pittsburgh Supercomputer Center supported by the NSF, on the
Parallel Computing Facility of the AMNH, and on an Ultrasparc III cluster
generously donated by Sun Microsystems.

\clearpage
\begin{deluxetable}{lcccccc}
\label{tab1}
\tablecolumns{7}
\tablecaption{Galaxy Models}
\tablehead{\colhead{Model} & \colhead{$V_{200}$\tablenotemark{a}} &
\colhead{$R_{200}$\tablenotemark{b}}&
\colhead{$M_{200}$\tablenotemark{c}}& \colhead{$m_{\rm
d}$\tablenotemark{d}} & \colhead{$f_{\rm g}$\tablenotemark{e}} &
\colhead{$R_{\rm d}$\tablenotemark{f}}}  
\startdata
G50-1  &   50.0   & 71.43  &  4.15  & 0.05 & 0.2  &  1.41  \\
G50-2  &   50.0   & 71.43  &  4.15  & 0.05 & 0.5  &  1.41  \\
G50-3  &   50.0   & 71.43  &  4.15  & 0.05 & 0.9  &  1.41  \\
G50-4  &   50.0   & 71.43  &  4.15  & 0.10 & 0.9  &  1.07  \\
 & & & & & \\   	 			   
G100-1 &  100.0   & 142.86 &  33.22 & 0.05 &  0.2 &  2.81  \\
G100-2 &  100.0   & 142.86 &  33.22 & 0.05 &  0.5 &  2.81  \\
G100-3 &  100.0   & 142.86 &  33.22 & 0.05 &  0.9 &  2.81  \\
G100-4 &  100.0   & 142.86 &  33.22 & 0.10 &  0.9 &  2.14  \\	
 & & & & & \\	          	 			   
G120-3 &  120.0   & 171.43 &  57.4 & 0.05  &  0.9 &  3.38   \\
G120-4 &  120.0   & 171.43 &  57.4 & 0.10  &  0.9 &  2.57   \\
 & & & & & \\	          	 			   
G160-1 &  160.0 6 & 228.57 &  136.0 & 0.05 & 0.2  &  4.51  \\
G160-2 &  160.0 6 & 228.57 &  136.0 & 0.05 & 0.5  &  4.51  \\
G160-3 &  160.0 6 & 228.57 &  136.0 & 0.05 & 0.9  &  4.51  \\ 
G160-4 &  160.0 6 & 228.57 &  136.0 & 0.10 & 0.9  &  3.42  \\ 	
& & & & & \\			                  
G220-1 &  220.0 0 & 314.29 &  353.7 & 0.05 & 0.2  &  6.20  \\
G220-2 &  220.0 0 & 314.29 &  353.7 & 0.05 & 0.5  &  6.20  \\
G220-3 &  220.0 0 & 314.29 &  353.7 & 0.05 & 0.9  &  6.20  \\
G220-4 &  220.0 0 & 314.29 &  353.7 & 0.10 & 0.9  &  4.71  \\
\enddata 
\tablenotetext{a}{Rotational velocity in km s$^{-1}$ at virial radius
$R_{200}$.} 
\tablenotetext{b}{Virial radius in kpc where the overdensity is 200.}
\tablenotetext{c}{Virial mass of the galaxy in 10$^{10} M_{\odot}$.}
\tablenotetext{d}{Fraction of total halo mass in disk.}
\tablenotetext{e}{Fraction of disk mass in gas.}
\tablenotetext{f}{Radial disk scale length in kpc where stellar
  surface density drops by $e^{-1}$.}
\end{deluxetable}

\clearpage
\begin{deluxetable}{lccccc}
\label{tab2}
\tablecolumns{6}
\tablecaption{Numerical Parameters 
for High Temperature Models\tablenotemark{a}}
\tablehead{\colhead{Model} & \colhead{$N_{\rm tot}$\tablenotemark{b}} &
\colhead{$\epsilon_{\rm g}$\tablenotemark{c}} &
\colhead{$m_{\rm g}$\tablenotemark{d}} & \colhead{$Q_{\rm
sg}$(HT)\tablenotemark{e}} & \colhead{$N_{\rm k}$(HT)\tablenotemark{f}}}  
\startdata
G50-1  & 1.0   & 10 &  0.08  &  1.45  &  465.8 \\
G50-2  & 1.0   & 10 &  0.21  &  1.53  &  177.8 \\
G50-3  & 1.0   & 10 &  0.37  &  1.52  &  100.7 \\
G50-4  & 1.0   & 10 &  0.75  &  0.82  &   49.7 \\ 
 & & & & & \\
G100-1 & 1.0   & 10 &  0.66  &  1.27  &  56.5 \\
G100-2 & 1.0   & 10 &  1.65  &  1.07  &  22.6 \\
G100-3 & 1.0   & 10 &  2.97  &  0.82  &  12.5 \\
G100-4 & 1.0   & 20 &  5.94  &  0.42  &  6.3  \\
 & & & & & \\
G120-3 & 1.0   & 20 &  5.17  &  0.68  &  7.2  \\
G120-4 & 1.0   & 30 &  10.3  &  0.35  &  3.6  \\     	  	     
 & & & & & \\
G160-1 & 1.0   & 20 &  2.72  &  1.34  &  13.7 \\
G160-2 & 1.0   & 20 &  6.80  &  0.89  &  5.5  \\
G160-3 & 1.0   & 30 &  12.2  &  0.52  &  3.1  \\ 
G160-4 & 1.5   & 40 &  16.3  &  0.26  &  2.3  \\         	  	     
 & & & & & \\
G220-1 & 1.0   & 20 &  7.07  &  1.11  &  5.3  \\
G220-2 & 1.2   & 30 &  14.8  &  0.66  &  2.5  \\
G220-3 & 2.0   & 40 &  15.9  &  0.38  &  2.3  \\
G220-4 & 4.0   & 40 &  16.0  &  0.19  &  2.3  \\
\enddata 
\tablenotetext{a}{For models with effective sound speed $c_s = 15$ km
s$^{-1}$}
\tablenotetext{b}{Total particle number divided by $10^6$.}
\tablenotetext{c}{Gravitational softening length of gas in pc.}
\tablenotetext{d}{Gas particle mass in units of $10^4 M_{\odot}$.} 
\tablenotetext{e}{Minimum initial $Q_{\rm sg}$ for high $T$ model}
\tablenotetext{f}{Kernel number $N_{\rm k}$ for high $T$ model, $M_{\rm J} =
1.49\times 10^7 M_{\odot}$. }
\end{deluxetable}

\clearpage
\begin{deluxetable}{lccccc}
\label{tab3}
\tablecolumns{6}
\tablecaption{Numerical Parameters
for Low Temperature Models\tablenotemark{a}}
\tablehead{\colhead{Model} & \colhead{$N_{\rm tot}$} & \colhead{$\epsilon_{\rm
g}$} & \colhead{$m_{\rm g}$} & \colhead{$Q_{\rm sg}$(LT)} & \colhead{$N_{\rm
k}$(LT)}} 
\startdata
G50-1  & 1.0   & 10 &  0.08  &  1.22  & 29.8 \\
G50-2  & 1.0   & 10 &  0.21  &  0.94  & 11.4 \\
G50-3  & 1.0   & 10 &  0.37  &  0.65  & 6.4  \\
G50-4  & 1.0   & 10 &  0.75  &  0.33  & 3.2  \\          	    
G100-1 & 6.4   & 7  &  0.66  &  1.08  & 24.0 \\          			    
G220-1 & 6.4   & 15 &  7.07  &  0.65  & 2.3  \\
\enddata 
\tablenotetext{a}{Same as Table~2 but for low temperature models with $c_{\rm
s}$ = 6 km s$^{-1}$, the corresponding Jeans mass is $M_{\rm J} = 9.54\times
10^5 M_{\odot}$.} 
\end{deluxetable}

\clearpage
\begin{deluxetable}{lcc}
\label{tab4}
\tablecolumns{3}
\tablecaption{Resolution Study}
\tablehead{\colhead{$N_{\rm k}$} & \colhead{G100-1} &
\colhead{G220-1}}  
\startdata
R1  & 0.38  & 0.035 \\
R8  & 3.0   & 0.28  \\
R64 & 24.0  & 2.3
\enddata 
\end{deluxetable}

\clearpage

\clearpage

\clearpage

\clearpage

\clearpage

\clearpage

\clearpage

\clearpage

\clearpage

\clearpage

\clearpage

\clearpage

\clearpage

\clearpage

\clearpage

\clearpage

\clearpage

\clearpage

\clearpage

\clearpage

\clearpage

\clearpage

\end{document}